\documentclass[sigconf,nonacm]{acmart}
\def\GerryPreprintClassLoaded{1}
\ifdefined\GerryPreprintClassLoaded\else
\documentclass[sigconf,nonacm]{acmart}
\fi
\usepackage{booktabs,tabularx,graphicx,microtype,etoolbox,needspace,float,xurl,longtable,tikz,alphalph}
\hypersetup{hidelinks,pdfauthor={Jiayuan Liu, Mingyu Guo, Vincent Conitzer},pdftitle={Feedback Gerrymandering: Steering Learned Judgments with Honest Feedback},pdfsubject={Human feedback collection and learned judgments}}
\graphicspath{{figures/}}
\usetikzlibrary{arrows.meta,positioning}
\newcommand{\F}{\mathcal F}

\newtheorem{theorem}{Theorem}
\floatstyle{ruled}\newfloat{algorithm}{tbp}{loa}\floatname{algorithm}{Algorithm}

 \setcopyright{none}
\title{Feedback Gerrymandering: Steering Learned Judgments with Honest Feedback}
\newcommand{\GerryAuthorLine}{\textbf{Jiayuan Liu}\textsuperscript{1,2}, \textbf{Mingyu Guo}\textsuperscript{3}, \textbf{Vincent Conitzer}\textsuperscript{1,2}}
\author{\GerryAuthorLine}
\renewcommand{\authors}{Jiayuan Liu, Mingyu Guo, Vincent Conitzer}
\affiliation[obeypunctuation=true]{\institution{\textsuperscript{1}Carnegie Mellon University}\\
  \institution{\textsuperscript{2}Foundations of Cooperative AI Lab (FOCAL)}\\
  \institution{\textsuperscript{3}Adelaide University}\city{}\country{}}
\renewcommand{\shortauthors}{Liu, Guo, and Conitzer}
 
\begin{document}
\begin{abstract}
A model trained to represent a community's judgments depends on who is asked to evaluate which items. When participant time is limited, each person can judge only a subset. Does preserving everyone's workload and every item's number of judgments prevent the collector from steering what the model learns? We study \emph{feedback gerrymandering}: assigning people to items using past judgments before their current judgments are known. All judgments remain truthful, and the learning procedure stays fixed. Experiments on conversational-AI safety and perceived hazards in videos show that assignment choices can change learned scores and decisions. Controls that reassign histories between people or shuffle their association with content identify which historical information enables this influence. In the video-rating task, fixing some assignments through a rule that does not use judgments reduces the gap between models steered toward higher and lower scores, without reducing participation. The effects of content quotas and score calibration depend on the task. These interventions respectively restrict each person's mix of items and adjust model scores using separate reference judgments. Theory and recorded-feedback audits characterize attainable training targets and reassignment costs. For organizations that learn from community feedback, participation counts describe how much people contribute. Auditing whose judgments shape each kind of content also requires examining their assignments.
 \end{abstract}
\maketitle
\hypersetup{pdfauthor={Jiayuan Liu, Mingyu Guo, Vincent Conitzer},pdfsubject={Human feedback collection and learned judgments}}
\pdfinfo{/Author (Jiayuan Liu, Mingyu Guo, Vincent Conitzer)}
\section{Introduction}
\label{sec:introduction}
Imagine an organization asking community members to judge the safety of an AI assistant's responses. It wants a model that reflects the community's judgments of new responses. There are many responses, but each participant has limited time and can judge only a subset. The organization fixes each person's workload and specifies how many evaluators, and from which demographic groups, must judge each response. A \emph{collector} assigns people to tasks. A fixed learning procedure trains a \emph{shared scorer}, which predicts judgments from response content. Could the collector steer this model through assignment alone, while everyone judges truthfully and training proceeds unchanged?

Consider two participants who can each judge one response, about health or politics. Past judgments suggest Alice is more likely than Bob to label health responses unsafe, and Bob more likely than Alice to label political responses unsafe. Initially, Alice judges politics and Bob health. If both are eligible for either task and share a demographic category, exchanging their tasks preserves participation counts. If these patterns persist, the exchange makes unsafe judgments more likely on both responses. Both participants remain truthful. A scorer trained on these judgments may assign higher unsafe scores to new responses.

\begin{figure*}[t]
\centering
\includegraphics[width=\textwidth]{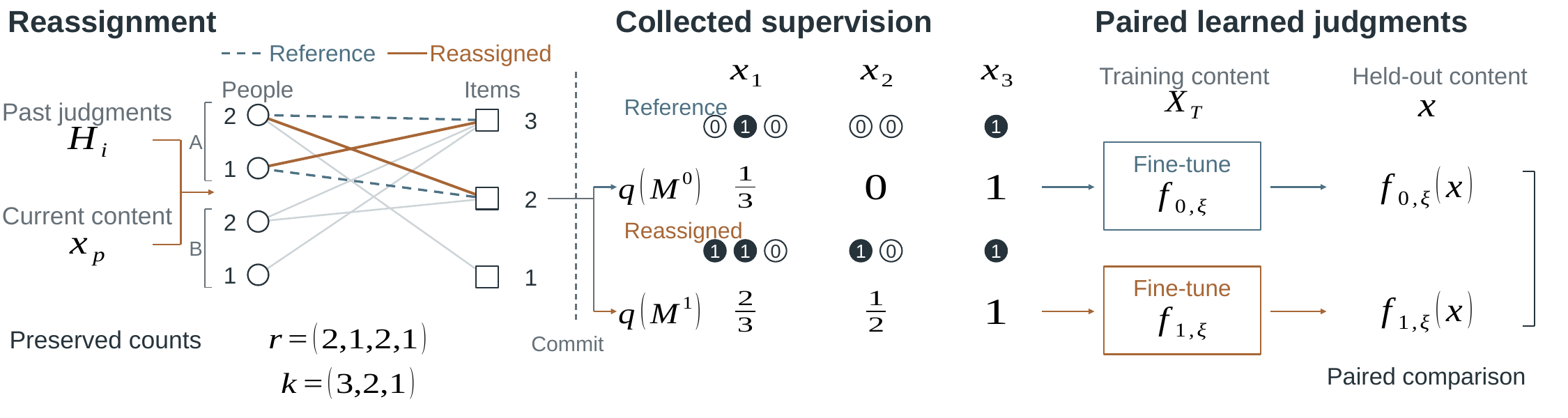}
\caption{\textbf{Preserved participation, different supervision.} Reference $M^0$ and reassignment $M^1$ share gray edges. Blue and ochre edges exchange two tasks within group A, preserving workloads $r=(2,1,2,1)$, panel sizes $k=(3,2,1)$, and each item's A/B counts. Person $i$'s history $H_i$ and item $p$'s content $x_p$ guide selection before current judgments are known. Selected people provide their own truthful judgments, yielding the illustrated $q(M^0)=(1/3,0,1)$ and $q(M^1)=(2/3,1/2,1)$. These train separate scorers $f_{j,\xi}:=f_{M^j,\xi}$, $j\in\{0,1\}$, with common content $X_T$, recipe, initial weights and randomness $\xi$. Evaluation uses the same held-out content $x$. The example specifies training targets, while Sections~\ref{sec:effects}--\ref{sec:governance} measure learned effects.}
 \label{fig:teaser}
\Description{A reference assignment and a history-guided reassignment preserve each person's and each item's original, nonuniform counts. After the assignment is fixed, the selected truthful judgments yield different training targets. Each target vector trains a separate shared scorer under the same learning procedure. Evaluation compares the scorers on the same held-out items.}
\end{figure*}

An organization may check participation counts to ensure that the collected feedback reflects the community. What can this audit establish if it leaves task assignments unchecked? Human judgments train language models to follow instructions and evaluate generated responses \cite{ouyang2022instructions}. Collective-input initiatives seek to represent a wider range of voices in training \cite{huang2024collective}. Social-choice approaches ask how those voices should be combined \cite{conitzer2024socialchoice,ge2024axioms}. The collector controls an earlier decision, determining whose judgment is obtained for each item. A model can therefore follow the intended training procedure while learning from systematically different community input.

We call strategic control of these assignments \emph{feedback gerrymandering}. Electoral gerrymandering changes voter grouping to influence outcomes \cite{dey2023priced}, while information gerrymandering changes interactions through which opinions influence decisions \cite{stewart2019information}. Here, a \emph{panel} is the set of people assigned to one item. People and items stay fixed, but each person evaluates only their assigned subset. Reassignment preserves workloads, panel sizes and required composition. The collector controls assignments, not judgments or training.

The collector chooses assignments before anyone judges the current items. It can use past judgments and the content of the new items, but not their unknown judgments. People's preferences vary with content \cite{zhang2025diverging}, so history can help predict whom to ask. Our \emph{history-residual predictor} first predicts judgments from the pooled historical data. It then adjusts each person's prediction using how their past judgments differed from those predictions on similar items. This auxiliary predictor guides assignment and is separate from the shared scorer. Once assignments are fixed, the mean of the selected people's truthful judgments on each item becomes the shared scorer's training target (Figure~\ref{fig:teaser}).

We test the mechanism offline using DICES judgments of conversational-AI safety \cite{dices} and Road ratings of perceived hazards in videos \cite{song2024road}. Each assignment selects existing judgments to simulate limited participation. We keep each model type's training recipe fixed and compare assignments using shared training states. Historical replay comparisons are identified separately. Public reference assignments obey the same constraints without using judgments. Both datasets test \emph{upward} and \emph{downward} routing, intended to raise or lower unsafe scores on DICES and hazard scores on Road. DICES effects vary across designs and reference panels, sometimes changing scores without changing decisions. Road scores move in both intended directions.

An organization may delegate scheduling to a collector who knows participants' availability and expertise. Public reservation constrains this discretion by fixing some assignments through a rule that does not use judgments. The collector fills the remaining slots under the participation requirements. On Road, partial reservation brings the upward- and downward-trained models' scores closer together at unchanged workloads and panel sizes. A second rule groups items by content similarity and fixes how many items each person evaluates \emph{within each group}. Control rules preserve these counts and group sizes while moving items between groups. Comparing the rules tests whether grouping similar content helps constrain the collector. We also test content quotas on text and, separately, score adjustments fitted using independent reference judgments.

Supporting theory and recorded-feedback audits characterize which majority targets truthful assignments can realize and the cost of preserving workloads while reaching them. Earlier work studies truthful omission, partial elicitation, annotator selection, and network-biased feedback \cite{barash2021omission,dey2018,acal,fan2026graphpref}. We connect these collection choices to the behavior of a shared model through three questions: what assignments can achieve and cost, which historical information enables influence, and when collection rules constrain it. For organizations that use community feedback to judge new content, the resulting audit target is concrete: who is asked to evaluate what, as well as how much each person contributes.
 \section{Collection setting and experimental protocol}
\label{sec:protocol}
In this section, we formalize the collector's assignment choices, the information available to it, and the offline comparisons used to measure effects on the shared scorer.
\subsection{Assigning truthful evaluators under fixed participation counts}
Let $I$ be the set of evaluators and $T$ the set of training items (questions). We index evaluators by $i$ and items by $p$, and write $x_p$ for an item's content. The set $\Omega\subseteq I\times T$ specifies allowed person--item assignments. For each eligible pair, $y_{ip}\in[0,1]$ is the judgment person $i$ would truthfully give item $p$. $Y=(y_{ip})$ denotes these judgments. An assignment $M=(m_{ip})$ has $m_{ip}=1$ when person $i$ is assigned to item $p$, and zero otherwise. A feasible reference assignment $M^0=(m^0_{ip})$ fixes each person's workload $r_i\ge0$ and each item's panel size $k_p>0$. Where composition requirements apply, $\mathcal G$ partitions people into known groups, or \emph{strata}. Let $g(i)$ be person $i$'s group and $k_{pg}$ item $p$'s reference count from group $g$. The allowed assignments form $\mathcal F$, preserving these counts:
\begin{equation}
\begin{aligned}
\sum_p m_{ip}&=\sum_p m^0_{ip}=r_i &&\forall i,\\
\sum_i m_{ip}&=\sum_i m^0_{ip}=k_p &&\forall p,\\
\sum_{i:g(i)=g}m_{ip}&=\sum_{i:g(i)=g}m^0_{ip}=k_{pg} &&\forall (p,g)\text{ specified}
\end{aligned}
\label{eq:quotas}
\end{equation}
All sums use eligible pairs in $\Omega$. Panel size does not determine group composition. The last line preserves each specified group count. Specifying all groups' counts also fixes their total. ``Honest'' refers to evaluators' truthful judgments. Balanced reassignments preserve these reference counts. Section~\ref{sec:governance} also restricts particular assignments or each person's counts within groups of \emph{items}.

\paragraph{Information and learning.} We use separate item sets for history $H$, training $T$, calibration $C$ where available, and evaluation $E$. The collector uses judgments on $H$ and current item content to choose assignment $M$ before seeing judgments on $T$. After training, reference judgments on $C$ are used to adjust scores. Evaluation on $E$ measures outcomes under the fixed scoring rule. The selected judgments on each training item have mean
\begin{equation}
q_p(M)=\frac{1}{k_p}\sum_{i:(i,p)\in\Omega}m_{ip}y_{ip}.
\label{eq:supervision}
\end{equation}
Each mean is the item's training target. Let $X_T=(x_p)_{p\in T}$ denote the training items' content, $q(M)=(q_p(M))_{p\in T}$ their targets, and $\mathcal A_\xi$ the common learning procedure with initial state and randomness $\xi$. It fits the shared scorer $f_{M,\xi}=\mathcal A_\xi(X_T,q(M))$. Comparisons at the same $\xi$ share that state and randomness. The scorer learns from content and mean judgments, without evaluator IDs. At evaluation it receives only new item content. Training uses positive-judgment fractions or mean ratings directly, rather than majority labels.

\subsection{Historical routing and controls}
The history-residual predictor combines a population prediction with a personal adjustment. Kernel ridge regression uses content similarity to fit the population prediction, with a penalty that limits model complexity. To estimate personal adjustments, we divide historical items into two sets. For each set, we fit a population predictor on the other set and predict judgments of the held-out items. An evaluator's recorded judgment minus this prediction is their \emph{residual}. For a current item, the personal adjustment combines that evaluator's residuals on similar historical items. It shrinks toward zero when little similar history is available. Appendix~\ref{app:method-walkthrough} gives the algorithm. Let $\widehat y_{ip}$ predict evaluator $i$'s judgment of item $p$. Upward ($+$) and downward ($-$) routing respectively maximize and minimize the total predicted feedback selected:
\begin{equation}
M^{\pm}\in\arg\max_{M\in\mathcal F}\;\pm\sum_{(i,p)\in\Omega}m_{ip}\widehat y_{ip}.
\label{eq:routing}
\end{equation}
This objective sums predictions over selected person--item pairs. Equivalently, it weights each item's predicted mean by its panel size. Predictions guide assignment, while the assigned people's \emph{real} judgments train the separate shared scorer. Its scores on held-out items measure the effect after learning.

A constant added to all predictions for person $i$ enters the objective $r_i$ times under every assignment. A constant added to all predictions for item $p$ similarly enters $k_p$ times. These contributions are fixed, so they leave the maximizing assignment unchanged. Selection therefore depends on how individual judgments vary with content. Appendix~\ref{app:quota-algebra} gives this identity for the actual optimized coefficients and explains clipping and additional content quotas.

Public baselines maximize the sum of independent random weights on eligible pairs under the same constraints, without using judgments. This differs from sampling feasible assignments with equal probability. Original DICES comparisons use two public assignments, an expanded comparison six, and Road four. The identity control predicts with another person's complete history, swapping histories within DICES gender groups or the Road cohort. The selected person's own current judgment still trains the scorer. A second control (Section~\ref{sec:history}) keeps each person's residuals but shuffles which historical items they correspond to.

\subsection{Data, learners, and evaluation}
\begin{table*}[t]
\centering
\caption{Recorded-feedback designs. $H/T/C/E$ counts items, not judgments. A dash denotes no independent calibration set. Each design preserves its listed per-person workload and per-item panel size across conditions. Panel composition specifies how many evaluators from each demographic group answer each item. S1 and S2 are two fixed splits of DICES350. BASE's training items are a subset of LARGER's.}
\label{tab:designs}
\begin{tabular}{llclll}
\toprule Data/design & People & $H/T/C/E$ & Judgments/person & Judgments/item & Panel composition\\\midrule
DICES350 S1/S2 &104&100/104/--/146&3&3&Gender quotas\\
DICES990 BASE &144&100/144/96/361&3&3&Locale$\times$gender\\
DICES990 LARGER &144&100/432/96/361&9&3&Locale$\times$gender\\
Road &48&288/288/192/588&18 / 36&3 / 6&No demographic quota\\\bottomrule
\end{tabular}
\end{table*}
\paragraph{Simulating limited participation.} Our datasets contain each person's recorded judgments of many items. We simulate limited participation by selecting, for each person, judgments on only an assigned subset of the training items. These subsets are chosen jointly to preserve workloads, panel sizes and applicable group counts in Equation~\ref{eq:quotas}. The collector uses history and current item content, with current, calibration and evaluation judgments hidden. Once assignment $M$ is fixed, only the selected current judgments are retrieved as training targets, unchanged from their recorded values. Appendix~\ref{app:method-walkthrough} explains the access checks.

DICES records disagreement over conversational-AI safety \cite{dices}. Its filtered 350-item release provides a complete judgment matrix for 104 evaluators. In each fixed split, S1 or S2, 104 people each judge three of the 104 training items, with three judgments per item: 312 selected judgments and 10,400 historical judgments. Yes maps to UNSAFE (one) and No/Unsure to zero. DICES990 has missing person--item judgments, so only recorded pairs are eligible. It includes 144 people, 6,499 observed historical judgments and 432/1,296 selected current judgments in BASE/LARGER. The designs share $H,C,E$ but differ in training content, workload and available assignments (Table~\ref{tab:designs}).

Road contains all 48 people's perceived-hazard ratings of 1,356 video excerpts \cite{song2024road}, normalized to $[0,1]$. The assignment selects each person's ratings of 18 of the 288 training videos, with three people per video: 864 selected ratings, alongside 13,824 historical ratings and 576 calibration judgments (three per $C$ item, denoted C3). Evaluation uses 219 groups defined by video source. Each file is represented by four frames decoded from its first 333\,ms. This input convention and one filename/source mapping are documented in Appendix~\ref{app:road-calibration}. All conditions use the same inputs and grouping.

DICES fine-tunes a Qwen3.5-2B scorer with low-rank adaptation (LoRA), which represents weight updates as products of two smaller matrices. Binary cross-entropy (BCE) penalizes disagreement between the predicted UNSAFE probability and each item's positive-judgment fraction. A separate comparison uses Mistral-7B-Instruct-v0.1. Road adapts the final stages of ResNet50 and ConvNeXt and trains a single-score output layer. Both reuse the same ResNet-based history predictions, holding the assignment procedure fixed. A training \emph{stream} specifies initialization, optimizer, schedule, data order, and random state. Fully paired comparisons restore these values across conditions. Appendix~\ref{app:scope-ledger} distinguishes these comparisons from historical replays with incomplete state records. Evaluation reads scores from the trained models, with their parameters fixed. Each text model assigns a score to a supplied response. Appendix~\ref{app:training-overview} gives the recipes and distinguishes selected training judgments from historical information.

\paragraph{Measuring influence.} We compute each model's metrics before taking means across runs or public assignments. For text, $z_{M,\xi}(x)$ is the logit, or log-odds of an unsafe judgment, and $z_{M,\xi}(x)>0$ flags an item as UNSAFE. Road uses the rating $f_{M,\xi}(x)$, with $f_{M,\xi}(x)>0.5$ as a secondary high-hazard flag. Rate differences are percentage points (pp). Road score differences use a 100-point scale. Each directional comparison subtracts the named reference condition: upward routing aims for positive changes, and downward routing aims for negative changes. These measure influence on learned judgments. Prediction quality measures agreement with reference judgments and is evaluated separately. Model outputs before calibration are called raw scores. Appendix~\ref{sec:closure-statistics} defines uncertainty and error measures, and Appendix~\ref{app:scope-ledger} identifies the paired comparisons.
 \section{Empirical effects on learned judgments}
\label{sec:effects}
Can the choice of panels change a model that learns from honest feedback? We compare history-guided and public assignments, measuring the resulting models' predictions on held-out items and their agreement with reference judgments.

\subsection{Conversational safety: shifts and reference-panel variation}
\begin{table}[t]
\centering
\caption{DICES upward history-guided routing versus six additional public assignments, at unsafe logit $>0$. Intervals resample conversation groups, whole training streams and the six fixed public assignments, using the same draws across conditions. Appendix~\ref{app:public-history} separates the initial two-public comparisons from later replays in the additional publics' training environment.}
\label{tab:dices}
\begin{tabular}{lcr}
\toprule Design & Streams & Flag change (pp) [95\% interval]\\\midrule
S1 &2&$+63.13\ [48.48,76.65]$\\
S2 &2&$+5.02\ [0.05,12.70]$\\
BASE &3&$+8.88\ [5.02,13.05]$\\
LARGER &3&$+6.11\ [-2.58,14.06]$\\\bottomrule
\end{tabular}
\end{table}
History-guided collection changes unsafe-flag rates while preserving participation and panel composition (Table~\ref{tab:dices}). All four estimates relative to the six-public-assignment mean follow the intended upward direction. S1 has the largest shift, and BASE's interval is entirely positive. S2's percentile interval has a near-zero lower endpoint. Its statistical test is sensitive to how uncertainty is calculated, as detailed in Appendix~\ref{app:public-history}. LARGER also has a positive estimate, with an interval that includes zero.

\paragraph{Why compare several public assignments?} Preserving every person's and item's count still permits many public assignments. In LARGER, the initial two-public comparison is $-5.54\ [-16.99,3.62]$ pp. Replaying the history-guided shared scorer in the additional publics' training environment changes the contrast by only $+0.092$ pp, while changing the public reference contributes $+11.557$ pp. The expanded estimate therefore compares history-guided routing with a different set of reference panels. Both comparisons are reported in Appendix~\ref{app:public-history}. Across the six additional public assignments, models disagree on 5.48\%, 0.23\%, 1.48\%, and 13.31\% of decisions on average in S1, S2, BASE, and LARGER. Each model appears in several pairwise comparisons, so those comparisons are statistically dependent.

\paragraph{A second language-model learner.} Mistral-7B-Instruct-v0.1 yields shifts of $+36.99\ [28.32,46.05]$ pp on S1 and $+2.40\ [0.04,6.81]$ pp on S2 against the original two-public reference, using two training streams shared across conditions. Both tested text learners show the intended direction, with estimates reported against their respective public references. Appendix~\ref{app:original-scopes} also reports the additional splits and their controls.

\paragraph{Directional influence and predictive quality.} In the original S1 comparison, public and history-guided flag rates are 15.64\% and 72.37\%. In S2 they are 0\% and 5.02\%. The receiver operating characteristic (ROC) curve plots the fraction of reference-UNSAFE items correctly flagged against the fraction of reference-SAFE items incorrectly flagged as the decision threshold varies. Its area (AUROC) summarizes how well scores rank UNSAFE items above SAFE items. Higher values indicate better ranking. In the original DICES990 comparisons, history-guided minus public AUROC is $-0.0597$ for BASE and $-0.0499$ for LARGER. These negative AUROC differences quantify the ranking-quality cost of changing which judgments train the model. Appendix~\ref{sec:histogram-effects} reports AUROC values, losses, error rates, and scoring-rule sensitivity.

\subsection{Video hazard ratings: steering scores in both directions}
On Road, we test both upward and downward assignments using the same recorded ratings and fixed panel sizes. With ConvNeXt and three-person panels, upward routing raises raw predicted ratings by $3.58\ [2.43,4.85]$ points relative to the four-public mean. Downward routing lowers them by $3.46\ [2.57,4.43]$ points. The brackets give 95\% confidence intervals. The corresponding flag changes are $+12.47$ and $-9.47$ pp (Table~\ref{tab:road}).

\begin{table*}[t]
\centering
\caption{Road ConvNeXt with three-person panels. Signed method-minus-comparator effects use a 100-point rating scale or flag percentage points. Intervals resample video-source groups and training streams. Public comparisons also resample the reference assignments. Both down contrasts are negative in the intended direction.}
\label{tab:road}
\begin{tabular}{llrr}
\toprule Direction & Comparator & Score points [95\% interval] & Flag change (pp) [95\% interval]\\\midrule
Up & Public mean & $+3.58\ [2.43,4.85]$ & $+12.47\ [8.64,16.93]$\\
Down & Public mean & $-3.46\ [-4.43,-2.57]$ & $-9.47\ [-13.69,-6.01]$\\
Up & Identity control & $+4.83\ [3.67,6.12]$ & $+15.14\ [11.13,19.64]$\\
Down & Identity control & $-4.19\ [-5.08,-3.31]$ & $-14.12\ [-17.83,-10.67]$\\\bottomrule
\end{tabular}
\end{table*}

ConvNeXt's public models predict hazard more accurately than assigning every item the historical mean rating: mean squared error (MSE, lower is better) is 0.04189 versus 0.04965. Up/down MSEs are 0.04848/0.04242. At the same 0.5 threshold, upward routing identifies more reference-high-hazard items but also increases false positives. Downward routing reduces false positives but misses more reference-high-hazard items.

Both routing directions are tested for ResNet and ConvNeXt with panels of 3 and 6 people (Appendix~\ref{app:road-historypanel22}). With six-person panels, ConvNeXt shifts upward by $+3.87\ [2.96,4.90]$ points and downward by $-2.47\ [-3.11,-1.88]$ relative to the public mean. Public MSE is 0.04178. The original ResNet pipeline gives $+5.28\ [3.72,6.83]$ and $-3.72\ [-4.99,-2.40]$ points at panel three, while its public MSE is 0.05304. The comparison spans complete learning pipelines and feedback budgets. The principal governance analysis uses ConvNeXt, whose public models have lower MSE than ResNet's.
 \section{Which historical information enables routing?}
\label{sec:history}
Which information makes history-guided routing effective: whose history is used, how much is linked to that person, or which content their judgments concern? We vary these links while keeping all current judgments attached to their actual owners.

\paragraph{Linking history to the right person.} The identity control gives each evaluator another person's complete history before selecting panels. On Road, using the correct histories rather than this control changes scores by $+4.83\ [3.67,6.12]$ points for upward routing and $-4.19\ [-5.08,-3.31]$ for downward routing (Table~\ref{tab:road}). In both conditions, the shared scorer receives item content and mean ratings, without evaluator IDs. The collector's assignments determine which real ratings it learns from. Appendix~\ref{app:road-identity21} reports the additional fixed history reassignments and calibrated comparisons.

\paragraph{Fixed historical data, varying personal information.} DICES S1 keeps the anonymous pooled judgments on 100 historical items (H100) fixed. We vary how many of each evaluator's records are linked to their identity, from 10 to 100. This increases the UNSAFE-flag rate by $+48.86\ [25.34,70.55]$ pp. The same comparison changes false-positive and false-negative rates by $+48.12$ and $-51.19$ pp, and soft-label BCE by $+0.0683\ [0.0157,0.1093]$. Linking more of the existing history to individuals therefore increases directional influence and changes both error types (Appendix~\ref{app:gaps-text}).

On Road, we hold the pooled history of 288 items (H288), the public models, and the initial ConvNeXt state fixed. Each person's linked records follow one fixed order, with the first 16 contained in the first 64 and then all 288. Define \emph{separation} as the mean upward-model score minus the mean downward-model score on held-out items. Before calibration, it is 2.31, 4.91, and 7.04 points. Before learning, the mean human rating selected by the upward assignment minus that selected by the downward assignment is 2.28, 4.06, and 7.41 points. Both increase along this sequence on the 100-point scale. At 16 records, the intervals comparing correct and reassigned identities include zero (Appendix~\ref{app:road-historypanel22}).

\paragraph{Keeping the person, changing historical content correspondence.} Reassigning whole histories also changes which evaluator appears to give consistently higher ratings. A narrower control keeps each person's history but shuffles which item is associated with each residual, the difference between a past judgment and its population prediction. Shuffling occurs separately within the two historical sets used in Section~\ref{sec:protocol}. Every residual value and its frequency are preserved. The content features, population predictions and current ratings are unchanged. The predictor also retains its rules for shrinking personal adjustments toward zero and clipping forecasts to the allowed range. The two shuffles are content maps A and B in Table~\ref{tab:hc-main}, fixed before examining model outputs.

\begin{table}[t]
\centering
\caption{Historical content correspondence at fixed feedback counts (Road, ConvNeXt/panel 3). Score separation is upward minus downward on the 100-point scale. Conditions keep identities and personal residual distributions. 95\% intervals resample evaluation groups and paired streams, conditional on the fixed maps.}
\label{tab:hc-main}
\begin{tabular}{lr}\toprule
History & Score points [95\% interval]\\\midrule
Intact & $+7.04\ [5.42,8.89]$\\
Content map A & $-0.89\ [-1.68,-0.15]$\\
Content map B & $+0.42\ [-0.63,1.42]$\\
Mean of A and B & $-0.23\ [-0.95,0.39]$\\\midrule
Intact $-$ mean & $+7.27\ [5.28,9.56]$\\\bottomrule
\end{tabular}
\end{table}

Table~\ref{tab:hc-main} reports both maps and their mean. Correctly linking residuals to content produces 7.27 points more separation than the control mean, with a 95\% interval of $[5.28,9.56]$. The difference is positive in all three training streams. Retaining the link between residuals and their historical content enables this additional separation, with each person's residual distribution held fixed. Appendix~\ref{app:gaps-context} reports stream-level and calibrated comparisons.

\paragraph{From prediction to collected supervision.} A forecast matters through the real judgment it helps select. In this Road design, everyone is eligible for every item and workloads are equal. Reassigning whole histories only permutes the people associated with each item's forecasts. It preserves the best assignment objective, including the integer rounding used by the solver, while changing whose real judgment is selected. Appendix~\ref{app:stage-diagnosis} compares the predicted and actual gaps in collected judgments and attributes their difference to prediction error on the selected pairs. This traces how historical information changes supervision before affecting the shared scorer.
 \section{Governing the destination of feedback}
\label{sec:governance}
We next ask whether rules about who evaluates what can reduce directional influence while preserving every participant's workload and every item's panel size. The interventions restrict the collector's available assignments, with the learning recipe fixed. Scoring-time calibration is evaluated separately.

\subsection{Public reservation and content coverage}
\begin{table*}[t]
\centering
\caption{Public reservation at the same annotation count. Raw up--down score separation on Road, with 95\% source-group/paired-stream intervals. Each learner is compared with and without reservation using the same training streams. Intervals for the change are computed from the paired differences, not by subtracting the separate interval endpoints.}
\label{tab:reservation-core}
\begin{tabular}{lrrr}\toprule
Learner & No public reservation & 50\% public reservation & Paired change (rating points)\\\midrule
ResNet50 & $9.00\ [7.02,11.04]$ & $3.47\ [1.95,4.99]$ & $-5.52\ [-7.65,-3.61]$\\
ConvNeXt & $7.04\ [5.42,8.89]$ & $2.76\ [1.73,3.89]$ & $-4.28\ [-5.12,-3.48]$\\\bottomrule
\end{tabular}
\end{table*}

\paragraph{Reserving public assignments.} A public rule that does not use judgments fixes half the person--item pairs in an original public assignment. The collector completes the remaining assignments. Each new pair must fit both the evaluator's remaining workload and the item's remaining panel slots. Table~\ref{tab:reservation-core} measures the resulting gap between the upward- and downward-trained models. The gap falls for both learners, with all 864 judgments still collected. Appendix~\ref{app:reservation-material} explains how the quotas restrict the remaining choices.

\paragraph{Content quotas and a structurally equivalent control.} The \emph{feature rule} groups Road items into four buckets by content similarity, using features from historical items. Each evaluator judges the same number of items from each bucket as in the original public assignment. A \emph{sham rule} is a control that moves items between buckets while preserving bucket sizes and each person's bucket counts. Every feasible feature-rule assignment has a corresponding feasible sham assignment. Videos, judgments and predictions remain on their original items. The comparison therefore changes which content is grouped together while preserving the constraint structure.

The feature rule gives 4.99 points of separation versus 6.54 under the original sham, a paired difference of $-1.55\ [-2.64,-0.57]$. A separately fixed comparison against two additional shams is $-2.02\ [-2.89,-1.24]$. In this Road design, grouping similar content reduces separation relative to the sham rules. The remaining separation measures the directional influence still available under each rule. Appendix~\ref{app:conditional-road} reports the individual shams, quality, and scoring rules within each comparison's original scope.

\subsection{Assignment rules and historical information}
\begin{figure}[t]
\centering
\includegraphics[width=\linewidth]{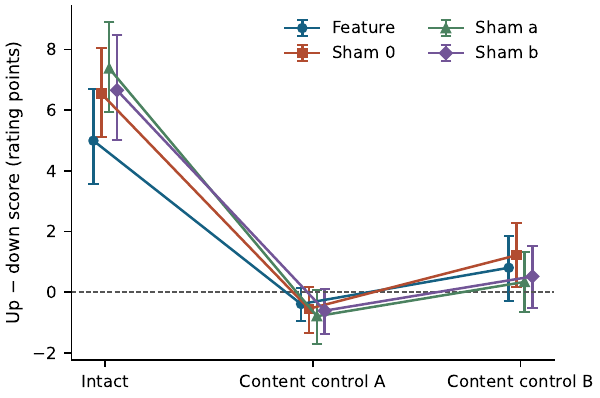}
\caption{Road raw up--down separation for each assignment rule and history condition, with 95\% intervals. Appendix~\ref{app:final-iteration} gives the numerical values and the interaction calculated from these conditions.}
\label{fig:history-policy}
\Description{The feature assignment rule has lower separation than the sham assignment rules with intact history. This difference changes under both historical-content maps.}
\end{figure}

On Road, does requiring the collector to satisfy content-group quotas reduce score separation more when personal histories retain their original content correspondence? An \emph{assignment rule} specifies item groups and how many items each evaluator must judge in each. We test the feature and three sham assignment rules with intact history and both within-person content maps, without public reservation. Let $\rho$ index assignment rules and $\eta\in\{\mathrm{intact},\mathrm{HC}_A,\mathrm{HC}_B\}$ index history conditions. The separation $S_\rho(\eta)$ is the mean upward-model score minus the mean downward-model score, on the 100-point scale. Subscripts $0,A,B$ identify the sham rules. We first compare the feature rule with the sham rules under each history condition. We then compare that difference between intact and shuffled histories:
\begin{equation}
\begin{aligned}
D(\eta)&=S_{\mathrm{feature}}(\eta)-\tfrac13\!\sum_{j\in\{0,A,B\}}S_{\mathrm{sham}\,j}(\eta),\\
I_R&=D(\mathrm{intact})-\tfrac12[D(\mathrm{HC}_A)+D(\mathrm{HC}_B)].
\end{aligned}
\label{eq:interaction}
\end{equation}
The feature--sham difference is $-1.87$ points under intact history, compared with $+0.26$ and $+0.11$ under the content maps. The change in this difference, $I_R$, is $-2.05\ [-3.36,-0.85]$ points and is negative in all three streams (Table~\ref{tab:joint}). Figure~\ref{fig:history-policy} shows the underlying up--down separations. Content-based quotas reduce separation more when residuals remain linked to their original historical content. This interaction includes the change in predictive information caused by shuffling that link.

\paragraph{Content quotas on text.} Both DICES350 splits use two content buckets defined from history alone. The shams exchange bucket memberships only between items with the same panel-size, demographic and eligibility requirements, preserving each person's bucket counts. Each split's primary, $D_{S1}$ or $D_{S2}$, is the feature-rule up--down unsafe-logit separation minus the mean separation under the two shams. This comparison keeps histories intact. Road additionally varies historical-content correspondence to estimate the interaction $I_R$.

\begin{table*}[t]
\centering
\caption{The three prespecified governance comparisons. Negative values indicate reduced up--down separation. Intervals are pointwise 95\% intervals for the fixed cohorts and maps. Joint statistical tests are reported in Appendix~\ref{app:final-iteration}.}
\label{tab:joint}
\begin{tabular}{llr}
\toprule Contrast & Unit & Estimate [95\% interval]\\\midrule
Road history--assignment rule $I_R$ & Rating points & $-2.05\ [-3.36,-0.85]$\\
Text feature--sham $D_{S1}$ & Unsafe logit & $+0.113\ [-0.010,0.273]$\\
Text feature--sham $D_{S2}$ & Unsafe logit & $+0.175\ [-0.109,0.359]$\\\bottomrule
\end{tabular}
\end{table*}
The text point estimates are $+0.113$ and $+0.175$ logits, with intervals spanning zero (Table~\ref{tab:joint}). Secondary comparisons show greater feature-rule separation: $+23.74\ [7.88,39.43]$ pp for S2 flags and $+0.032\ [0.006,0.066]$ for S1 probability. Before training, the upward-minus-downward positive-judgment counts under feature/sham-A/sham-B are 36/39/33 for S1 and 52/47/39 for S2. The feature rule thus leaves S1's gap equal to the sham mean and increases S2's. These supervision differences help explain why the effects of these assignment rules depend on how the chosen content groups align with real disagreement. History-guided routing and the rules intended to constrain it answer separate questions. S1 and S2 have no separate calibration set, so their scores remain uncalibrated.

\subsection{Collection policy and scoring policy}
Road scores can be used raw or adjusted using independent calibration judgments. Mean-offset calibration shifts each model's scores so that its mean on $C$ equals a common reference. For two conditions, let $\Delta_E^{\mathrm{raw}}$ and $\Delta_C^{\mathrm{raw}}$ be their raw mean-score differences on $E$ and $C$, and $\Delta_E^{\mathrm{cal}}$ their difference after this adjustment. With common within-set weights and no clipping,
\[
\Delta_E^{\mathrm{cal}}=\Delta_E^{\mathrm{raw}}-\Delta_C^{\mathrm{raw}}.
\]
With ConvNeXt/panel 3, mean offsets retain flag contrasts of $+5.63\ [2.51,9.80]$ and $-3.84\ [-6.67,-0.90]$ pp relative to the public-model mean. Affine calibration also fits a nonnegative scale factor and gives $-1.08\ [-5.32,3.89]$ and $-1.98\ [-6.33,3.69]$ pp. All four panel-by-direction affine intervals include zero. Appendix~\ref{app:road-retention21} gives both rules, both panel sizes, and prediction quality.

The Road interaction becomes $+1.16\ [-1.04,3.44]$ after per-model offsets and $+1.14\ [-1.04,3.14]$ after affine calibration. Each rule uses $C$ alone and is refitted within saved $C$ resamples. Calibration changes the measured interaction from a raw-score reduction to a positive point estimate with an interval spanning zero. Public models continue to flag some items as high hazard under both calibration rules. Appendix~\ref{app:final-iteration} reports prediction errors and the fraction of items on which decisions differ under these scoring rules.
 \section{Feasibility and the cost of honest influence}
\label{sec:costs}
\paragraph{Direct majorities and learned decisions.} How much can assignment alone change, and at what cost? We develop theory for strict majority: more than half the selected judgments support the desired answer on each binary item or candidate pair. These majorities summarize collected judgments. A shared scorer can predict differently because its parameters and pretraining link learning across items. \emph{Attainable supervision} is the set of per-item mean judgments $q(M)$ that feasible assignments can supply. Theory studies their majority patterns and costs, while experiments measure the resulting learned scores (Appendix~\ref{app:assumption-map}).

\paragraph{Reachability before cost.} Fix a feasible reference assignment $M^0$ and a binary vector $t=(t_p)_{p\in T}$ specifying every item's desired strict majority. $C_{\mathrm{flip}}(M^0,t)$ is the fewest selected binary judgments that must be flipped, at fixed $M^0$, to realize all of $t$. Reassignment preserves truthful judgments and minimizes the number of newly selected pairs:
\[
C_{\mathrm{bal}}(M^0,t)=\min_{\substack{M\in\mathcal F:\,M\text{ realizes }t}}
\tfrac12\|M-M^0\|_1,
\]
where the norm sums absolute entry differences and an infeasible minimum is $+\infty$. A two-person task exchange costs two. Strict majority needs $h_p=\lfloor k_p/2\rfloor+1$ supporters. Let $a_p^0$ be the number of initially selected judgments supporting $t_p$. The additional support needed is $v_p=\max\{h_p-a_p^0,0\}$, giving $C_{\mathrm{flip}}=\sum_pv_p$. Reassignment may be impossible because too few eligible people support the target, or their assignments cannot jointly meet the quotas. Our \emph{audits} are offline optimization checks using all recorded judgments, including current labels hidden from the history-guided collector (Appendix~\ref{app:intervention-cost}).

\paragraph{Costs under the original data constraints.} We solve 120 fixed DICES problems under the original eligibility and quotas. Four designs each use eight predefined majority targets and both public starts, P1 and P2 (64 problems). Each design also supplies 12 fixed items. Starting from P1, each local problem reverses one item's majority and preserves all others (48). The remaining eight reproduce a history-guided assignment's per-item positive counts, as described below. There are 29 certified infeasibilities and 91 reachable cases, including four with zero cost. In the 87 positive-cost cases, minimum reassignment cost divided by minimum flip cost lies between one and two, using the same target for both interventions. All 43 reachable one-item changes have $C_{\mathrm{bal}}/C_{\mathrm{flip}}=2$.

Exact-supervision costs, $C^{\mathrm{exact}}_{\mathrm{flip}}$ and $C^{\mathrm{exact}}_{\mathrm{bal}}$, minimize flips and replacements while matching \emph{every item's positive count}, not just its majority. With panel sizes fixed, positive counts determine the mean-judgment training targets $q_p$. Equal majorities can still give different targets. Each design supplies one target count vector from its upward history-guided assignment. Starting from P1 and P2 gives eight cost problems. Six have equal minima (Table~\ref{tab:exact-cost-core}). A separate audit asks whether honest assignment can reproduce 16 count vectors obtained by editing judgments. The eight S1/S2 vectors are reachable at costs of 63--66 edges, whereas the eight BASE/LARGER vectors are infeasible. Appendix~\ref{app:honest-realizability} reports all cases, and Appendix~\ref{app:cost-learning} connects these targets to learner inputs.

\begin{table}[t]
\centering
\caption{Eight exact-supervision cost problems within the 120-case audit: four designs, each with one history-guided target and two public starts. Each cell lists the minimum costs from P1/P2 to that design's target. Costs count binary flips or newly selected truthful pairs.}
\label{tab:exact-cost-core}
\begin{tabular}{lrr}
\toprule Design & $C^{\mathrm{exact}}_{\mathrm{flip}}$ & $C^{\mathrm{exact}}_{\mathrm{bal}}$\\\midrule
S1 &94/95&94/95\\
S2 &84/73&84/73\\
BASE &77/80&83/84\\
LARGER &252/284&252/284\\\bottomrule
\end{tabular}
\end{table}

\paragraph{Matching counts versus matching training inputs.} The learner uses each item's content and training target together. Swapping targets between items preserves how often each target value occurs, but changes the content--target pairs used for training. An offline selector uses current judgments to maximize the selected positive total within a replacement budget. Two truthful reassignment controls match its actual cost and the number of items undergoing each initial-to-final count change, but assign these changes to different items. The selector-minus-control differences in held-out UNSAFE-flag rate are $-4.34$, $-0.46$, $0.00$, and $+4.34$ pp for S1, S2, BASE, and LARGER. All four primary intervals include zero. BASE has $2.03\ [0.89,3.36]$\% gross decision disagreement at zero mean net change, and LARGER has $7.94\ [4.61,12.07]$\%. Items gaining and losing flags can cancel in the net rate. Appendix~\ref{app:honest-realizability} gives the four-test family, calibrated quality, and editing controls.

We reuse a run only when both interventions match every item's target, content, weight and order, the complete initial state, and the numerical recipe and runtime (Appendix~\ref{app:cost-learning}). Conditions passing these checks are \emph{exact-input aliases}: they share one training computation and reuse its resulting model.

\paragraph{Capacity even with uniform counts.} Can the same people's honest answers yield different collective judgments solely through assignment? Consider $s\ge3$ alternatives and $d=\binom{s}{2}$ pairs. Each person has a fixed strict ranking and can compare every pair. We count only answers to directly assigned pairs. Fix a reference ranking $R$. A target tournament $t=(t_e)_{e=1}^d$ specifies the desired majority direction on every pair: $t_e=+1$ agrees with $R$, and $t_e=-1$ reverses it. These directions may form a cycle. A pair's signed margin counts answers agreeing with $R$ minus those opposing it.
\needspace{8\baselineskip}
\begin{theorem}[Balanced direct-query realization]\label{thm:new-balanced}
Let $L=\lceil d/2\rceil$ and $N=d(4L+1)$. For any $R$, there exists a profile $P$ of $N$ strict rankings with both properties: (i) its full-population margin is $d$ on every pair, and (ii) for every workload $r\in\{1,\ldots,d-1\}$ and target $t$, an assignment gives each person exactly $r$ distinct pairs and each pair exactly $k=r(4L+1)$ truthful judgments, with selected margin $2Lt_e+r$ on pair $e$.
\end{theorem}
Since $2L\ge d>r$, this margin has sign $t_e$. As in classical majority constructions \cite{mcgarvey1953}, rankings are constructed to control pairwise margins. Here the same $P$ also supports every target and incomplete workload under exact counts, including $r=d-1$. Appendix~\ref{app:balanced-proof} proves this joint property and states the margin and transitive-inference boundaries.

\needspace{9\baselineskip}
\paragraph{Preserving workloads can require many exchanges.} The next theorem shows how restoring workloads can make replacement costs grow far beyond flip costs.
\begin{theorem}[Unbounded cost of balance]\label{thm:unbounded-cost}
For every integer $m\ge2$, there exists a complete-support binary-feedback instance with $3m$ evaluators and $3m$ items, workload and panel size three, and a feasible starting assignment from which an all-one-majority target is attainable with $C_{\mathrm{flip}}=3$ and $C_{\mathrm{bal}}=3m$. Thus $C_{\mathrm{bal}}/C_{\mathrm{flip}}$ has no instance-independent constant upper bound even when both interventions can reach the target.
\end{theorem}
The proof is in Appendix~\ref{app:unbounded-proof}.

\paragraph{Constructed schedules allow local exchanges.} Write $M_r(u)$ for the \emph{canonical schedule} produced by Theorem~\ref{thm:new-balanced}'s construction for target $u\in\{-1,+1\}^d$ and workload $r$. Its helper pairs allow task exchanges that restore workloads locally. Other balanced assignments with majority $u$ need not retain those pairs.
\begin{theorem}[Tight bound for the constructed schedules]\label{thm:cost-of-balance}
For the fixed electorate of Theorem~\ref{thm:new-balanced}, starting from $M_r(u)$ with $1\le r<d$ under its complete eligibility and exact row/panel counts, every target tournament $t$ satisfies
$C_{\mathrm{flip}}\le C_{\mathrm{bal}}\le2C_{\mathrm{flip}}$.
The upper bound is attained when $u$ and $t$ differ on one comparison.
\end{theorem}
Appendix~\ref{app:canonical-proof} gives the proof, a sufficient condition for local exchanges, and a noncanonical start in the same 27-person electorate with costs one and three. The factor-two guarantee therefore depends on the constructed starting schedule, not just on preserving participation counts.
 \section{Related work}
\label{sec:related}
\paragraph{Feedback collection and aggregation.} Human-feedback alignment learns from judgments \cite{ouyang2022instructions,huang2024collective}, while social-choice approaches aggregate differing preferences \cite{conitzer2024socialchoice,ge2024axioms,kleinebuening2025strategyproof}. DICES and preference studies document this variation \cite{dices,zhang2025diverging}. ACAL selects annotators using histories and content, trains soft-label predictors, and evaluates DICES \cite{acal}. HyPER routes human versus AI feedback, PILAF selects comparisons, and CoPL learns personalized preferences \cite{hyper,feng2025pilaf,choi2025copl}. Our collector uses history strategically under exact workload, panel and composition constraints.

\paragraph{Gerrymandering and electoral control.} Electoral gerrymandering changes voter grouping, whereas information gerrymandering changes what participants observe \cite{dey2023priced,stewart2019information}. Electoral control can add, delete or partition voters or candidates \cite{bartholdi1992hard}. Bribery changes reported votes \cite{faliszewski2009microbribery}. We instead preserve participants, items and their counts. Dey studies the selection of compatible partial reports from a fixed profile under minimum pairwise coverage \cite{dey2018}. Our capacity result adds simultaneous exact counts across targets and workloads, building on majority and abstention constructions \cite{mcgarvey1953,nehringpivato2011,knoblauch2013}.

\paragraph{Omission and editing.} Truthful omission and adversarial missingness influence learning \cite{barash2021omission,koyuncu2023omission}. Preference poisoning, policy teaching and minimum-cost label flipping intervene on feedback or learned policies \cite{poison,wu2025preferencepoisoning,nika2025policy,kusaka2026costminimized}. Clean-label reward poisoning preserves labels but changes content \cite{duan2025badreward}. We compare the cost of reaching the same majority or exact-supervision target with content held fixed. This complements work on ranking-swap costs and the welfare cost of fairness \cite{elkind2009swap,bertsimas2011price}.

\paragraph{Governing collection and use.} Randomized reviewer assignment limits strategic influence \cite{jecmen2020,xu2019}. Fair sortition seeks fair selection probabilities under demographic quotas \cite{flanigan2021fair}. We fix realized counts across items and vary who judges which content. Balanced sampling fixes auxiliary totals, while fairness gerrymandering concerns subgroup outputs \cite{devilleTille2004balanced,kearns2018gerrymandering}. Graph-Preference Learning studies inclusion-weighted welfare under network-sampled feedback \cite{fan2026graphpref}. Selection-bias correction and calibration act on observations or scores \cite{schnabel2016treatments,guo}. Our reservation and content-quota tests act on collection.

\section{Discussion and conclusion}
\label{sec:conclusion}
For an organization learning from community feedback, participation counts and task assignments provide complementary information about whose judgments shape the model. Our controls trace that influence through personal history, selected judgments and shared-model predictions. Public reservation reduces Road score gaps at unchanged workloads. The effect of content quotas depends on how the groups align with patterns of disagreement. Calibration changes which effects remain in the adjusted scores and decisions.

Collection audits should therefore record assignment rules and information access alongside participation counts, and assess model quality under the stated scoring rules.
 \label{sec:core-end}
\section*{Acknowledgment}
Jiayuan Liu and Vincent Conitzer thank the Cooperative AI Foundation, Macroscopic Ventures (formerly Polaris Ventures / the Center for Emerging Risk Research) and Jaan Tallinn's donor-advised fund at Founders Pledge for financial support.

\section*{AI Use Statement}
OpenAI ChatGPT and Codex assisted with language editing, \LaTeX{} formatting, and implementing, running, and checking experiment code under the authors' direction. The authors are responsible for the final manuscript, experiments, and references.

\clearpage
\appendix
\renewcommand{\thesection}{\AlphAlph{\value{section}}}
\label{sec:supp-start}

\section*{Guide to the appendix}
The main paper and this appendix describe the same collection problem: changing who evaluates which item while preserving participation counts. Appendix A gives the proofs and audits the cost of feasible reassignment. Appendix B explains the assignment and learning procedures, model configurations, and statistical comparisons. Appendix C reports conversational-AI safety results, and Appendix D reports perceived-hazard ratings of videos. Appendix E compares reassignment with answer editing at fixed targets or budgets. Appendix F tests rules that restrict the collector's assignment choices.

\paragraph{Judgments, assignments and learning.} Evaluators $i\in I$ judge training items $p\in T$. A judgment can be a binary answer or a numerical rating. $H,C,E$ are the history, calibration, and evaluation item sets. The eligible person--item pairs are $\Omega\subseteq I\times T$, and their fixed judgments are $Y=(y_{ip})$. An assignment matrix, also called a \emph{mask}, is $M=(m_{ip})\in\mathcal F$. Its nonzero entries specify who evaluates which item. The feasible set $\mathcal F$ preserves each evaluator's workload $r_i$, each item's panel size $k_p$, and applicable group counts $k_{pg}$ relative to the reference assignment $M^0$. Throughout, \emph{native} constraints mean all constraints of the actual design, including eligibility and demographic composition. An \emph{auditor} may read current judgments to measure what is feasible, whereas the history-guided collector selects assignments without them.

The target $q_p(M)=k_p^{-1}\sum_{i:(i,p)\in\Omega}m_{ip}y_{ip}$ is the mean judgment of the people assigned to item $p$. $q(M)$ lists these training targets in item order. A training \emph{stream} specifies the initial state and randomness $\xi$ shared by the conditions being compared. This includes the state of the pseudorandom number generator (RNG), which determines the sequence of random draws during training. The resulting shared scorer $f_{M,\xi}(x)$ receives only item content at evaluation. \emph{Raw} scores are the model's outputs before calibration, which adjusts scores using the separate reference set $C$. Thresholding a score produces a binary decision, or \emph{flag}.

\paragraph{Two types of group constraint.} Participant groups, called \emph{strata}, describe which people form an item's panel. Content groups, also called \emph{buckets}, instead divide the items and specify how many items from each group a person evaluates. Keeping a panel's total size does not preserve its demographic composition. For example, consider two panels and four people who each evaluate one item. Both panels may initially contain one person from each of two demographic groups. Reassigning the same people can put one whole group in each panel while preserving every workload and panel size. The item--stratum constraints exclude this change. If all strata partitioning a panel have prescribed counts, summing those counts implies the total panel size. The total-size constraint remains applicable when composition is unspecified or only partly specified.

\paragraph{Condition labels.} In tables, H or Hist denotes history-guided collection, P a public reference assignment, and G the control that gives each person someone else's history for selection. The selected judgments still come from the people actually assigned. H100 and H288 denote histories of 100 and 288 items. Public subscripts identify assignments, and map labels identify fixed reassignments of historical information or content groups. Numerical stream labels identify runs that share training state across conditions. All comparisons concern the same recorded cohort.

\section{Mathematical capacity and intervention costs}
\begingroup\let\GerrySavedSubsection\subsection\let\section\GerrySavedSubsection\let\subsection\subsubsection
\section{Proof of Theorem~\ref{thm:new-balanced} and its scope}
\label{app:balanced-proof}
The proof runs from ``Local construction'' through ``Full profile and fixed quantifiers.'' The remaining paragraphs discuss extensions and conditions for applying the construction.

Let $s\ge3$ be the number of common alternatives, $d=\binom{s}{2}$ the number of unordered pairs, $L=\lceil d/2\rceil$, and $N=d(4L+1)$. Fix a strict reference order $R$. The profile $P$ is the list of $N$ individual orders. Each voter answers $r$ pairs, each pair receives $k=r(4L+1)$ answers, and $t=(t_j)_{j=0}^{d-1}\in\{-1,+1\}^d$ gives desired directions relative to $R$. Write $D_P(e)$ and $D_{P,M}(e)$ for the full-profile and selected supporting-minus-opposing margins on pair $e$.

The electorate has strict, fixed orders over a common alternative set. Every voter can answer every unordered pair. A response mask records which comparisons were directly requested, and the aggregation rule counts those direct responses only. It does not first close each voter's observed relation transitively. The quantity $r$ is the number of distinct directly asked pairs, not the number of comparisons logically determined from their answers or private history. The same electorate serves every target and every $1\le r<d$. The masks, not the electorate, vary.

\paragraph{Local construction.}
Let the $d$ unordered pairs be $e_0,\ldots,e_{d-1}$ in a fixed cyclic order. For each $e_j=\{a,b\}$, oriented by $R$, take $U_j=(a,b,T_j)$ and $V_j=(b,a,T_j)$, where $T_j$ is any fixed ordering of the remaining alternatives, and their reverses $\bar U_j,\bar V_j$. The orders $U_j,V_j$ differ only on $e_j$. Reversal negates every vote. On $e_j$, voters $U_j,\bar V_j$ contribute $+2$, while $V_j,\bar U_j$ contribute $-2$. On any other pair, either of these voter pairs contributes zero. Thus the appropriate pair can answer $e_j$ and the other pair the assistant $e_{j+1}$ without changing either query's direct-response count.

\paragraph{Exact row degrees.}
Ask all four voters the additional set $B_j=\{e_{j+2},\ldots,e_{j+r}\}$. This set is empty at $r=1$. For $r<d$, it contains $r-1$ distinct queries, neither $e_j$ nor $e_{j+1}$. Each voter therefore answers exactly one switch query and $r-1$ common queries, for total $r$. On every common query the block's four votes cancel.

\paragraph{Exact column degrees.}
Use $L$ copies of every block. Query $e_j$ receives $2L$ primary responses, $2L$ assistant responses and $4L(r-1)$ common responses, hence $4Lr$ in total. Its only nonzero block contribution is $2Lt_j$. Add $d$ voters with order $R$, assigning reference voter $h$ to $e_h,\ldots,e_{h+r-1}$. Each reference voter answers $r$ queries and each query receives $r$ reference votes. The resulting electorate has $N=d(4L+1)$ voters, each column degree is $k=r(4L+1)$, and the observed margin is $2Lt_j+r$. Since $2L\ge d>r$, no margin is zero and every sign is $t_j$.

\paragraph{Full profile and fixed quantifiers.}
The block voters cancel pairwise when everyone is observed, while the reference voters contribute $d$ in the direction of $R$. Thus $D_P(e)=d$ for each pair. The electorate was chosen without reference to $r$ or $t$. When $r=d$, every voter must answer every query and the mask is all ones. This proves Theorem~\ref{thm:new-balanced}. Its population bound is sufficient, not optimal.

\paragraph{Public strata.}
Replicate the entire electorate in each public stratum and apply the same construction within every stratum. Each person still answers $r$ queries and every query receives $k$ votes from each stratum. All stratum margins agree with $t$. This gives a stratified existence construction for these replicated group quotas.

\paragraph{A perfect-history implementation.}
Introduce historical anchors $h_1,\ldots,h_{s-1}$, and let $\pi_i(j)$ be voter $i$'s alternative at rank $j$. Extend the active order as $\pi_i(1)>h_1>\pi_i(2)>\cdots>h_{s-1}>\pi_i(s)$. Past comparisons of every active alternative with every anchor identify its active rank by the number of anchors below it. Every voter provides the same $s(s-1)$ past comparisons, none an active--active current query. The assignment can then be computed from this private history before direct current answers are disclosed. Stable identifying history makes this construction implementable before current-label disclosure. Empirical history prediction is measured separately. If the aggregator also uses the histories, the direct-only theorem does not apply.

\paragraph{A necessary margin boundary.}
For any profile of $N$ voters and schedule with $k$ answers on a pair, $D_{P,M}(e)-D_P(e)$ is the negative sum of the $N-k$ omitted votes. Hence $|D_{P,M}(e)-D_P(e)|\le N-k$. A full margin of magnitude at least $N-k$ cannot be strictly reversed, although equality can permit a tie. In the construction the normalized full margin is only $1/(4L+1)$. The near-complete-coverage construction uses small full-profile margins. The inequality states the protection supplied by a larger margin.

\paragraph{Why transitive inference is a different model.}
In the $s=3$, $r=2$ reverse-target example, direct margins are $(-2,-2,-2)$ with 18 direct votes on each pair. Adding each voter's transitively implied answers yields margins $(0,-3,-4)$ with comparability counts $(20, 21, 20)$. The strict reverse target therefore fails under that rule. This is a finite counterexample to extending the construction, not a proof that completion-based aggregation is generally immune. Complete ranking bundles and scalar neural predictions are likewise different objects.

\subsection{Necessary-outcome certificates under partial historical orders}
Let $\mathcal H_i$ be a strict partial order, and let compatible profiles be the Cartesian product of the voters' sets of linear extensions. For a target pair $a>b$, let $C_{ie}=1$ exactly when the transitive closure of $\mathcal H_i$ entails $a>b$, and zero otherwise. Here $e=\{a,b\}$ indexes that query, $M_{ie}$ selects its direct response, and $k_e$ is its panel size. A fixed schedule guarantees the target strict majority graph at every compatible profile if and only if $\sum_i M_{ie}C_{ie}\ge\lfloor k_e/2\rfloor+1$ for every query.

The condition is sufficient because the certified votes already form a strict majority. For necessity, choose any query with fewer certified supporters. Every selected uncertified voter has a compatible linear extension reversing the target comparison. Otherwise that comparison would be entailed by $\mathcal H_i$. Choosing these extensions separately across voters is permitted by the Cartesian-product information set and gives a compatible profile where the target query fails. One failed query suffices to refute a guarantee of the full graph. This proves the characterization. An absent certificate does not rule out average or realized gains, and a high Hist probability is not a logical certificate.

With only row/column/eligibility constraints, the existence test is a lower-bound integral-flow problem with certified and uncertified channels per query. Crossing demographic constraints require the actual constrained integer formulation or a separate network justification. For possible outcomes, independently possible orientations need not admit a common transitive completion: an empty single-voter history permits each edge of a three-cycle separately but no total order realizing the full cycle.
 \endgroup
\begingroup\let\GerrySavedSubsection\subsection\let\section\GerrySavedSubsection\let\subsection\subsubsection
\section{Proofs of Theorems~\ref{thm:unbounded-cost} and~\ref{thm:cost-of-balance}, cost bounds, and data audits}
\label{app:intervention-cost}
This section gives the proofs of Theorems~\ref{thm:unbounded-cost} and~\ref{thm:cost-of-balance}. It then examines how the starting assignment affects cost, gives methods for computing minimum costs, and connects exact supervision to the DICES audit.
We use $C_{\mathrm{flip}}$ and $C_{\mathrm{bal}}$ for intervention costs, to distinguish these operation counts from empirical score changes. Throughout this appendix, judgments $y_{ip}\in\{0, 1\}$ are fixed on the eligible support $\Omega$, the starting binary assignment $M^0$ is feasible, and every panel size $k_p$ is positive. The basic model preserves every row sum $r_i$ and column sum $k_p$ of $M^0$, as well as eligibility. These counts may be nonuniform. Balance here refers to preservation, not equality between different rows or columns. All person--item sums in this section range over eligible pairs in $\Omega$. Here $\mathbf1\{\cdot\}$ denotes the indicator of a condition, and $\|\cdot\|_1$ sums absolute matrix entries. A target $t$ specifies a strict majority at every item, including the items whose initial majority already agrees. Write $h_p=\lfloor k_p/2\rfloor+1$ and $a_p^0=\sum_i M^0_{ip}\mathbf 1\{y_{ip}=t_p\}$.

\paragraph{Two edit units.}
The cost of honest reassignment is the endpoint distance $c(M,M^0)=\sum_{ip}(1-M^0_{ip})M_{ip}=\tfrac12\|M-M^0\|_1$. It counts new selected person--item edges, not the length of a sequence of pairwise swaps. Row and panel equalities hold at the final assignment. A coordinated cycle can be executed simultaneously. Direct editing instead fixes $M^0$ and charges one per reversed selected binary answer. It permits independent pairwise-answer edits, rather than requiring an arbitrary modified ballot to remain a linear order. This comparator belongs to the microbribery tradition~\cite{faliszewski2009microbribery}. Candidate-position swap bribery uses a different unit~\cite{elkind2009swap}.

\paragraph{Deficits and elementary lower bounds.}
Write $(v)_+=\max\{v,0\}$ and let $v_p=(h_p-a_p^0)_+$. Each selected-answer edit can improve only one item's support by one, and reversing exactly $v_p$ opposing answers suffices. Hence $C_{\mathrm{flip}}=\sum_p v_p$. Each honest reassignment must introduce at least $v_p$ new supporters at deficient item $p$, yielding $C_{\mathrm{bal}}\geq C_{\mathrm{flip}}$. Moreover, if $a$ previously unselected people enter $p$, then $a$ old assignments leave $p$ and each entrant must lose an assignment elsewhere to preserve their own row sum. These disjoint deletion sets have total size at least $2a$. Consequently,
\[
 C_{\mathrm{bal}}\geq\max\left\{\sum_pv_p,\; 2\max_pv_p\right\}.
\]
An infeasible honest target has cost $+\infty$. If all deficits vanish, both costs are zero. The ratio is defined only for positive flip cost.

\paragraph{Truthful supply and reachability.}\label{app:supply-proof}
For an item $p$ and target label $t_p$, let $a_{pg}(t_p)$ count eligible truthful supporters in stratum $g$. The strata here form a partition, with exact item-level counts $k_{pg}$. A feasible starting assignment ensures that each stratum contains enough eligible people to fill its quota. Within a single item, its greatest possible target support is
\[
 U_p(t_p)=\sum_g\min\{k_{pg},a_{pg}(t_p)\}.
\]
Each summand is an upper bound because a panel can use neither more than its quota nor more supporters than exist. It is attained locally by selecting as many supporters as possible in each stratum and filling remaining slots with opponents. Thus $U_p(t_p)<h_p$ certifies honest infeasibility, even after individual workload constraints are removed. If all eligible responses on a three-person item are zero, two selected-answer flips attain majority one, but every honest panel still reports zero. This gives $C_{\mathrm{flip}}=2$ and $C_{\mathrm{bal}}=+\infty$ for that change. Meeting every local bound leaves the separate question of a joint assignment respecting individual workloads. With no strata, take a single group.

\paragraph{Proof of Theorem~\ref{thm:unbounded-cost}: an unbounded cost ratio.}\label{app:unbounded-proof}
For any $m\geq2$, form $m$ blocks $B_i$, each containing three items, and $3m$ evaluators: core people $c_i$, local people $a_i$, and fillers $z_i$. Support is complete and indices are cyclic. Person $a_i$ answers one on exactly $B_i$. Every filler answers zero everywhere. For $i>0$, core person $c_i$ answers one on $B_i\cup B_{i+1}$. Person $c_0$ answers one only on $B_1$. All other potential answers are zero. Initially assign $c_i,a_i,z_i$ to all three items of $B_i$. Every row and column has degree three. The all-one-majority target already holds outside $B_0$ and needs one answer edit at each of its three items, so $C_{\mathrm{flip}}=3$.
Only the $2m$ core/local people can provide a positive answer and each has capacity three. Meeting all $3m$ target majorities requires at least $6m$ positive answers, equal to this entire capacity. Every core/local assignment must therefore go to a positive edge, with exactly two positives per item. Each $a_i$ is forced to its own block, and $c_0$ is forced to $B_1$. This fills the core demand of $B_1$, forcing $c_1$ to $B_2$, and inductively forces the whole core cycle. Exactly $3m$ core assignments change. Leaving all locals and fillers in their original blocks attains the target at that cost. Hence $C_{\mathrm{bal}}=3m$ and the ratio is $m$. All instances in this family are finite, every person--item pairing is eligible, and the target has the explicit honest realization above. The growing ratio comes from the length of the forced compensation cycle while row and panel sizes remain three. The items are general binary questions. The ordinal profile in Theorem~\ref{thm:new-balanced} is a different domain. The result compares the two optimal costs and is compatible with exact polynomial-time computation in the basic network model below.

\paragraph{Proof of Theorem~\ref{thm:cost-of-balance}: short repairs in the constructed schedules.}\label{app:canonical-proof}
Let $d=\binom{s}{2}$ and $L=\lceil d/2\rceil$. The electorate has the finite size $N=d(4L+1)$. For initial tournament $u\in\{-1,+1\}^d$, the starting schedule is exactly $M_r(u)$, constructed in Appendix~\ref{app:balanced-proof}. The endpoint needs only the target majority $t$, rather than all margins or selected edges of $M_r(t)$. In each four-voter copy for primary comparison $e_j$, the two voters with orders $U_j,V_j$ differ only on $e_j$, as do the two reversed orders. These are the two \emph{switchable pairs}. The complementary pairs used to create the initial positive or negative margin in Theorem~\ref{thm:new-balanced} are a different pairing of the same four voters. Under the canonical schedule $M_r(u)$, one member of each such pair answers $e_j$ and the other answers helper $e_{j+1}$. Both answer the same $r-1$ remaining comparisons. Exchanging their primary/helper assignments introduces exactly two edges, changes exactly one primary answer, and leaves the helper's answer unchanged. It preserves every row and panel count.
There are $2L$ such independent exchanges on each primary. A change from initial positive to desired negative requires $v_j=L+\lfloor r/2\rfloor+1\leq2L$ exchanges. A change from negative to positive requires $v_j=L-\lceil r/2\rceil+1\leq2L$. These formulas use the original margins $2Lu_j+r$ and strict, rather than weak, majorities. Performing exactly $v_j$ exchanges on each deficient primary costs $2\sum_jv_j$. Primary blocks have disjoint voters, and helper contributions are unchanged, so all target comparisons hold together. Combining this construction with the lower bound proves $C_{\mathrm{flip}}\leq C_{\mathrm{bal}}\leq2C_{\mathrm{flip}}$.
When the target differs on only one comparison, $2\max_jv_j=2C_{\mathrm{flip}}$ gives the matching lower bound. The factor two is therefore tight. For example, $s=4,d=6,L=3,r=5$ yields a negative initial primary margin $-1$. Turning only that comparison positive costs one answer edit and exactly two assignment replacements. In this canonical construction the corresponding answer edits can also be chosen as adjacent-order swaps in distinct voters, so this example does not rely on introducing cyclic individual reports.

This completes the proof of Theorem~\ref{thm:cost-of-balance}. The following example explains why its starting-assignment condition matters.

\paragraph{The same electorate with a different starting schedule.}\label{app:noncanonical}
For the manuscript's three-alternative profile, use the following allocation of the same seven $ABC$ voters and four voters of each other order:
\begin{center}
\begin{tabular}{lrrrr}
\hline
Order & People & $AB$ & $AC$ & $BC$\\\hline
$ABC$ & 7 & 2 & 5 & 0\\
$ACB$ & 4 & 2 & 0 & 2\\
$BAC$ & 4 & 0 & 0 & 4\\
$BCA$ & 4 & 4 & 0 & 0\\
$CAB$ & 4 & 0 & 1 & 3\\
$CBA$ & 4 & 1 & 3 & 0\\\hline
Total & 27 & 9 & 9 & 9\\\hline
\end{tabular}
\end{center}
Every person answers one comparison and each panel contains nine people. Full-electorate margins remain $(3, 3, 3)$, while direct margins are $(-1,+1,-1)$. The target $C>B>A$ changes only $AC$, giving $C_{\mathrm{flip}}=1$. Any cost-two balanced repair is a two-person exchange. To improve $AC$ it must replace an $ABC$ voter there by a $C>A$ voter elsewhere. Such entrants are $BCA/CBA$ at $AB$, or $CAB$ at $BC$. In the first case the transferred $ABC$ vote breaks the one-vote $B>A$ majority. In the second it breaks the one-vote $C>B$ majority. No cost-two repair attains the full target, and a positive balanced change cannot cost one. The cycle $ABC:AC\to AB$, $ACB:AB\to BC$, $CAB:BC\to AC$ leaves the other two aggregate comparisons unchanged and repairs $AC$. Thus $C_{\mathrm{bal}}=3$. The factor two therefore depends on the constructed starting schedule as well as the electorate. Whether a larger constant holds uniformly over arbitrary balanced starts across this special ordinal profile family is a separate question. Theorem~\ref{thm:unbounded-cost} establishes the unbounded result for general binary items.

\paragraph{A sufficient local-compensation condition.}\label{app:local-compensation}
The short-repair argument can also be stated without using the full canonical construction. For every deficient item $p$, suppose there are $v_p$ switches $(a,b,h)$, where $h\ne p$, satisfying
\[
 \begin{gathered}
 M^0_{ap}=M^0_{bh}=1,\qquad M^0_{bp}=M^0_{ah}=0,\\
 y_{ap}\ne t_p,\quad y_{bp}=t_p,\quad y_{ah}=y_{bh}.
\end{gathered}
\]
Require the two inserted edges $(b,p),(a,h)$ to be eligible, every switch to preserve all applicable composition quotas, and the evaluators used by different switches to be disjoint across all deficient items. For ordinary partition strata, using $a,b$ from the same stratum suffices for quota preservation. Replace $(a,p),(b,h)$ by $(b,p),(a,h)$ in every switch. Each evaluator loses one assignment and gains one, each panel retains its size, target support at $p$ increases by one, and the helper's label count is unchanged. Disjointness makes these operations jointly valid in the binary mask. Their total cost is $2\sum_p v_p$, establishing $C_{\mathrm{bal}}\le2C_{\mathrm{flip}}$ under this explicit sufficient condition. This is a sufficient repair structure, rather than a necessary condition for an inexpensive realization. The canonical schedule supplies these switches with $U_j,V_j$ and their reverses.

\paragraph{Dropping workload constraints.}
Retain the support and each panel size, but remove all constraints coupling a person's participation across items. If the target remains attainable, each deficient panel can replace exactly $v_p$ opposing selected people with unselected supporters, independently of other panels. Thus the minimum relaxed replacement cost is $C_{\mathrm{flip}}$. Fixed nonoverlapping evaluator-stratum counts inside each item admit the same argument by distributing the replacements over cells with available supportive slack. This identity is conditional on sufficient local supply and on dropping all individual cross-item constraints. It is not a statement that any target is attainable. On commonly attainable nonzero-cost targets, $C_{\mathrm{bal}}/C_{\mathrm{flip}}$ measures the extra edit cost of preserving workloads, rather than the welfare-efficiency loss studied under a different price-of-fairness objective~\cite{bertsimas2011price}.

\paragraph{Integral-flow implementation.}
For the basic model, create a supply node of $r_i$ for each evaluator. Each item has a target-support node $g_p$ and a demand node $p$. An eligible person--item edge points to $g_p$ when its answer agrees with $t_p$, and otherwise to $p$. Its capacity is one and its cost is $1-M^0_{ip}$. The arc $g_p\to p$ has lower bound $h_p$ and upper bound $k_p$, and item node $p$ has demand $k_p$. Feasible integral flows correspond exactly to target-realizing masks and their costs are $c(M,M^0)$. Standard network integrality guarantees the existence of an integral optimum whenever the instance is feasible. Convex combinations of optimal solutions may still be fractional. Additional crossing quota rows require their own integral-network justification. The native cost audits retain them in a binary mixed-integer formulation with explicit bounds and solver statuses.

\paragraph{A finite basic-model bound.}
Let $N$ be the number of evaluators and $B=\sum_i r_i$. For a feasible target, $C_{\mathrm{bal}}\leq\min\{B,N C_{\mathrm{flip}}\}$. Increase the baseline flow on each deficient $g_p\to p$ arc by $v_p$, producing $C_{\mathrm{flip}}$ units of imbalance while respecting all arc bounds. A feasible target flow provides a residual repair. The initial residual costs are nonnegative: reversing old selected edges costs zero, using previously unselected edges costs one, and auxiliary arcs cost zero. Decompose a repair into integral paths and remove cycles. Each simple unit path visits at most $N$ evaluator nodes and costs at most $N$, proving the bound. Its linear dependence is attained for panel-one identity assignments with one defective diagonal and a forced all-positive cyclic matching. This size-dependent bound is compatible with the unbounded ratio in Theorem~\ref{thm:unbounded-cost}.

\paragraph{Exact supervision connects the audit to existing learners.}
A target majority fixes only a sign and need not fix the soft label. Let $c_p^0=\sum_i M^0_{ip}y_{ip}$ be the initial positive count. For an existing honest assignment $M_H$, instead prescribe $c_p^H=\sum_iM_{H,ip}y_{ip}$ at every item. Independent binary answer editing reproduces this vector at cost
\[
 C^{\mathrm{exact}}_{\mathrm{flip}}=\sum_p|c_p^H-c_p^0|.
\]
Replacing majority inequalities by these exact count equalities gives the closest balanced honest witness, which is always feasible because $M_H$ itself is available. Its cost is at most $c(M_H,M^0)$. If both witnesses produce identical actual content/aggregate-label/weight/order tensors and share the initial state and RNG, they have the same losses and updates. This must be checked against the actual loader: equal aggregate counts alone do not imply identical individual-vote minibatch trajectories. Duplicate training of identical learner inputs is unnecessary. The comparison is a full-information supervision-cost audit, with the information needed to certify the recorded supervision, rather than a bound on held-out model behavior.

\paragraph{Cost audit on human feedback.} We specified 120 DICES \cite{dices} questions before solving: 64 original menu targets (four designs, two original public starts and eight targets), 48 one-item changes to the P1 majority, chosen by a fixed ordering of item identifiers (12 per design), and eight exact per-item positive-count vectors from existing Hist-up assignments. All 120 problems have certified optimal solutions or infeasibility certificates. The native optimization preserved the observed support, individual workloads, item panels and original gender or locale-by-gender quotas. Integral flow solved the row/column/eligibility relaxation. A binary mixed-integer model retained the crossing native quotas. Every optimal mask was independently checked against original labels and all quotas. Twenty-four menu cases fail the row-relaxed native feasibility check. Five local cases are feasible after dropping strata in the basic model but infeasible even without row couplings when the actual per-item strata are retained. The table below summarizes all fixed requests and their certified status.
\paragraph{Constraint signatures.} $C_{\mathrm{basic}}$ preserves row sums, panels and eligible support but drops strata. $C_{\mathrm{native}}$ also preserves the original item-stratum quotas. The reported row-relaxed native quantity $R_{\mathrm{native\text{-}local}}$ preserves panels, support and those strata, but drops individual cross-item workloads. Both the basic and row-relaxed native feasible sets contain the native set. Neither is asserted to contain the other. This distinction explains the local cases in BASE and LARGER where the basic model is feasible but the row-relaxed native model is infeasible.
\begin{center}\small
\begin{tabular}{lrrrr}\toprule
Design & \shortstack{Menu\\opt./16} & \shortstack{Local\\opt./12} & \shortstack{Exact\\opt./2} & \shortstack{Native\\infeas.}\\\midrule
S1 & 16 & 12 & 2 & 0\\
S2 & 16 & 12 & 2 & 0\\
BASE & 4 & 10 & 2 & 14\\
LARGER & 4 & 9 & 2 & 15\\\bottomrule
\end{tabular}
\end{center}
For every one of the 43 reachable local changes, $C_{\mathrm{bal}}/C_{\mathrm{flip}}=2$. $C_{\mathrm{flip}}$ is one or two and $C_{\mathrm{bal}}$ is two or four. Across the 87 reachable positive-cost problems, including menu and exact-count cases, the finite observed ratio ranges from one to two. These are the observed ratios within the fixed roster's feasible positive-cost subset. Theorem~\ref{thm:unbounded-cost} characterizes the broader worst case. When $C_{\mathrm{flip}}=0$ the starting majority already meets the target and the ratio is left undefined.
The table above covers all 120 requests, including every native-infeasible case. Zero-cost and infeasible cases are counted separately from finite positive-cost ratios.
\paragraph{Existing-learner bridge.} The exact-count target fixes the positive-vote count of every T item to that of the existing Hist-up assignment. In the order S1, S2, BASE and LARGER, the P1/P2 cost pairs $(C_{\mathrm{flip}},C_{\mathrm{bal}})$ are $(94, 94)/(95, 95)$, $(84, 84)/(73, 73)$, $(77, 83)/(80, 84)$ and $(252,252)/(284,284)$. The corresponding original Hist assignment distances from public are 298/297, 300/300, 370/370 and 1115/1117 selected-edge replacements. The complete per-item $q$ vector, item content, uniform aggregate-label weights, training batches and initial learner state match those of the original honest Hist-up model. These conditions therefore reuse the existing model, with the human records unchanged. Thus these are prices of reproducing the existing aggregate training input, not measured new neural outcomes. No ratio of these counts to held-out flag percentage points is meaningful.
\paragraph{From assignment edits to identical supervision.} For a fixed binary feedback table and starting assignment $M^0$, write $c_p(M)=\sum_i M_{ip}y_{ip}$. Each item retains its panel size, so its inserted and deleted edge sets have equal size. Consequently $\sum_p|c_p(M)-c_p(M^0)|\leq\sum_{ip}(1-M^0_{ip})M_{ip}$. The left side is the minimum number of selected binary-answer edits needed to reproduce the complete count vector of $M$ at the fixed assignment $M^0$. Every honest supervision vector attainable with at most $b$ new assignment edges therefore has a direct-edit clone using at most $b$ flips. This elementary action-set inclusion concerns aggregate supervision. When actual content, labels, weights, order, learner state and execution recipe coincide, one trajectory can serve both roles. The measured performance of a specified editing rule is not the optimum over all editing policies.
 \endgroup
\begingroup\let\GerrySavedSubsection\subsection\let\section\GerrySavedSubsection\let\subsection\subsubsection
\section{Which theoretical assumptions are tested?}\label{app:assumption-map}
The theory asks which direct-majority outcomes assignments can produce and at what cost. Theorem~\ref{thm:new-balanced} uses strict pairwise majorities of fixed preference orders. Theorems~\ref{thm:unbounded-cost} and~\ref{thm:cost-of-balance} compare the minimum costs of realizing specified strict majorities. A majority on each pair is different from a learned score or a single winner over all alternatives. The experiments retain the collection constraints and measure how changed training targets affect the fitted shared scorer. The table distinguishes the assumptions used by the constructions from the properties checked in the data.

\begin{table*}[tp]
\centering\small
\begin{tabular}{p{0.27\linewidth}p{0.27\linewidth}p{0.37\linewidth}}\toprule
Assumption or conclusion & Role in the theory & Empirical status\\\midrule
Fixed honest judgments & A voter's order does not change when queries are reassigned. & Honest-assignment experiments select recorded judgments without changing their values. The answer-editing controls in Appendix~\ref{app:placement-controls} use separate counterfactual copies.\\
Preserved person/item counts & General model: each row $r_i$ and column $k_p$ matches its reference. The constructions explicitly use uniform $r,k$. & The measured designs use uniform budgets within each condition and check preservation in loaders. DICES retains item strata. Road \cite{song2024road} uses no demographic quotas.\\
Complete eligible support & Every voter can answer every comparison. & DICES350 and Road have complete observed training matrices. DICES990 is sparse and is analyzed on its actual support. Its audits test attainability under the observed eligibility constraints.\\
Strict orders on a common candidate set & Makes target tournaments and transitive individual preferences meaningful. & The ordinal profiles satisfy this assumption. Safety and hazard experiments instead measure judgments of individual items.\\
Direct-only aggregation & Unasked implications and private history add no direct votes. & Used by the construction. The closure counterexample shows this is material. Neural scalars and complete-ranking bundles are different aggregation objects.\\
Perfectly identifying stable history & Implements the construction before current direct answers are revealed. & A theoretical information-rich special case. Access isolation and identity controls test the use of imperfect historical information.\\
Any target graph, same electorate & A worst-case capacity statement. & The finite ordinal checks implement it. The 32 DICES menu targets test selected binary-item outcomes under native support.\\
Transfer to held-out model behavior & A separate empirical outcome. & Learning comparisons use the experiment-specific state and recipe records described in Appendix~\ref{app:scope-ledger}. Held-out scores, prediction errors and calibrated decisions measure the consequences of changing their selected judgments.\\\bottomrule
\end{tabular}
\end{table*}

\paragraph{From direct aggregation to learned aggregation.} Feasible assignments determine which vectors of mean judgments can be collected. For fixed binary responses $Y$ and feasible set $\mathcal F$, this attainable supervision set is $\mathcal Q(Y,\mathcal F)=\{q(M):M\in\mathcal F\}$. A complete strict-majority vector requires every item to have a majority, so first restrict to $\mathcal Q_{\mathrm{strict}}=\{q\in\mathcal Q:q_p\ne1/2\text{ for all }p\}$. For $q$ in this restricted set, define $b_p(q)=\mathbf1\{q_p>1/2\}$ and $b(q)=(b_p(q))_p$. Excluding ties avoids treating an evenly split panel as a strict victory for label zero. The theoretical target belongs to $b(\mathcal Q_{\mathrm{strict}})$. The native three-person-panel majority audits have no tied panels.

For learning, let $\mathcal A_\xi$ denote the fixed training procedure under initial state and randomness $\xi$, let $X_T=(x_p)_{p\in T}$, and write $f_{M,\xi}=\mathcal A_\xi(X_T,q(M))$. The fitted function shares parameters across items and carries information from pretraining. Its scores depend on the whole training input and optimization trajectory. Subsequent calibration and thresholding determine decisions. A training-item majority need not determine the learned decision on that item, and an evaluation item has no corresponding collected training panel. This distinction holds even when the entire optimization is deterministic. Variability across fine-tuning runs provides an additional source of variation \cite{mosbach2021stability}. Evaluation scores fixed checkpoints rather than sampling generated answers. Appendix~\ref{app:scope-ledger} specifies the state matching available for each comparison.

\paragraph{A deterministic illustration.} Consider an intercept-only binary classifier that predicts one probability $a\in(0,1)$ for two equally weighted items. Binary cross-entropy is minimized at the mean soft target. Targets $(2/3,0)$ give $a=1/3$: the first item's strict positive majority yields a negative thresholded prediction. Targets $(1/3,1)$ give $a=2/3$, producing the reverse mismatch. This analytic example isolates parameter sharing. The paired experiments measure the responses of the actual learners.
 \endgroup
\subsection{Finite construction checks and the native majority menu}\label{app:menu}
Independent finite checks cover 9,552 masks, six stratum replications, 2,192 fixed-mask history cases and 640 flow/enumeration cases. 24 frozen partial-history instances and the named closure counterexample also pass. These checks complement the complete proofs above. The real-data menu asks which of a fixed list of majority targets can be reached under the observed eligibility, workload, panel and demographic constraints.
{\small
\begin{table*}[tp]
\centering
\small
\caption{Eight pre-fixed T-only majority targets per design. Columns count verified feasible (F), locally certified infeasible (I), and unresolved (U) targets. These label-aware auditor masks are never used to train a model.}
\label{tab:menu}
\begin{tabular}{lrrrr}
\toprule Design & People / T items & F & I & U \\ \midrule

S1 & 104 / 104 & 8 & 0 & 0 \\
S2 & 104 / 104 & 8 & 0 & 0 \\
BASE & 144 / 144 & 2 & 6 & 0 \\
LARGER & 144 / 432 & 2 & 6 & 0 \\
\bottomrule\end{tabular}
\end{table*}
}
 All 32 targets were fixed before solving: 20 have explicit assignments satisfying every constraint, called feasible witnesses, and 12 have local supply certificates showing why they cannot be reached. An infeasibility certificate identifies an item with fewer than two eligible target-supporting votes under its stratum quotas. It rules out that full target while leaving other targets available for audit.

\section{Methods, pairing and evidence scopes}
\label{app:reproducibility}
\begingroup\let\GerrySavedSubsection\subsection\let\section\GerrySavedSubsection\let\subsection\subsubsection
\section{Algorithm and interpretation guide}\label{app:method-walkthrough}
Algorithm~\ref{alg:history-routing} follows the collection process from historical information to model evaluation. The collector first predicts individual judgments and chooses an assignment. The selected real judgments then train the shared scorer. These steps use different information. A \emph{condition} specifies the assignment rule, historical information and routing direction. A \emph{stream} identifies training state and randomness, while a \emph{split} assigns items to $H,T,C,E$. A public draw provides a reference assignment within the same split. Stream numbers index the paired training states within each experiment.

\begin{algorithm}[ht]
\caption{History-guided collection and shared-scorer training}\label{alg:history-routing}
\begin{enumerate}
\item Read historical judgments, current item content and eligibility. Fit the population predictor using H only. Estimate personal residuals by cross-fitting, which predicts each historical item without using that item's judgments.
\item Clip forecasts to the allowed range and quantize them for the assignment solver. Preserve each person's workload, each item's panel size and applicable demographic composition. Add any public reservation or per-person content-group quotas.
\item Solve each specified assignment rule without reading current judgments. Check all participation and eligibility constraints and verify that changing hidden labels cannot affect selection. Save the fixed assignment.
\item Retrieve each assigned person's recorded judgment of their assigned item. Compute the mean judgment on each item to obtain its training target $q_p$.
\item Restore each stream's complete common state and train a separate shared scorer for each assignment on item contents and targets, without evaluator IDs. Reuse a model only under certified complete input/state identity.
\item Evaluate the shared scorer on E. If C exists, fit the specified score adjustment on C and apply it to E. Compute each model's metrics before averaging, and use the saved resampling records for the specified comparisons.
\end{enumerate}
\end{algorithm}
Step 5 describes complete-state comparisons. Appendix~\ref{app:scope-ledger} separately identifies historical replays with incomplete state records.

\paragraph{Checking the information boundary.} Splitting items defines which judgments the collector may use. Access checks enforce that restriction by blocking current, calibration and evaluation judgments during assignment. Dummy-label tests replace hidden current judgments with artificial values and verify that the assignment stays unchanged. The real selected judgments are retrieved only after assignments are fixed. The separate cost and opportunity audits use current judgments to compute attainable training targets and their costs. They answer what assignment could achieve with full information, while the collector predicts from history.

\paragraph{History-residual prediction.} ``Hist'' labels history-guided collection using this predictor in tables. The predictor starts with a population estimate from kernel ridge regression. It then adjusts that estimate using the evaluator's past prediction errors, giving more weight to errors on similar content. Each error compares a recorded judgment with a prediction fitted without that item's judgments. This procedure forecasts whom to ask. The separate shared scorer learns from the real judgments selected by the resulting assignment.

Let $H_i\subseteq H$ contain the historical items observed for evaluator $i$. For historical item $h\in H$ with content $x_h$, let $n_h$ be its number of judgments, $\bar y_h$ their mean, $\bar y=(\bar y_h)_{h\in H}$, and $\mu_H$ the overall historical judgment mean. Let $K(\cdot,\cdot)$ be the fixed population kernel, $K_{HH}=(K(x_h,x_{h'}))_{h,h'\in H}$ its historical matrix, $K_{xH}=(K(x,x_h))_{h\in H}$ its query row, $\mathbf1$ the all-one vector, and $\operatorname{diag}(1/n_h)$ the diagonal matrix of inverse counts. The fitted coefficients $\beta$ and population prediction $g_H$ are
\[
\begin{gathered}
\beta=(K_{HH}+\operatorname{diag}(1/n_h))^{-1}(\bar y-\mu_H\mathbf1),\\
g_H(x)=\mu_H+K_{xH}\beta.
\end{gathered}
\]
Content similarity determines how strongly a past judgment informs a new prediction. Let $v_x^{\rm raw}$ be the fixed content vector and $v_x=v_x^{\rm raw}/\max\{\|v_x^{\rm raw}\|,10^{-12}\}$ its L2 normalization, where $\|\cdot\|$ is Euclidean and $v_h=v_{x_h}$. Let $b_H$ be the median squared distance between distinct H items, floored at $10^{-10}$. If H contains a single item, its bandwidth is set to one. For content $x$ and historical item $h$, the population kernel is $K(x,h)=\exp(-\|v_x-v_h\|^2/b_H)$, with unit amplitude. The personal kernel $\kappa$ below is identical. The floor also applies when the median is zero. To estimate personal prediction errors, a fixed partition based only on item identifiers divides H into two sets, keeping all judgments of an item together. Let $B(h)$ be the set containing item $h$. The model $g_{H\setminus B(h)}$ is fitted using only judgments from the other set. Define residual $e_{ih}=y_{ih}-g_{H\setminus B(h)}(x_h)$ and forecast $\widehat y_{ix}$:
\[
\begin{gathered}
\kappa(x,h)=\exp(-\|v_x-v_h\|^2/b_H),\\
\widehat y_{ix}=\operatorname{clip}_{[0,1]}\left[g_H(x)+
\frac{\sum_{h\in H_i}\kappa(x,h)e_{ih}}{1+\sum_{h\in H_i}\kappa(x,h)}\right].
\end{gathered}
\]
Clipping restricts the forecast to $[0,1]$. The added one in the denominator shrinks the personal adjustment toward zero when little similar history is available.

\paragraph{Integer objectives and identity controls.} The assignment solver multiplies each forecast by a fixed scale and rounds it to an integer. This conversion is called quantization. Original DICES flow costs use $\operatorname{rint}(10^6\widehat y)$. Original Road uses a scale of $10^9$, with sign reversed for maximization. NumPy's rint rounds to the nearest integer, resolving rounding ties toward the even integer. The fixed edge order and solver determine which assignment is returned when objectives tie. There is no additional random tie-breaking objective, and several masks may achieve the same value. The DICES identity control reassigns whole histories within gender groups even when assignment quotas also preserve locale--gender composition. Road reassigns histories within its cohort. In both cases, current judgments remain attached to the people who provided them.

\begin{center}\small
\begin{tabular}{p{0.22\linewidth}p{0.69\linewidth}}\toprule
Quantity & Meaning and measurement stage\\\midrule
$\widehat y_{ip}$ & Forecast used by the collector to choose an eligible person--item pair.\\
$q_p(M)$ & Mean of selected real judgments on training item $p$, used as the shared scorer's training target.\\
$f_{M,\xi}(x)$ & Score for a new item's content $x$, from the model trained under assignment $M$ and training state/randomness $\xi$.\\
Net / gross flags & Net change is the fraction newly flagged minus the fraction no longer flagged. Gross disagreement is the fraction whose flags differ from the comparator.\\
Reservation width $W$ & Difference between the greatest and least achievable positive-judgment rates on T after designated assignments are fixed.\\\bottomrule
\end{tabular}
\end{center}

The ordinal construction and its examples are given in Appendix~\ref{app:balanced-proof}.
 \endgroup
\subsection{Learner recipes and information budgets}\label{app:training-overview}
DICES \cite{dices} uses Qwen3.5-2B with a scalar head prepared on historical items before the assignment comparison. Low-rank adaptation (LoRA) uses rank 16, scaling parameter 32 and dropout probability 0.05. The learning rate is $10^{-5}$, the effective batch size is 8, and the input length cap is 2,048 tokens. The primary endpoint is two epochs, following the beginning of the fixed 32-epoch cosine schedule. Parameters and scoring with batch size one use 32-bit floating point (FP32). Training computation uses bfloat16 (BF16). The frozen text representation is the final nonpadding hidden state used by the scalar head.

Mistral-7B-Instruct-v0.1 uses its own frozen features and history-prepared head. Its physical batch size is one instead of Qwen's four, with the same effective batch size eight and two-epoch recipe. LoRA targets its seven attention and multilayer-perceptron projections. Qwen additionally targets hybrid attention. Mistral's pad token is \texttt{<unk>}, distinct from the end-of-sequence token. Within each model type, compared conditions share the recorded training states.

Road \cite{song2024road} uses 2,048-dimensional ResNet50 ImageNet-1K V2 features. The features of four frames are averaged after the pretrained evaluation transform, then normalized for routing. The original learner fine-tunes ResNet50 layer4 and its scalar head for four epochs/144 updates, with frozen batch-normalization statistics. ConvNeXt fine-tunes its final stage and head for two epochs/72 updates. It uses FP32, batch size eight, AdamW learning rates $10^{-5}/10^{-4}$, weight decay 0.01 and a constant schedule.
The current collection budget counts selected judgments on $T$. Historical information is counted separately. DICES350 uses 10,400 H judgments and 312 selected T judgments per split. DICES990 uses 6,499 observed H judgments and 432/1,296 selected T judgments, with 288 additional calibration judgments (C3: three judgments per calibration item). Road uses 13,824 H judgments, 864 selected T judgments (1,728 for panel 6), and 576 C3 judgments. The full calibration reference, C48, uses all 48 ratings per calibration item, totaling 9,216. Evaluation references are used only to measure outcomes. Complete-state comparisons restore weights, optimizer, scheduler, pseudorandom-generator and data-loading order across conditions. Appendix~\ref{app:scope-ledger} separately identifies historical replays with incomplete state records. An \emph{exact-input alias} names an additional condition whose complete inputs and initial state match an existing run. It reuses that run and adds no independent model.

\subsection{Endpoints and uncertainty}\label{sec:closure-statistics}\label{sec:closure-scope}
Metrics and threshold decisions are computed for each model before averaging. A reference judgment is the recorded label or mean rating used to evaluate a prediction. Positive denotes UNSAFE for text and high hazard for Road. Bracketed uncertainty ranges are confidence intervals (CIs), except where a table explicitly identifies reference-label choice ranges. A false positive is a positive prediction on a reference-negative item, and a false negative is the reverse. FPR and FNR are their rates within the corresponding reference classes. Balanced accuracy (BA) averages the fractions correctly classified within the positive and negative reference classes. MSE and RMSE are mean squared error and its square root. The receiver operating characteristic (ROC) curve plots the true-positive rate ($1-\mathrm{FNR}$) against FPR as the decision threshold varies. Its area, AUROC (also AUC), summarizes how well scores rank reference-positive items above reference-negative items. Higher values indicate better ranking. BCE is binary cross-entropy, a loss that compares predicted positive probabilities with reference labels or label fractions. Lower values indicate better probability prediction. \emph{Acquisition} counts items flagged by the method being evaluated but not its named comparator, and \emph{loss} counts the reverse. Gross disagreement adds these counts, while net change subtracts losses from acquisitions. Tables express them as rates over evaluation items. When each model selects its $K$ highest-scored items, the selected counts are equal. Set replacement counts how many previously selected items are replaced. Precision is the fraction of selected items that are reference-positive, and recall is the fraction of all reference-positive items that are selected. Average precision (AP) summarizes precision as the threshold moves through the ranked items. Majority ties are excluded only when a binary reference label is required for classification or ranking.
Original DICES primaries resample groups of related conversations and training streams, keeping the two public assignments fixed. Later summaries also draw reference assignments with replacement from the finite public set. This adds variation in which reference assignments are drawn, with the participant cohort held fixed. The 15 pairs among six publics share models and are statistically dependent. A resample can draw the same public assignment twice, yielding zero disagreement for that pair. With model outputs fixed, the resulting bootstrap mean is $5/6$ of the point estimate over distinct public pairs. This is a property of the specified resampling procedure.

The reported intervals and tests summarize uncertainty in different ways. Let $B$ be the number of saved resamples, $\Delta_b^*$ the contrast in draw $b=1,\ldots,B$, $\widehat\Delta$ its observed estimate, and $\delta_0$ a hypothesized contrast. The centered two-sided diagnostic is $p(\delta_0)=[1+\sum_{b=1}^B\mathbf1\{|\Delta_b^*-\widehat\Delta|\ge|\widehat\Delta-\delta_0|\}]/(B+1)$. The \emph{inverse accepted set} contains all hypothesized contrasts that this test does not reject, namely those satisfying $p(\delta_0)>0.05$. A percentile interval instead uses quantiles of the resampled estimates, so its endpoints need not agree with that set. Appendix~\ref{app:public-history} reports all three for S2. An interval spanning zero includes both positive and negative effects and therefore expresses uncertainty about direction. A prespecified set of primary comparisons is a test family. Holm correction adjusts its tests for multiple comparisons without changing pointwise interval coverage. Each adjustment covers its named family. Later controls reuse the same cohort and evaluation items. Rounded zeros reflect the displayed precision. Exact equality of inputs, logits or decisions is stated separately in Appendix~\ref{app:cost-learning}.
Road uses its saved 2,000 resampling weights for groups of videos sharing a source and for entire training streams. Public contrasts also resample from the fixed public-assignment set. DICES cost/placement and text governance use the saved 5,000 conversation-group and whole-stream weights. Assignments and control maps remain fixed throughout. A fixed-C analysis holds the fitted calibration rule constant. A C/E-refit analysis resamples calibration and evaluation items, refitting the rule using only the calibration draw. These procedures respectively exclude or include uncertainty from estimating calibration. Designs without an independent C set report calibration as unavailable.

\subsection{Comparison scopes}\label{app:scope-ledger}
The two-public reference experiment uses three streams in each DICES design. The six-public follow-up has two available streams for S1/S2 and three for BASE/LARGER: the 12 unavailable public-model slots remain unobserved. A replay reruns the history-guided learner in the training environment used for the additional public models. The public-reference replay uses the recorded initial tensors and order with newly fixed training randomness. Appendix~\ref{app:public-history} reports these comparisons separately and identifies the training-state records available for each. Earlier missing random states remain unavailable.

The cost audit reports all 120 targets, including 29 native infeasibilities. The joint governance family contains the Road interaction and both text-policy comparisons. It reuses 24 intact Road reference roles and adds 48 Road and 72 text trajectories under three paired training streams per design. The reported intervals describe variation within the recorded cohort and fixed maps. A trajectory is a model-training run, so these counts measure computational comparisons within that cohort.

\section{Text learning: historical information and public variation}
\label{app:original-scopes}
\subsection{Two-public reference designs and linked-history controls}
S1/S2 are two fixed partitions of the same complete 104-person release, each with 100 history, 104 training and 146 evaluation items. Each selects 312 real judgments, with three judgments per person, three per item and the original per-item gender quotas. DICES990 contains judgments for only some person--item pairs in a fixed 144-person cohort. Its designs share 100 history, 96 calibration and 361 evaluation items (H100/C96/E361). BASE uses 144 training items (T144), a subset of LARGER's 432 (T432). The larger design also changes workloads and feasible assignments. Recorded Yes/No/Unsure judgments map to 1/0/0. The table reports original Qwen results and the separate Mistral comparison, with training states shared within each model type.
\begin{table*}[t]
\caption{Raw H--public flag change at the original two-epoch endpoint. Intervals resample conversation groups and training streams jointly across conditions within each design, keeping the recorded participant cohort fixed.}
\label{tab:main}
\begin{center}
\small
\begin{tabular}{lrrr}
\toprule
Setting / learner & Seeds & $\Delta$ flag (pp) & 95\% interval \\
\midrule
350 S1 / Qwen & 3 & +56.74 & [43.81, 67.36] \\
350 S2 / Qwen & 3 & +5.02 & [0.60, 11.16] \\
350 S1 / Mistral & 2 & +36.99 & [28.32, 46.05] \\
350 S2 / Mistral & 2 & +2.40 & [0.04, 6.81] \\
990 BASE / Qwen & 3 & +9.60 & [5.64, 14.16] \\
990 LARGER / Qwen & 3 & -5.54 & [-16.99, 3.62] \\
\bottomrule
\end{tabular}
\end{center}
\end{table*}
 \subsection{Personal history at a fixed pooled-data budget}\label{app:gaps-text}
The table compares the original two-epoch Hist models' uncalibrated UNSAFE-flag rates with the mean public-reference rate, averaging over the available training seeds. The conditions with 100 and 10 person-linked historical items use the same anonymous pooled history H100 but give the predictor different amounts of person-linked history. The identity control assigns whole historical profiles to different people for selection. Selected current judgments still come from their actual owners. The table includes the public references, both available history levels, and the identity and direction controls under the same Hist predictor. False-positive and false-negative rates accompany the directional changes.

\begin{center}
\small
\begin{tabular}{llrrrr}
\toprule
Split & Condition & Seeds & $\Delta$flag pp & FPR\% & FNR\% \\
\midrule
S1 & Public1 & 3 & -15.64 & 0.00 & 100.00 \\
S1 & Public2 & 3 & +15.64 & 22.61 & 40.48 \\
S1 & \shortstack[l]{Personal 100 /\\pooled 100} & 3 & +56.74 & 66.09 & 4.76 \\
S1 & \shortstack[l]{Personal 10 /\\pooled 100} & 3 & +7.88 & 17.97 & 55.95 \\
S1 & Identity shuffle 100 & 3 & -14.27 & 1.16 & 97.62 \\
S2 & Public1 & 3 & +0.00 & 0.00 & 100.00 \\
S2 & Public2 & 3 & +0.00 & 0.00 & 100.00 \\
S2 & \shortstack[l]{Personal 100 /\\pooled 100} & 3 & +5.02 & 2.59 & 84.52 \\
S2 & Identity shuffle 100 & 3 & +0.00 & 0.00 & 100.00 \\
S2 & Hist down (100 items) & 3 & +0.00 & 0.00 & 100.00 \\
\bottomrule
\end{tabular}
\end{center}
 {\small
\begin{table*}[tp]
\centering
\small
\caption{Direct S1 personal-history contrast at fixed pooled anonymous H100, using the original three paired seeds and joint item/seed intervals. Flag changes measure directional influence. Error rates and loss measure quality. This personally linked 100-versus-10 comparison was run on S1.}
\label{tab:personal-direct}
\begin{tabular}{lrr}
\toprule
100 versus 10 historical items & Difference & Original joint 95\% interval \\
\midrule

UNSAFE flags (pp) & +48.86 & [25.34, 70.55] \\
False-positive rate (pp) & +48.12 & [24.09, 68.93] \\
False-negative rate (pp) & $-51.19$ & [$-80.77$, $-23.33$] \\
Soft-label BCE & +0.0683 & [0.0157, 0.1093] \\
\bottomrule
\end{tabular}
\end{table*}
}
 S1's comparison of 100 versus 10 person-linked historical items keeps pooled anonymous H100 fixed, isolating the value of linking more records to the people who supplied them. S1's original public mean is 15.64\% flagged, versus 0\% in S2. Hist selects 137/120 positive judgments out of 312 in S1/S2. These judgments determine training targets, while learned scores and the decision threshold determine flags. Original S2's three downward models and both publics classify every item as SAFE. Their net flag change is 0 pp despite a 0.09944 drop in mean predicted UNSAFE probability.

The downward-routing reference experiment compares P1 with Hist-down on S1, BASE and LARGER using newly fixed training randomness. S1/BASE have zero net change and zero decision disagreement. LARGER has net change -3.878 pp and gross disagreement 11.265\%. Its initialization/group interval [-7.480, 0.277] includes variation from training initialization and evaluation groups. The fixed-model/group-only interval [-6.740,-1.300] resamples evaluation groups while holding the fitted models fixed. The first asks how the estimate varies across both training and evaluation, whereas the second concerns these particular fitted models.
Two additional splits, A and B, use the same release, with their original two-primary upward Holm sensitivity 0.0010 each. The additional-public comparison adds four public assignments and compares downward routing with its identity control. The table distinguishes these reference assignments from the original two-public comparison and reports both routing directions.
{\small
\begin{table*}[tp]
\centering
\small
\caption{Additional text splits, raw signed history-minus-public flags (pp). Original upward basic intervals and the later public-panel basic intervals retain different scopes. Downward rows are secondary. They do not inherit the upward family adjustment. All models use three original paired streams and no independent C.}
\begin{tabular}{llrr}
\toprule Split & Direction & Original two publics [95\%] & Later four publics [95\%]\\\midrule

A & up & $+49.89\ [40.71, 60.81]$ & $+58.79\ [42.58, 76.22]$\\
A & down & $-10.62\ [-17.53,-2.32]$ & $-1.71\ [-3.65, 1.48]$\\
B & up & $+37.21\ [16.57, 56.98]$ & $+44.52\ [23.18, 64.16]$\\
B & down & $-8.90\ [-12.53,-4.79]$ & $-1.60\ [-3.20, 0.82]$\\\bottomrule\end{tabular}
\end{table*}
}
The later downward Hist-minus-identity-control contrast is -9.82 [-14.64,-4.29] pp in A and 0 pp in B, where both models flag every item SAFE. Original upward FPR rises from 8.82/6.03\% to 54.06/38.51\% in A/B while recall rises from 19.75/18.39\% to 90.12/74.71\%. Downward FNR is 98.77/100\%. These error rates describe the prediction-quality consequences alongside the directional effects and their sensitivity to the public reference.

\Needspace{12\baselineskip}
\subsection{Public assignments and training-environment comparisons}\label{app:public-history}\label{sec:public-audit}
{\small
\begin{table*}[tp]
\centering
\small
\caption{Raw flag contrasts (pp) by public-reference set and training environment. The original primary uses 3 seeds. Columns restricted to jointly available seeds use 2 for S1/S2 and 3 for BASE/LARGER. The original-Hist/additional-public comparison uses additional public models trained in a later environment. The final column replays Hist in that same environment. Appendix~\ref{sec:closure-scope} defines each uncertainty scope.}
\label{tab:scope-public}
\begin{tabular}{lrrrr}
\toprule
Design & Original primary & \shortstack{Two publics /\\available streams} & \shortstack{Six publics /\\original Hist} & \shortstack{Six publics /\\replayed Hist} \\
\midrule

S1 & +56.74 & +52.91 & +63.13 & +63.13 \\
S2 & +5.02 & +5.14 & +5.02 & +5.02 \\
BASE & +9.60 & +9.60 & +8.79 & +8.88 \\
LARGER & -5.54 & -5.54 & +6.02 & +6.11 \\
\bottomrule
\end{tabular}
\end{table*}
}
 \paragraph{Separating a changed reference from a changed learner.}
For LARGER, the later comparison changes both the training environment and the public reference. Let $h_o,h_r$ be the old and replayed history-guided flag rates, and let $p_2,p_6$ be the original two-public and later six-public mean rates. Their contributions separate as
\[
(h_r-p_6)-(h_o-p_2)=(h_r-h_o)+(p_2-p_6).
\]
The recorded components are $+0.092$ and $+11.557$ percentage points. Keeping the two-public comparator during replay gives $-5.45$ rather than $-5.54$. Switching to six publics gives $+6.11$. Each estimate compares the history-guided model with its own set of public reference models. Each contrast uses its own resampling calculation, preserving the dependence among the models involved.

Each column identifies its comparator, available streams and interval, including when point estimates agree after rounding. Original H/public quality and calibration in the additional publics' training environment appear separately below. Replaying the models in that environment changed neither S1 nor S2 decisions. BASE/LARGER H flags each increase by 0.092 pp. Across 14 H/P1 replays, the maximum parameter and E-logit differences are 0.00113 and 0.07591, respectively. They quantify sensitivity to the execution environment. Original G was not replayed. For the missing original public runs, known H flag counts give deterministic bounds on the complete mean contrast: [36.15, 69.48] pp for S1 and [-28.39, 4.95] pp for S2 as the missing public mean ranges over [0, 1]. These bounds describe possible missing-run contributions rather than sampling uncertainty.
{\small
\begin{table*}[tp]
\centering
\small
\caption{Inference for the original Hist model versus additional public models, in flag pp. The table reports the percentile interval, the interval obtained by inverting the centered test, and that test's $p$ value for the same fixed comparison. The original primary remains separately reported.}
\begin{tabular}{lrrr}
\toprule
Design & Percentile 95\% & Inverted 95\% & Centered $p$ \\
\midrule

S1 & [48.483, 76.647] & [49.075, 77.180] & 0.0002 \\
S2 & [0.055, 12.697] & [-1.423, 11.469] & 0.0956 \\
BASE & [5.143, 12.831] & [4.980, 12.594] & 0.0002 \\
LARGER & [-2.588, 13.815] & [-2.323, 14.358] & 0.1342 \\
\bottomrule
\end{tabular}
\end{table*}
}
 Mean public--public disagreement is 5.48\%, 0.23\%, 1.48\%, 13.31\% in S1/S2/BASE/LARGER. The original LARGER H--public gross 18.37\% is compared with original P1--P2 gross 19.39\%, to distinguish ordinary reference variation from directional movement. The six-public display conditions on the six frozen random-priority assignments.
The public-reference replay restores the original trainable tensors and data order, then uses newly fixed randomness after initialization. Nine conditions per split share this complete state: Hist, the two original public assignments and six additional public assignments. The newly fixed random stream defines a separate comparison, with the earlier missing random state still unavailable. Its intervals resample E conversation groups while holding the fitted models and public assignments fixed.
{\small
\begin{table*}[tp]
\centering
\small
\caption{The public-reference replay, using shared newly fixed training randomness. Four rows have conditional intervals and two report descriptive means. R denotes the six additional public assignments. Public--public (PP) gross averages disagreement between public-model pairs, which share models. T+ mean is the mean number of positive training judgments selected by the public assignments. The all-eight summaries report arithmetic means, with CI and PP-gross unavailable.}
\label{tab:paired18-contrasts}
\begin{tabular}{llrrrr}
\toprule
Split & Public set & Net pp [conditional CI] & Gross \% & PP gross \% & T+ mean \\
\midrule

S1 & Six R & +75.11 [68.33, 81.53] & 75.11 & 10.32 & 107.833 \\
S1 & $P_1,P_2$ & +64.38 [58.16, 70.69] & 64.38 & 35.62 & 109.500 \\
S1 & All eight (desc.) & +72.43 [NA] & 72.43 & NA & 108.250 \\
S2 & Six R & +4.68 [1.44, 8.44] & 4.68 & 0.23 & 97.667 \\
S2 & $P_1,P_2$ & +4.79 [1.45, 8.57] & 4.79 & 0.00 & 101.500 \\
S2 & All eight (desc.) & +4.71 [NA] & 4.71 & NA & 98.625 \\
\bottomrule
\end{tabular}
\end{table*}
}
 In this block S1 H FPR is 78.26\% and recall 96.43\%, versus six-public means 4.64\% and 15.48\%. Soft BCE is 0.74387 versus 0.61996. S2 H FPR is 1.72\%, FNR 82.14\%, and BCE 0.61221 versus six-public 0.63996. The block contains no new G, histogram control or C.

\subsection{Label placement, ranking and calibrated decisions}\label{sec:histogram-effects}
Two auditor controls per DICES990 budget use recorded current judgments to change which items receive positive supervision. They preserve native quotas and the number of training items receiving 0/1/2/3 positive votes. Every selected judgment still comes from its actual owner. This count of items at each positive-vote value is the \emph{histogram}. It stays fixed while the assignment changes which items receive those values. The comparison includes the zero BASE mean and the lower control--public rate in LARGER.
{\small
\begin{table*}[tp]
\centering
\small
\caption{Auditor histogram controls, selected using current judgments: raw flag and AUROC contrasts. H is history-guided routing. The control mean averages the histogram controls, and the public mean averages the public models. The original H--public primary is reported separately.}
\begin{tabular}{llrr}
\toprule
Budget & Contrast & $\Delta$ raw flag pp [95\% CI] & $\Delta$AUROC [95\% CI] \\
\midrule

BASE & H minus control mean & +0.00 [-11.27, +10.41] & -0.0317 [-0.0910, +0.0214] \\
BASE & Control minus public mean & +9.60 [+1.97, +17.62] & -0.0280 [-0.0548, -0.0017] \\
LARGER & H minus control mean & +17.31 [+12.78, +22.18] & -0.0138 [-0.0720, +0.0477] \\
LARGER & Control minus public mean & -22.85 [-34.96, -11.91] & -0.0362 [-0.0883, +0.0141] \\
\bottomrule
\end{tabular}
\end{table*}
}
The control mean averages the two fixed histogram controls after computing each model's endpoint. The diagnostic intervals resample paired groups and seeds, conditional on those two placements. The following tables give ranking contrasts and source-C decision and error profiles. This label-aware comparison isolates the consequence of assigning the same collection of target values to different items. Its inference concerns these fixed honest placements.
 BASE's H--control differences across streams are -11.7729,+10.6648,+1.1080 pp. In LARGER, H exceeds the control by +17.31 pp, while the control--public difference is -22.85 pp. Adding these differences recovers the original H--public value of -5.54 pp. With uniform item weights and the same total positive supervision, the two target sets contribute equally to any constant component of the model's logits. Their association with different content can still change BCE and its gradients.
{\small
\begin{table*}[tp]
\centering
\small
\caption{Original DICES990 ranking comparisons. These secondary diagnostics measure changes in ordering alongside the primary flag-rate effects.}
\begin{tabular}{llrrr}
\toprule
Budget & Contrast & $\Delta$AUROC & 95\% interval & Holm $p$ \\
\midrule

BASE & H minus public mean & -0.0597 & [-0.1038, -0.0191] & 0.0234 \\
BASE & H-G & -0.0670 & [-0.1248, -0.0125] & 0.0408 \\
LARGER & H minus public mean & -0.0499 & [-0.0963, -0.0063] & 0.0408 \\
LARGER & H-G & -0.0659 & [-0.1076, -0.0288] & 0.0112 \\
\bottomrule
\end{tabular}
\end{table*}
}
The other designs show how ranking quality varies with the learner and split: 350 Qwen S1/S2 AUROC differences are -0.0004/+0.0085, while Mistral S1/S2 are +0.0643/+0.0087. The 990 rank reference has 354 non-tied items (41 positive, 313 negative). None of the 5000 exponential-weight draws lacks a class. The displayed four-test ranking family concerns AUROC.
 Constant offsets preserve each model's ranks, AUROC, average precision (AP) and fixed top-$K$ set. Comparing models' ranked sets therefore tests a different aspect of influence from changing their flag rates. The AUROC family is a secondary four-test diagnostic, separate from raw flag primaries. The original top10/top20 rules select $K=37/73$ items in DICES990.
{\small
\begin{table*}[tp]
\centering
\small
\caption{Text calibration: net H--public flags (pp), gross decision disagreement (\%), and H prediction quality. Common uses the P1/P2 C offset, while threshold rows use per-model C offsets with UNSAFE probability $p=1/(1+\exp(-z))$. Metrics are computed per model before averaging, and original calibration is retained during replay.}
\label{tab:calibration-scopes}
\begin{tabular}{llrrrrr}
\toprule
Design & Rule & Net pp & Gross\% & H FPR\% & H FNR\% & H BCE \\
\midrule

\multicolumn{7}{l}{\emph{Original H / original two P}} \\
BASE & raw & +9.603 & 9.603 & 7.03 & 74.80 & 0.540 \\
BASE & common & +5.540 & 5.540 & 3.41 & 80.49 & 0.533 \\
BASE & $p>0.5$ & +0.092 & 0.092 & 0.00 & 99.19 & 0.536 \\
BASE & $p>0.2$ & -1.247 & 16.297 & 38.13 & 21.14 & 0.536 \\
LARGER & raw & -5.540 & 18.375 & 15.76 & 59.35 & 0.541 \\
LARGER & common & +0.000 & 5.448 & 2.56 & 86.18 & 0.546 \\
LARGER & $p>0.5$ & +0.277 & 4.894 & 1.60 & 81.30 & 0.539 \\
LARGER & $p>0.2$ & -1.293 & 13.204 & 39.72 & 24.39 & 0.539 \\
\midrule
\multicolumn{7}{l}{\emph{Same-runtime H / new six R}} \\
BASE & raw & +8.880 & 9.557 & 7.03 & 73.98 & 0.540 \\
BASE & common & +5.525 & 5.925 & 3.62 & 79.67 & 0.534 \\
BASE & $p>0.5$ & +0.092 & 0.092 & 0.00 & 99.19 & 0.536 \\
BASE & $p>0.2$ & -0.970 & 22.176 & 38.13 & 20.33 & 0.536 \\
LARGER & raw & +6.110 & 14.758 & 15.87 & 59.35 & 0.541 \\
LARGER & common & +1.554 & 3.894 & 2.56 & 86.18 & 0.546 \\
LARGER & $p>0.5$ & +0.646 & 3.509 & 1.60 & 81.30 & 0.539 \\
LARGER & $p>0.2$ & -1.216 & 12.758 & 39.83 & 24.39 & 0.539 \\
\bottomrule
\end{tabular}
\end{table*}
}
 The table reports the net change in flag rate, the fraction of decisions that differ, and prediction errors. Text calibration adds a constant to a model's unsafe logits, shifting its predicted probabilities. A common offset uses one constant fitted from the original P1/P2 predictions on C and applies it to every condition. A per-model offset fits a separate constant for each model using its own C predictions and the C reference judgments. All independent logistic offsets use C only, bounds [-8, 8] and ridge 0.001. The fixed 0.2 probability threshold is evaluated separately. Another original rule sets a threshold for 80\% recall on C. It reports NA when C contains too few reference-positive or reference-negative items to fit that rule. Reporting both disagreement and error rates shows which decisions change even when net effects are small or nearly all predictions are SAFE. S1/S2 calibration is NA.

\subsection{External text and reference provenance}\label{app:external-text}
The 24 original BASE/LARGER P1/P2/H/G scorers classify 4,096 question--answer pairs from BeaverTails \cite{ji2023beavertails}, grouped under 2,849 prompts. They score existing text, without generating answers or training on this dataset. The release contains repeated question--answer pairs. We retain each distinct pair once, together with the safety label on its first occurrence in the released file. These are the \emph{first-copy labels}. A fixed ordering chosen without labels selects a prefix of whole prompt groups. Scoring uses the original tokenizer and text serialization, FP32 with batch size one, and the 2,048-token limit. No item is truncated.

Calibrated scores use the offsets fitted in the original DICES experiments. The public-offset rule applies one common logit shift. The per-model rule applies a shift fitted separately for each model. Here \emph{source C} means the DICES calibration set, distinct from the external BeaverTails items. Both rules transfer unchanged to BeaverTails. Raw H--public forms the two-test primary family. Both budgets also report H--G, G--public, public--public, calibration and gross disagreement. Appendix~\ref{app:external-reference-audit} examines reference-label choices for duplicate pairs while keeping these model outputs fixed.
{\small
\begin{table*}[tp]
\centering
\small
\caption{BeaverTails raw history-minus-public UNSAFE-flag changes and their prespecified uncertainty sensitivities. Positive changes mean more items are flagged UNSAFE. All intervals are in percentage points.}
\begin{tabular}{lrrrr}
\toprule Budget & Group+seed [95\%] & Group-only [95\%] & Centered $p$ & Holm $p$\\\midrule

BASE & $+6.45\ [+3.55,+8.63]$ & $+6.45\ [+5.71,+7.22]$ & 0.0002 & 0.0004\\
LARGER & $-7.66\ [-14.56,-3.38]$ & $-7.66\ [-8.39,-6.95]$ & 0.0070 & 0.0070\\
\bottomrule\end{tabular}
\end{table*}
}
{\small
\begin{table*}[tp]
\centering
\small
\caption{BeaverTails history-minus-comparator effects under each scoring rule. The public offset is one common logit shift, while per-model offsets are fitted separately. Both use the original DICES calibration set and remain fixed on BeaverTails. Logits use native units, and net/gross changes are percentage points.}
\begin{tabular}{lllll}
\toprule
Budget / rule & Contrast & Logit & Net flags & Gross flags\\\midrule

BASE / Raw & $H-P$ & $+0.43\ [+0.40,+0.46]$ & $+6.45\ [+3.55,+8.63]$ & $+6.45\ [+3.55,+8.63]$\\
BASE / Raw & $H-G$ & $+0.56\ [+0.46,+0.63]$ & $+6.49\ [+3.55,+8.72]$ & $+6.49\ [+3.55,+8.72]$\\
BASE / Public offset & $H-P$ & $+0.43\ [+0.40,+0.46]$ & $+2.49\ [+1.70,+3.34]$ & $+2.49\ [+1.70,+3.34]$\\
BASE / Public offset & $H-G$ & $+0.56\ [+0.46,+0.63]$ & $+2.49\ [+1.70,+3.34]$ & $+2.49\ [+1.70,+3.34]$\\
BASE / Per-model offset & $H-P$ & $-0.12\ [-0.14,-0.10]$ & $-0.04\ [-0.14,+0.00]$ & $+0.04\ [+0.00,+0.14]$\\
BASE / Per-model offset & $H-G$ & $+0.05\ [-0.02,+0.15]$ & $+0.00\ [+0.00,+0.00]$ & $+0.00\ [+0.00,+0.00]$\\
LARGER / Raw & $H-P$ & $-0.18\ [-0.38,-0.06]$ & $-7.66\ [-14.56,-3.38]$ & $+10.99\ [+8.40,+14.86]$\\
LARGER / Raw & $H-G$ & $-0.07\ [-0.96,+0.40]$ & $-15.52\ [-57.46,+7.76]$ & $+23.19\ [+3.73,+57.46]$\\
LARGER / Public offset & $H-P$ & $-0.18\ [-0.38,-0.06]$ & $-0.20\ [-0.57,+0.05]$ & $+0.50\ [+0.01,+1.24]$\\
LARGER / Public offset & $H-G$ & $-0.07\ [-0.96,+0.40]$ & $-13.00\ [-38.60,+0.16]$ & $+13.09\ [+0.00,+38.60]$\\
LARGER / Per-model offset & $H-P$ & $-0.16\ [-0.30,+0.06]$ & $-0.45\ [-0.93,+0.04]$ & $+0.60\ [+0.16,+1.07]$\\
LARGER / Per-model offset & $H-G$ & $-0.46\ [-0.68,-0.24]$ & $-2.96\ [-7.86,-0.17]$ & $+2.99\ [+0.24,+7.86]$\\
\bottomrule\end{tabular}
\end{table*}
}
{\small
\begin{table*}[tp]
\centering
\small
\caption{BeaverTails raw prediction quality. Each distinct question--answer pair uses the safety label on its first released occurrence. BCE measures error in predicted UNSAFE probabilities (lower is better). AUROC measures how well scores rank reference-UNSAFE items above reference-SAFE items (higher is better). Flag/FPR/FNR are percentages. Metrics are computed for each model before averaging.}
\begin{tabular}{lllllll}
\toprule
Budget & Model & Flag & FPR & FNR & BCE & AUROC\\\midrule

BASE & P1 & 0.13 & 0.06 & 99.81 & 0.809 & 0.675\\
BASE & P2 & 0.10 & 0.02 & 99.84 & 0.810 & 0.702\\
BASE & H & 6.56 & 4.10 & 91.56 & 0.726 & 0.639\\
BASE & G & 0.07 & 0.02 & 99.90 & 0.844 & 0.711\\
LARGER & P1 & 12.87 & 5.02 & 81.12 & 0.707 & 0.722\\
LARGER & P2 & 14.26 & 7.72 & 80.75 & 0.723 & 0.705\\
LARGER & H & 5.91 & 4.17 & 92.76 & 0.771 & 0.655\\
LARGER & G & 21.43 & 16.65 & 74.92 & 0.782 & 0.686\\
\bottomrule\end{tabular}
\end{table*}
}
Offsets fitted on DICES C give near-all-SAFE decisions on BeaverTails in both budgets. The quality and disagreement measures report the consequences of applying this fixed score adjustment to the external items.
 \begingroup\let\GerrySavedSubsection\subsection\let\section\GerrySavedSubsection\let\subsection\subsubsection
\section{External-reference provenance and fixed-score sensitivity}\label{app:external-reference-audit}
This audit examines the labels attached to the same 4,096 distinct question--answer pairs, holding all model scores and UNSAFE flags fixed. The primary evaluation uses the label on each pair's first occurrence in the released file. The classification release contains 33,396 rows but 11,088 distinct byte-exact question--answer pairs, leaving 22,308 repeated rows after the first copies. All 11,088 distinct pairs occur more than once. Among their duplicate groups, 3,075 contain conflicting Boolean safety labels. In the selected sample, 1,136 pair identities have such conflicts and 2,960 do not. The dataset's Boolean safety labels are converted to UNSAFE labels by reversing true and false. Each score is matched to the same question--answer pair and its source label. Unicode compatibility normalization (NFKC), which standardizes equivalent character forms, and whitespace normalization change some strings but do not merge previously distinct pairs. Tokenization also produces no identical token sequences from byte-distinct pairs. These counts measure repeated records in the release. The number of independent evaluators behind each pair and the reasons for conflicting labels are unspecified.

Accuracy measures agreement with the chosen reference labels. Brier loss is the mean squared error of the predicted UNSAFE probabilities, with lower values indicating better prediction. We compare first-copy labels with two alternative references: averaging evaluation over all released copies of each pair, and retaining only pairs whose copies agree. Every distinct pair receives equal weight within the stated subset.
{\small
\begin{table*}[tp]
\centering
\small
\caption{Reference-label sensitivity on the same BeaverTails predictions: history minus fixed-public-mean. First-copy uses each pair's first released label. Soft-copy evaluates each prediction against all copies' labels and takes the mean. Consistent-only retains the 2,960/4,096 pairs whose labels agree. Accuracy differences are percentage points (positive is better), and Brier differences are squared-probability loss (negative is better). Choice ranges vary conflicting labels identically for all models and are not confidence intervals. Per-model offsets remain those fitted on DICES C.}
\begin{tabular}{lrrrrrr}
\toprule Budget / rule & First-copy acc. & Soft-copy acc. & Consistent acc. & Acc. choice range & First-copy Brier & Brier choice range\\\midrule

BASE / Raw & +2.93 & +2.78 & +3.83 & [+1.43,+4.10] & -0.038 & [-0.064,-0.011]\\
BASE / Per-model offset & -0.04 & -0.03 & -0.05 & [-0.04,-0.02] & +0.019 & [+0.012,+0.025]\\
LARGER / Raw & -5.75 & -5.69 & -7.73 & [-7.48,-3.69] & +0.026 & [+0.014,+0.037]\\
LARGER / Per-model offset & -0.41 & -0.33 & -0.45 & [-0.45,-0.20] & +0.026 & [+0.018,+0.034]\\
\bottomrule\end{tabular}
\end{table*}
}
For an external question--answer pair, let $q$ be the fraction of its released copies labeled UNSAFE. The released-copy reference evaluates the same prediction against each copy's label and takes the mean loss. Its expected Brier loss includes the label variance $q(1-q)$, which is absent from squared error to the fraction itself. Changing the reference can change measured accuracy and Brier loss but cannot change the fixed models' UNSAFE flags. The choice ranges measure sensitivity to conflicting released labels, with the evaluated items and model predictions held fixed. These alternative references are used only for evaluation, without selecting a threshold, model, assignment or sample.
 \endgroup
 \section{Learning video hazard ratings from personal histories}
\subsection{Video inputs and the original learner}\label{sec:road-status}\label{app:road-calibration}
Road records perceived-hazard ratings from 48 people for 1,356 video files. The fixed split contains 288 history, 288 training, 192 calibration and 588 evaluation items (H288/T288/C192/E588). E contains 219 groups of videos sharing a source. A panel size of 3 selects 864 ratings, with 18 per person. A panel size of 6 selects 1,728 ratings, with 36 per person. These designs preserve workloads and panel sizes without demographic quotas. Every condition represents a video using four decoded frames from the file's first 333 ms. The original study reports 333 ms viewing. Our implementation uses the first-window convention consistently, while the original presentation code would be needed to identify its exact playback interval. Longer files contain changing frames after 333 ms. Grouping follows the documented filename-to-source mapping, including one assumed filename alias.
The original Road experiment includes four public assignments and both Hist directions under three training streams. Frozen ResNet50 features supply the history predictor. The shared scorer fine-tunes layer4 and its head for 144 updates. Historical data alone determine its common four-epoch schedule before training or evaluation judgments are used. The reported raw-score Hist contrasts retain the original six-test Holm-adjusted $p=0.0060$ each. That family includes the original companion-selector comparisons, so its adjustment is unchanged for the Hist results shown here.
{
\begin{table*}[tp]
\centering
\normalsize
\caption{Road Hist contrasts on 588 held-out excerpts (219 registered source groups). Every row is Hist minus public. Intended downward movement is negative. Score points are $100\times$ native rating differences, not accuracy pp. Raw flag is a secondary $f_{M,\xi}(x)>0.5$ contrast. C-only per-model offsets attenuate mean steering. Percentile intervals retain the original source-group, stream and public-draw scope. The six-test adjustment from the original design is unchanged.}
\label{tab:road-terminal}
\begin{tabular}{lrrr}
\toprule
Direction & Raw score points [95\% CI] & Raw flag pp & Calibrated score points [95\% CI] \\
\midrule

Hist up & +5.28 [+3.720, +6.833] & +13.73 & -0.46 [-2.163, +1.114] \\
Hist down & -3.72 [-4.987, -2.397] & -6.28 & +0.18 [-1.422, +1.793] \\
\bottomrule
\end{tabular}
\end{table*}
}
 {\small
\begin{table*}[tp]
\centering
\small
\caption{Hist continuous contrasts after per-model C3 calibration, in signed Hist-minus-public rating points. The original 95\% interval conditions on C. The C/E-refitted interval resamples C/E groups, paired streams and public assignments, refitting the score adjustment on each C draw. These Hist rows retain the original comparison definitions.}
\begin{tabular}{lrrr}
\toprule Contrast & Point & Original fixed C & C/E refit \\ \midrule

Hist up $-$ public & -0.457 & [-2.163, 1.114] & [-3.340, 2.082] \\
Hist down $-$ public & +0.176 & [-1.422, 1.793] & [-2.154, 2.742] \\
\bottomrule\end{tabular}
\end{table*}
}
 Every contrast subtracts the named comparator from the method being evaluated. Downward points and interval endpoints use this same sign convention, converted together from the earlier direction-standardized display. The underlying results are unchanged. Intervals that hold calibration fixed and those that resample C and E are shown separately. Public raw MSE 0.05304 exceeds the historical-data constant predictor's 0.04965. The coefficient of determination, $R^2$, compares squared prediction error with that of predicting E's mean for every item. Its negative public-model value means that the latter has lower squared error. The following table reports prediction quality alongside directional changes.
{\small
\begin{table*}[tp]
\centering
\small
\caption{Road raw model-quality means over three matched streams. RMSE is in native 0--1 units. Rate columns are percentages. $R^2$ is the coefficient of determination. Spearman $\rho$ measures agreement between the rankings induced by predicted and reference scores. Higher values indicate closer ordering. Directional score changes are reported separately.}\label{tab:optional6-quality}
\begin{tabular}{lrrrrrr}
\toprule
Condition & RMSE & $R^2$ & $\rho$ & FPR & FNR & Bal. acc.\\\midrule

P1 & 0.2255 & $-0.0474$ & 0.3434 & 16.86 & 60.78 & 61.18\\
P2 & 0.2347 & $-0.1342$ & 0.3032 & 20.31 & 59.04 & 60.33\\
P3 & 0.2257 & $-0.0478$ & 0.3811 & 18.39 & 58.82 & 61.39\\
P4 & 0.2350 & $-0.1361$ & 0.3216 & 19.77 & 63.62 & 58.31\\
Hist up & 0.2523 & $-0.3098$ & 0.4151 & 31.65 & 44.23 & 62.06\\
G Hist up & 0.2350 & $-0.1367$ & 0.3208 & 18.93 & 59.91 & 60.58\\
Hist down & 0.2300 & $-0.0888$ & 0.2721 & 13.33 & 69.06 & 58.80\\
G Hist down & 0.2344 & $-0.1317$ & 0.3383 & 21.46 & 54.47 & 62.04\\\bottomrule
\end{tabular}
\end{table*}
}
 
\subsection{Identity linkage and equivariance}\label{app:road-optional6}\label{app:road-identity21}
The identity-control follow-up adds upward and downward G assignments under the original three complete training states. It is reported separately from the original Road experiment. Each permutation gives a person another person's historical profile for assignment selection. The selected current ratings still come from their actual owners. Comparing Hist directly with G measures the value of linking history to the correct person. Two later fixed permutations add 12 model controls. Another 14 fixed permutations receive only assignment-level CPU diagnostics. The tables include both directions and their calibrated counterparts.
{\small
\begin{table*}[tp]
\centering
\small
\caption{Road history-identity control. Each row is Hist minus its corresponding $G_{\mathrm{Hist}}$ control. A negative raw downward contrast has the intended direction. Score points and flag percentage points are distinct units. Raw intervals resample E source groups and training streams together across conditions. Per-model calibrated intervals also resample C and refit calibration. These later controls reuse the original evaluation set.}
\label{tab:road-m2-identity}
\begin{tabular}{llrr}
\toprule
Rule & Direction & Score points [95\% interval] & Flag pp [95\% interval]\\\midrule

Raw & Up & $+5.03\ [3.49, 6.57]$ & $+13.49\ [9.34, 17.67]$\\
Raw & Down & $-4.60\ [-5.93,-3.32]$ & $-9.81\ [-14.16,-5.86]$\\
Per-model C & Up & $-0.82\ [-3.56, 1.99]$ & $+3.46\ [-1.39, 7.98]$\\
Per-model C & Down & $+0.74\ [-1.65, 3.14]$ & $+0.40\ [-3.79, 5.04]$\\\bottomrule
\end{tabular}
\end{table*}
}
 {\small
\begin{table*}[tp]
\centering
\small
\caption{Additional identity controls A and B: actual Hist minus each control in both directions. Negative down is intended. Intervals reuse the stored 2000 E-source-group/whole-stream weights. Per-model C additionally reuses and refits the stored C draws. They condition on each fixed permutation, not a distribution of permutations.}\label{tab:identity21}
\begin{tabular}{lllrr}
\toprule
Permutation & Rule & Direction & Score points [95\% interval] & Flag pp [95\% interval]\\\midrule

Identity A & Raw & Up & $+5.967\ [4.235, 7.873]$ & $+13.492\ [9.224, 17.899]$\\
Identity A & Raw & Down & $-3.775\ [-4.933,-2.649]$ & $-5.669\ [-8.629,-2.478]$\\
Identity B & Raw & Up & $+6.931\ [4.394, 9.462]$ & $+18.367\ [11.650, 24.796]$\\
Identity B & Raw & Down & $-4.150\ [-5.753,-2.520]$ & $-8.447\ [-12.102,-4.950]$\\\midrule
Identity A & Per-model C & Up & $-0.759\ [-3.259, 1.917]$ & $+0.680\ [-3.558, 5.275]$\\
Identity A & Per-model C & Down & $+0.321\ [-1.488, 2.152]$ & $+1.304\ [-1.901, 5.158]$\\
Identity B & Per-model C & Up & $+0.211\ [-3.082, 3.563]$ & $+5.329\ [-0.163, 11.074]$\\
Identity B & Per-model C & Down & $-1.758\ [-3.863, 0.345]$ & $-3.968\ [-8.075, 0.000]$\\\bottomrule
\end{tabular}
\end{table*}
}
 All three raw-score runs have the intended signs for each tested identity map. In this Road design, every person can evaluate every item and workloads are equal. All 16 fixed profile permutations therefore preserve the collection of forecast values for each item and the optimal quantized assignment objective after people are relabeled. This one-to-one correspondence under relabeling is the \emph{equivariance} checked here. The selected people and their real judgments can still differ. More generally, the relabeling must preserve workloads, demographic requirements and eligibility. The reported uncertainty holds the tested maps fixed.

\subsection{Amount of personal history and learner--panel comparisons}\label{app:road-historypanel22}\label{app:road-reviewer21}
The history-amount comparison fixes pooled H288, the ConvNeXt state prepared on history, and the original public models. A fixed ordering chosen without labels determines which personal records the predictor can use: the first 16, then 64, then all 288. Each smaller set is contained in the next. Raw separation is the upward-trained model's mean score minus the downward-trained model's mean score. It is 2.31, 4.91, 7.04 points across the three history levels. The corresponding gaps in selected real ratings, before training, are 2.28, 4.06, 7.41 points. The tables report both directions against public and G, calibration, prediction quality and variation among public assignments. At 16 records, Hist--G score intervals and Hist--public flag intervals contain zero.
{\footnotesize
\begin{table*}[tp]
\centering
\footnotesize
\caption{Raw signed Hist minus comparator. Personal-history budgets keep pooled H288 fixed. Negative downward contrasts are intended. P averages four public models per stream. G uses the original whole-profile permutation. The 95\% intervals describe the post-result sensitivity on the same evaluation set. Shared reference models are reused.}\label{tab:hp22-raw}
\begin{tabular}{llrr}
\toprule Personal H / panel & Contrast & Score points [interval] & Flag pp [interval]\\\midrule

16 / panel 3 & Hist up--P & $+1.13\ [0.47, 1.82]$ & $+2.49\ [-0.88, 5.67]$ \\
16 / panel 3 & Hist down--P & $-1.18\ [-1.91,-0.45]$ & $-1.25\ [-4.36, 1.66]$ \\
16 / panel 3 & Hist up--G & $+0.42\ [-0.50, 1.34]$ & $+0.11\ [-3.45, 3.71]$ \\
16 / panel 3 & Hist down--G & $-0.64\ [-1.48, 0.16]$ & $-2.55\ [-5.42, 0.06]$ \\
64 / panel 3 & Hist up--P & $+1.21\ [0.43, 2.03]$ & $+3.34\ [-0.12, 6.78]$ \\
64 / panel 3 & Hist down--P & $-3.70\ [-4.47,-2.94]$ & $-7.09\ [-10.59,-4.23]$ \\
64 / panel 3 & Hist up--G & $+1.27\ [0.06, 2.45]$ & $+3.12\ [-1.00, 7.27]$ \\
64 / panel 3 & Hist down--G & $-2.43\ [-3.13,-1.73]$ & $-5.50\ [-8.36,-2.92]$ \\
288 / panel 3 (reused) & Hist up--P & $+3.58\ [2.43, 4.85]$ & $+12.47\ [8.64, 16.93]$ \\
288 / panel 3 (reused) & Hist down--P & $-3.46\ [-4.43,-2.57]$ & $-9.47\ [-13.69,-6.01]$ \\
288 / panel 3 (reused) & Hist up--G & $+4.83\ [3.67, 6.12]$ & $+15.14\ [11.13, 19.64]$ \\
288 / panel 3 (reused) & Hist down--G & $-4.19\ [-5.08,-3.31]$ & $-14.12\ [-17.83,-10.67]$ \\
288 / panel 6 & Hist up--P & $+3.87\ [2.96, 4.90]$ & $+12.12\ [8.83, 15.57]$ \\
288 / panel 6 & Hist down--P & $-2.47\ [-3.11,-1.88]$ & $-9.03\ [-12.14,-6.18]$ \\
288 / panel 6 & Hist up--G & $+4.22\ [3.22, 5.30]$ & $+15.42\ [11.68, 19.49]$ \\
288 / panel 6 & Hist down--G & $-3.44\ [-4.17,-2.73]$ & $-12.02\ [-15.58,-8.77]$ \\
\bottomrule\end{tabular}
\end{table*}
}

{\footnotesize
\begin{table*}[tp]
\centering
\footnotesize
\caption{Per-model C offset, refit within saved C/E resamples. Personal-history budgets keep pooled H288 fixed. Negative downward contrasts are intended. P averages four public models per stream. G uses the original whole-profile permutation. The 95\% intervals describe the post-result sensitivity on the same evaluation set. Shared reference models are reused.}\label{tab:hp22-per_model}
\begin{tabular}{llrr}
\toprule Personal H / panel & Contrast & Score points [interval] & Flag pp [interval]\\\midrule

16 / panel 3 & Hist up--P & $+0.21\ [-0.93, 1.40]$ & $+1.77\ [-0.62, 3.96]$ \\
16 / panel 3 & Hist down--P & $-0.17\ [-1.43, 1.22]$ & $+1.43\ [-1.45, 3.77]$ \\
16 / panel 3 & Hist up--G & $+0.32\ [-1.37, 2.07]$ & $+1.25\ [-2.43, 4.51]$ \\
16 / panel 3 & Hist down--G & $-0.34\ [-2.02, 1.39]$ & $-0.51\ [-4.39, 2.45]$ \\
64 / panel 3 & Hist up--P & $-0.48\ [-1.93, 0.93]$ & $+0.58\ [-2.08, 3.19]$ \\
64 / panel 3 & Hist down--P & $-0.31\ [-1.75, 1.10]$ & $-1.01\ [-3.65, 1.49]$ \\
64 / panel 3 & Hist up--G & $-0.54\ [-2.53, 1.44]$ & $-0.06\ [-4.03, 3.45]$ \\
64 / panel 3 & Hist down--G & $-0.47\ [-1.66, 0.76]$ & $-2.49\ [-5.20, 0.16]$ \\
288 / panel 3 (reused) & Hist up--P & $-0.64\ [-2.67, 1.23]$ & $+5.63\ [2.51, 9.80]$ \\
288 / panel 3 (reused) & Hist down--P & $-0.21\ [-1.79, 1.52]$ & $-3.84\ [-6.67,-0.90]$ \\
288 / panel 3 (reused) & Hist up--G & $-0.68\ [-2.66, 1.30]$ & $+5.95\ [2.72, 10.63]$ \\
288 / panel 3 (reused) & Hist down--G & $+0.12\ [-1.74, 1.87]$ & $-6.35\ [-9.85,-2.98]$ \\
288 / panel 6 & Hist up--P & $-0.44\ [-2.12, 1.16]$ & $+5.54\ [2.51, 8.79]$ \\
288 / panel 6 & Hist down--P & $-0.28\ [-1.38, 0.78]$ & $-4.21\ [-7.18,-1.79]$ \\
288 / panel 6 & Hist up--G & $-0.62\ [-2.44, 1.08]$ & $+6.24\ [2.94, 9.98]$ \\
288 / panel 6 & Hist down--G & $-0.15\ [-1.51, 1.20]$ & $-5.95\ [-9.36,-2.75]$ \\
\bottomrule\end{tabular}
\end{table*}
}

{\small
\begin{table*}[tp]
\centering
\small
\caption{Raw history-dose quality at every tested dose and panel. Each row keeps its original model reference. FPR/FNR/BA are percentages. MSE is squared native rating. The preceding table gives the matched offset-adjusted contrasts.}\label{tab:hp22-quality}
\begin{tabular}{llllll}
\toprule
Personal H/panel & Condition & MSE & FPR & FNR & BA\\\midrule

All H / panel 3 & Public & 0.04189 & 9.56 & 66.39 & 62.02\\
16 / panel 3 & Hist up & 0.04457 & 11.57 & 62.53 & 62.95\\
16 / panel 3 & Hist down & 0.04141 & 8.12 & 67.10 & 62.39\\
64 / panel 3 & Hist up & 0.04496 & 13.33 & 64.27 & 61.20\\
64 / panel 3 & Hist down & 0.04239 & 4.98 & 80.61 & 57.20\\
288 / panel 3 (reused) & Hist up & 0.04848 & 21.00 & 50.98 & 64.01\\
288 / panel 3 (reused) & Hist down & 0.04242 & 3.60 & 85.84 & 55.28\\
288 / panel 6 & Public & 0.04178 & 9.29 & 64.05 & 63.33\\
288 / panel 6 & Hist up & 0.04769 & 21.15 & 51.20 & 63.83\\
288 / panel 6 & Hist down & 0.04105 & 3.45 & 82.14 & 57.21\\
\bottomrule\end{tabular}
\end{table*}
}

{\footnotesize
\begin{table*}[tp]
\centering
\footnotesize
\caption{Gross disagreement: the percentage of items with different decisions in either direction. Each public model appears in several pairs, so the pairwise comparisons are dependent. The same public-model variation provides a reference for the history comparisons.}\label{tab:hp22-gross}
\begin{tabular}{llrrr}
\toprule Personal H / panel & Scoring & Hist up--P & Hist down--P & P--P\\\midrule

16 / panel 3 & Raw & 7.51 & 7.26 & 7.46 \\
16 / panel 3 & C offset & 4.24 & 4.97 & 4.58 \\
64 / panel 3 & Raw & 8.56 & 8.93 & 7.46 \\
64 / panel 3 & C offset & 4.95 & 5.46 & 4.58 \\
288 / panel 3 (reused) & Raw & 13.78 & 10.43 & 7.46 \\
288 / panel 3 (reused) & C offset & 7.67 & 5.65 & 4.58 \\
288 / panel 6 & Raw & 12.88 & 9.42 & 3.92 \\
288 / panel 6 & C offset & 6.90 & 4.80 & 2.75 \\
\bottomrule\end{tabular}
\end{table*}
}
 {\small
\begin{table*}[tp]
\centering
\small
\caption{Complete learner-by-panel raw public contrasts (method minus own public). Scores are rating points. Flags are pp. ResNet/panel 3 retains its original interval scope. ResNet/panel 6 uses its saved descriptive C/E/stream/public sensitivity. ConvNeXt intervals retain their original block. Each cell's interval applies to its stated learner, panel and comparator. Public MSE is computed per model before averaging.}\label{tab:learner-panel-complete}
\begin{tabular}{llrrr}
\toprule Learner/panel & Direction & Score [95\%] & Flags [95\%] & Public MSE\\\midrule

ResNet/3 & Up & $+5.28\ [3.72, 6.83]$ & $+13.73$ & 0.05304\\
ResNet/3 & Down & $-3.72\ [-4.99,-2.40]$ & $-6.28$ & 0.05304\\
ResNet/6 & Up & $+5.30\ [4.05, 6.65]$ & $+13.61\ [10.14, 17.19]$ & 0.05124\\
ResNet/6 & Down & $-2.34\ [-3.57,-1.21]$ & $-5.22\ [-9.19,-1.67]$ & 0.05124\\
ConvNeXt/3 & Up & $+3.58\ [2.43, 4.85]$ & $+12.47\ [8.64, 16.93]$ & 0.04189\\
ConvNeXt/3 & Down & $-3.46\ [-4.43,-2.57]$ & $-9.47\ [-13.69,-6.01]$ & 0.04189\\
ConvNeXt/6 & Up & $+3.87\ [2.96, 4.90]$ & $+12.12\ [8.83, 15.57]$ & 0.04178\\
ConvNeXt/6 & Down & $-2.47\ [-3.11,-1.88]$ & $-9.03\ [-12.14,-6.18]$ & 0.04178\\
\bottomrule\end{tabular}
\end{table*}
}
 The two-learner/two-panel comparison includes ResNet with panel size 6. Its upward/downward public contrasts are +5.30/-2.34 score points and +13.61/-5.22 flag pp. Its corresponding identity-control contrasts are +4.86/-4.29 points. Separation is 7.64, compared with 9.00 for ResNet with panel size 3. The paired change is -1.35 [-3.06, 0.36], a direction-consistent reduction with uncertainty about its sign. Its raw public MSE is 0.05124. ConvNeXt's public models at panel sizes 3/panel 6 have mean MSEs 0.04189/0.04178. All 12 individual public models per panel improve on the historical-data constant predictor's 0.04965. Each comparison includes the learner's architecture, preparation and training schedule.

\subsection{Within-person historical content}\label{app:gaps-context}
The historical-content controls, HC A and HC B, keep each person's identity and all their residual values. They shuffle which item is associated with each residual, separately within the two historical sets used to fit population predictions. The population predictor, features, shrinkage, clipping and original T judgments stay unchanged. Assignments are fixed before selected current judgments are retrieved. Twelve training trajectories use the original three ConvNeXt states and 72 updates. The primary comparison subtracts the mean separation under the two controls from separation under intact history. It is +7.27 [5.28, 9.56] points, with intact-history separation 7.04 and control mean -0.23. Uncertainty resamples the original E groups and training streams while holding these maps fixed.
Table~\ref{tab:hc-main} in the main paper gives the complete raw five-row comparison. The tables below provide individual-stream and scoring-rule results.
{\small
\begin{table*}[tp]
\centering
\small
\caption{Raw upward-minus-downward score separation (points) in all three shared training streams.}
\begin{tabular}{lrrr}
\toprule Condition & Stream 1 & Stream 2 & Stream 3\\\midrule

Original Hist & +7.85 & +6.72 & +6.54\\
Content map A & -0.53 & -1.19 & -0.94\\
Content map B & +0.25 & +0.34 & +0.67\\
Mean of A and B & -0.14 & -0.42 & -0.13\\
Hist minus A/B mean & +7.99 & +7.15 & +6.67\\
\bottomrule\end{tabular}
\end{table*}
}

\begin{table*}[tp]
\centering
\normalsize

\caption{Hist minus each content control: all directions and C3 calibration rules. Values are point [95\% interval]. Score units are points and decision units are pp.}
\begin{tabular}{llrrr}
\toprule Rule & Map/direction & Score & Net flags & Gross flags\\\midrule
Raw & Map A / up & $+4.33\ [+3.23,+5.62]$ & $+12.53\ [+8.97,+16.35]$ & $+14.23\ [+10.98,+17.93]$\\
Raw & Map A / down & $-3.59\ [-4.78,-2.56]$ & $-12.07\ [-16.51,-8.20]$ & $+13.66\ [+9.88,+17.82]$\\
Raw & Map B / up & $+3.49\ [+2.20,+4.85]$ & $+10.83\ [+6.35,+15.50]$ & $+13.55\ [+10.01,+17.65]$\\
Raw & Map B / down & $-3.13\ [-4.31,-2.03]$ & $-9.30\ [-13.68,-5.32]$ & $+12.13\ [+8.55,+16.12]$\\
Public common offset & Map A / up & $+4.33\ [+3.23,+5.62]$ & $+12.47\ [+8.45,+16.24]$ & $+12.70\ [+8.90,+17.14]$\\
Public common offset & Map A / down & $-3.59\ [-4.78,-2.56]$ & $-8.62\ [-13.38,-4.29]$ & $+9.30\ [+4.65,+14.21]$\\
Public common offset & Map B / up & $+3.49\ [+2.20,+4.85]$ & $+11.17\ [+6.63,+15.91]$ & $+13.32\ [+8.55,+17.76]$\\
Public common offset & Map B / down & $-3.13\ [-4.31,-2.03]$ & $-5.73\ [-10.49,-2.20]$ & $+7.31\ [+3.54,+12.74]$\\
Per-model offset & Map A / up & $-0.69\ [-2.59,+1.29]$ & $+4.25\ [+1.25,+8.31]$ & $+6.97\ [+4.36,+10.80]$\\
Per-model offset & Map A / down & $+0.21\ [-1.89,+2.20]$ & $-5.27\ [-9.28,-1.51]$ & $+7.20\ [+3.83,+11.71]$\\
Per-model offset & Map B / up & $-0.73\ [-2.74,+1.39]$ & $+4.20\ [+0.44,+8.87]$ & $+8.39\ [+4.92,+12.94]$\\
Per-model offset & Map B / down & $+0.26\ [-1.91,+2.21]$ & $-3.29\ [-7.67,+0.70]$ & $+7.48\ [+4.01,+12.33]$\\
Nonnegative affine & Map A / up & $-0.69\ [-2.23,+1.07]$ & $-1.19\ [-4.69,+4.24]$ & $+5.84\ [+2.43,+10.36]$\\
Nonnegative affine & Map A / down & $+0.27\ [-1.89,+2.12]$ & $-1.13\ [-6.63,+5.41]$ & $+6.92\ [+3.40,+11.75]$\\
Nonnegative affine & Map B / up & $-0.75\ [-2.39,+1.16]$ & $-2.10\ [-7.25,+4.13]$ & $+7.31\ [+3.97,+12.86]$\\
Nonnegative affine & Map B / down & $+0.32\ [-2.24,+2.34]$ & $+1.36\ [-5.33,+8.66]$ & $+9.18\ [+3.28,+14.22]$\\
\bottomrule\end{tabular}
\end{table*}

{\small
\begin{table*}[tp]
\centering
\small
\caption{Prediction quality for every within-person historical-content condition. MSE is shown under each C3 scoring rule. Raw FPR/FNR are percentages. The preceding table compares scores and decisions under these same conditions.}
\begin{tabular}{lllllll}
\toprule
Condition & Raw MSE & Common & Per-model & Affine & Raw FPR & Raw FNR\\\midrule

Public mean & 0.04189 & 0.04177 & 0.04178 & 0.04282 & 9.56 & 66.39\\
Hist up & 0.04848 & 0.04517 & 0.04570 & 0.04271 & 21.00 & 50.98\\
Hist down & 0.04242 & 0.04539 & 0.04281 & 0.04438 & 3.60 & 85.84\\
Map A up & 0.04227 & 0.04283 & 0.04243 & 0.04293 & 9.04 & 65.14\\
Map A down & 0.04267 & 0.04244 & 0.04268 & 0.04272 & 12.34 & 64.27\\
Map B up & 0.04352 & 0.04332 & 0.04332 & 0.04449 & 11.42 & 65.36\\
Map B down & 0.04064 & 0.04081 & 0.04087 & 0.04083 & 9.20 & 66.01\\
\bottomrule\end{tabular}
\end{table*}
}

 Mean-offset calibration adds a constant to each model's scores so that its mean on C matches the reference mean. After this adjustment, three of the four directional decision intervals exclude zero. All four affine-calibrated decision intervals contain zero. The controls preserve each person's residual distribution while changing its association with historical content. Thus the history-attribution comparison, the calibrated decisions and prediction quality describe different stages of the experiment.

\subsection{Learner retention and score calibration}\label{app:road-retention21}\label{app:oct01}
A separate ablation asks whether including historical items during training changes how the learner retains that information. It reuses the original ResNet assignments and complete states. Both conditions use 144 updates, adding 48 trajectories. ``Half-T'' multiplies current-item MSE by one half while retaining all selected T data. History replay adds one half of the H MSE. Batches contain eight T items and, for replay, eight H items. The two conditions therefore give the T loss the same weight. A bias-only control keeps the common prediction function fixed and analytically fits only its additive intercept to minimize T MSE. All conditions and four directional comparisons appear below. Replay improves fit on H. On E, the replay public models have MSE 0.05460, compared with 0.05304 for half-T and 0.04965 for the constant predictor. These measurements distinguish fitting historical judgments from predicting held-out judgments.
{\small
\begin{table*}[tp]
\centering
\small
\caption{Actual signed method minus comparator, raw rating points. Downward negatives indicate the intended direction. Original points provide a reference for the ablation. Their primary intervals are reported in the original-results table. Public MSE is averaged over individual models.}\label{tab:road-retention21}
\begin{tabular}{lrrrrr}
\toprule Learner & Hist$+$--P & Hist$-$--P & Hist$+$--G$+$ & Hist$-$--G$-$ & Public MSE\\\midrule

Original & $+5.280$ & $-3.716$ & $+5.032$ & $-4.602$ & 0.053044 \\
Half-T control & $+5.275$ & $-3.729$ & $+5.017$ & $-4.610$ & 0.053044 \\
Common-H replay & $+3.826$ & $-2.635$ & $+4.687$ & $-2.877$ & 0.054596 \\
Analytic bias only & $+4.720$ & $-2.691$ & $+5.179$ & $-3.387$ & 0.044966 \\
\bottomrule\end{tabular}
\end{table*}
}

{\footnotesize
\begin{table*}[tp]
\centering
\footnotesize
\caption{Common-H replay: raw and calibrated effects. Per-model offsets are refit on C in each saved C/E-group resample. Intervals also resample the three existing streams and four finite public draws. This post-result sensitivity uses the original evaluation set and pointwise inference.}\label{tab:road-retention21-ci}
\begin{tabular}{lrr}
\toprule Contrast & Raw score points [95\% interval] & Per-model-C score points [95\% interval]\\\midrule

Hist up--public & $+3.826\ [2.437, 5.290]$ & $+0.247\ [-1.973, 2.168]$ \\
Hist down--public & $-2.635\ [-3.967,-1.296]$ & $-0.096\ [-1.925, 1.821]$ \\
Hist up--G up & $+4.687\ [3.728, 5.657]$ & $-0.139\ [-2.174, 1.842]$ \\
Hist down--G down & $-2.877\ [-4.099,-1.464]$ & $+0.516\ [-1.356, 2.434]$ \\
\bottomrule\end{tabular}
\end{table*}
}
 Affine calibration fits a slope and offset using only the separate calibration set. For each Road model, let $f$ be its uncalibrated scores on C, $y$ the C reference ratings, and bars their means under the relevant resampling weights. Write covariance and variance on C as $\operatorname{Cov}_C$ and $\operatorname{Var}_C$. The nonnegative slope $a$ and offset $b$ are $ a=\max(0,\operatorname{Cov}_C(f,y)/\operatorname{Var}_C(f))$ and $ b=\bar y_C-a\bar f_C$. The fixed tolerance rule uses $ a=0$ when the score is constant. Predictions are unclipped $ af+b$, and flags use the strict 0.5 threshold. Each saved C draw refits the calibration. C3 and full48 provide different amounts of calibration information. All 96 point/reference fits have positive slopes and retain variation in fitted scores. All four primary affine flag intervals contain zero. The public-model quality results cover both panel sizes and every scoring rule.
{\small
\begin{table*}[tp]
\centering
\small
\caption{Mean-offset and nonnegative affine calibration. Signed Hist minus the four-public mean, in flag percentage points, at the unchanged threshold $f>0.5$. Both rules use C3 alone and are refitted inside the same 2,000 saved C/E draws. Marginal intervals describe each contrast's uncertainty under its scoring rule.}
\label{tab:oct01-affine}
\begin{tabular}{llr}
\toprule
Panel/direction & Rule & Flag pp [95\% interval]\\\midrule

3 / up & offset & $+5.63\ [2.51, 9.80]$\\
 & affine & $-1.08\ [-5.32, 3.89]$\\
3 / down & offset & $-3.84\ [-6.67,-0.90]$\\
 & affine & $-1.98\ [-6.33, 3.69]$\\
6 / up & offset & $+5.54\ [2.51, 8.79]$\\
 & affine & $-1.46\ [-5.12, 3.14]$\\
6 / down & offset & $-4.21\ [-7.18,-1.79]$\\
 & affine & $-1.23\ [-4.79, 3.17]$\\
\bottomrule\end{tabular}
\end{table*}
}
 {\small
\begin{table*}[tp]
\centering
\small
\caption{Public quality under the three C3 scoring rules, averaged over all four publics and three streams. Flag, FPR and FNR are percentages. Balanced accuracy is a fraction.}
\label{tab:oct01-quality}
\begin{tabular}{llrrrrr}
\toprule
Panel & Rule & MSE & Flag & FPR & FNR & Balanced acc.\\\midrule

3 & raw & 0.04189 & 15.82 & 9.56 & 66.39 & 0.620\\
3 & per model & 0.04178 & 8.77 & 4.25 & 78.38 & 0.587\\
3 & affine & 0.04282 & 11.22 & 6.09 & 74.18 & 0.599\\
6 & raw & 0.04178 & 16.23 & 9.29 & 64.05 & 0.633\\
6 & per model & 0.04167 & 9.03 & 4.25 & 77.40 & 0.592\\
6 & affine & 0.04289 & 11.72 & 6.03 & 72.11 & 0.609\\
\bottomrule\end{tabular}
\end{table*}
}
 With a common additive offset and no clipping, continuous mean differences are unchanged. Let $\Delta_E^{\rm raw},\Delta_C^{\rm raw}$ denote the same paired mean-score contrast on E and C, and $\Delta_E^{\rm cal}$ its value after per-model mean offsets. Under common within-set weights, $\Delta_E^{\rm cal}=\Delta_E^{\rm raw}-\Delta_C^{\rm raw}$. Each model retains its own ordering of items. Applying different offsets to different models can change comparisons between their scores. The tables distinguish both directions, both panel sizes, and the stated calibration rules.
 \section{From intervention costs to learning and supervision placement}
\subsection{Same majority targets and joint costs}\label{app:cost-learning}
For each design, we ask how cheaply P1 and P2 can reach the complete vector of item majorities produced by Hist-up. Items whose initial majority agrees with the target must retain that majority. Both minimum-cost answer editing and minimum-cost truthful reassignment succeed in all eight starting problems. Among truthful assignments with the same minimum cost, a fixed secondary objective minimizes the sum of absolute per-item differences from the positive counts produced by minimum-cost editing. The optimal truthful assignments match the editing count vector exactly at seven starts. At BASE/P1, the vectors differ by one vote. The same majorities can therefore coexist with different mean training targets and learned scores.
The joint-target audit reverses the first 2/4/8/12 items in each design's fixed ordering of 12 items while preserving every other P1 majority. Eleven cases are feasible and five are infeasible. Every infeasible prefix contains a question whose target is already impossible under its own eligibility and demographic quotas. Local supply alone therefore certifies these infeasibilities. In S1, reversing the twelve-item prefix costs 17 truthful replacements rather than the sum of 32 replacements for the separate single-item problems. Direct editing costs 16. A replacement that restores one person's workload can also help achieve another item's target. Optimizing the items together therefore reuses some of the replacements required by separate solutions.
{\small
\begin{table*}[tp]
\centering
\small
\caption{Minimum costs for the complete old Hist-up majority vector. Each cell lists P1/P2. The last column is the sum of absolute per-item positive-count differences between the honest and direct-edit endpoints, not a difference in majority outcomes. All sixteen endpoints attain their prescribed full vector.}
\label{tab:cost-learning-majority}
\begin{tabular}{lrrr}
\toprule Design & $C_{\rm flip}$ & $C_{\rm bal}$ & Count distance\\\midrule

S1 & 39/49 & 39/49 & 0/0\\
S2 & 39/36 & 39/36 & 0/0\\
BASE & 39/39 & 47/47 & 1/0\\
LARGER & 112/136 & 112/136 & 0/0\\\bottomrule
\end{tabular}
\end{table*}
}

{\small
\begin{table*}[tp]
\centering
\small
\caption{Native joint replacement cost / sum of old single-target costs. Prefixes and other-item majorities are fixed. ``Infeasible'' denotes the joint native problem. Unavailable single-target sums are not assigned zero.}
\label{tab:cost-learning-joint}
\begin{tabular}{lrrrr}
\toprule Design & 2 items & 4 items & 8 items & 12 items\\\midrule

S1 & 5/6 & 8/12 & 12/20 & 17/32\\
S2 & 8/8 & 13/16 & 17/26 & 24/40\\
BASE & 6/6 & Infeasible & Infeasible & Infeasible\\
LARGER & 6/6 & 10/12 & Infeasible & Infeasible\\\bottomrule
\end{tabular}
\end{table*}
}
 \paragraph{Input versus output identity.} We distinguish three kinds of equality. \emph{Input equality} requires identical token inputs and attention masks, target and weight tensors, batching and order, complete initial state, and numerical recipe and runtime. These are the learner's inputs, after the assignment has selected the judgments. \emph{Logit equality} requires identical saved logits for each item. \emph{Decision equality} requires identical thresholded decisions for each item. Verifying these equalities requires item-level outputs and input/state records, beyond an overall decision rate. The 44 verified exact-$q$ editing clones reuse existing models as aliases, without additional training. Let $c_p(M)=\sum_i m_{ip}y_{ip}$ be item $p$'s selected positive count. For an honest assignment $M$, $\sum_p|c_p(M)-c_p(P1)|\le\tfrac12\|M-P1\|_1$: direct editing can reproduce its per-item counts at no greater cost. The learning comparisons report the neural outcomes of the specified editing rules. The optimization selects training labels under each rule.

\subsection{Equal-budget routing and editing}\label{app:budget-grid}
The allowances are fixed at 5/10/20\% of the number of judgments in the starting assignment, rounded up: 16/32/63 edits for S1/S2, 22/44/87 for BASE and 65/130/260 for LARGER. Here an edit is one newly selected pair for reassignment or one flipped judgment for answer editing. H/G use historical forecasts to choose assignments before current T judgments are accessible. The separate auditor O then uses real T judgments to maximize the selected positive count under all native quotas. F, also labeled F0, keeps assignment P1 fixed and flips selected zero answers on a separate counterfactual copy. A predetermined ordering of eligible edits resolves ties among equally effective choices. Assignments with an edit budget are solved as binary programs. The allowance is an upper bound, not necessarily the cost actually used. O20 denotes O at the 20\% allowance and spends 63/63/86/260. Both O and F optimize the selected training-label total. Training the shared scorer then measures how those labels affect learned decisions.
The grid requests 192 condition/stream comparisons and 36 reference runs. Identical complete inputs and initial states reuse the same model, leaving 207 distinct training trajectories (188 newly trained, 19 reused exactly). The comparisons retain separate training-state sources: the downward-routing reference experiment, the supplementary reference experiment and the public-reference replay (Appendix~\ref{app:scope-ledger}). Earlier missing random states remain unavailable. Every model has complete evaluation scores, with independent calibration data for BASE/LARGER only. The primary family contains eight two-sided 20\% H--G/O--F tests, using the original 5,000 group/whole-stream weights and Holm adjustment. The other budget points complete the fixed grid and were not selected according to outcomes.
{\small
\begin{table*}[tp]
\centering
\small
\caption{Primary contrasts at the 20\% allowance, in raw UNSAFE decisions (pp). Conditions share three training streams, and intervals use the saved conversation-group and whole-stream draws. Holm adjustment covers the fixed eight-contrast family. H--G tests whether linking histories to the correct people matters. O--F compares truthful reassignment with answer editing when the auditor can read current T judgments.}
\label{tab:cost-learning-primary}
\begin{tabular}{lrrcrrc}
\toprule Design & H--G & 95\% interval & Holm $p$ & O--F & 95\% interval & Holm $p$\\\midrule

S1 & $+2.74$ & $[+0.00,+7.11]$ & $0.7475$ & $-3.20$ & $[-9.80,+1.00]$ & $1.0000$\\
S2 & $+0.00$ & $[+0.00,+0.00]$ & $1.0000$ & $+4.34$ & $[-1.84,+13.66]$ & $1.0000$\\
BASE & $+5.63$ & $[+2.66,+9.13]$ & $0.0140$ & $-3.23$ & $[-13.15,+4.90]$ & $1.0000$\\
LARGER & $+6.65$ & $[-4.03,+20.30]$ & $1.0000$ & $+17.54$ & $[+7.17,+28.55]$ & $0.0048$\\
\bottomrule\end{tabular}
\end{table*}
}
 \begin{figure*}[tp]
\centering\includegraphics[width=0.95\linewidth]{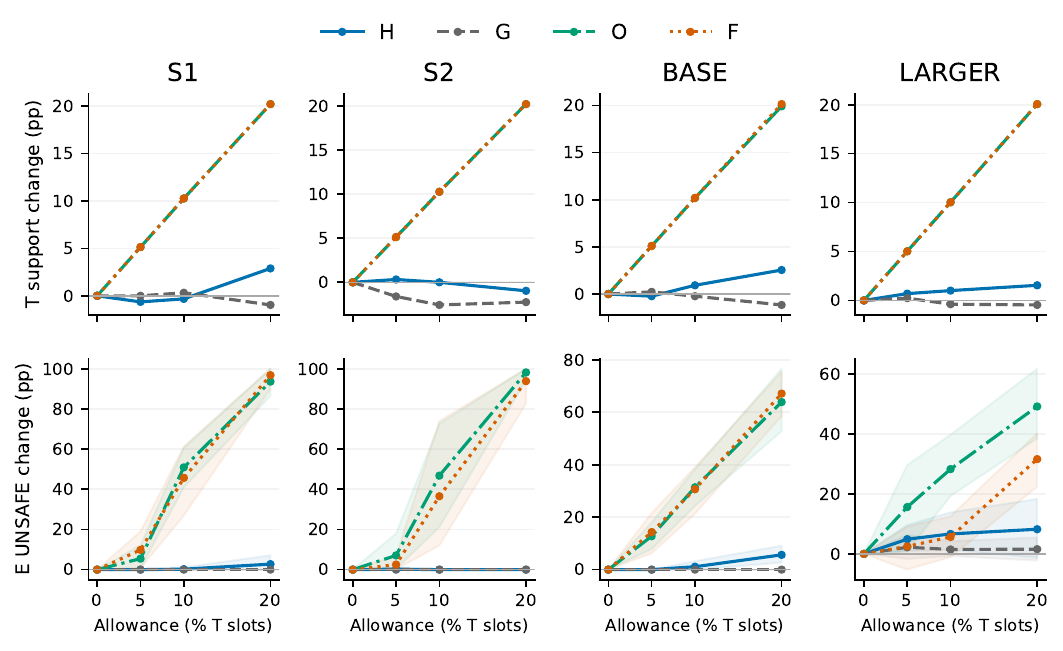}
\caption{The complete fixed 0/5/10/20\% budget grid for all four designs and arms, relative to P1 from the same training stream. Points and intervals come from the existing evaluated grid, without interpolation. The upper panels show selected training judgments, and the lower panels show model decisions on E. An allowance is the permitted cost, which can exceed the cost actually used. These descriptive intervals accompany the separate eight-test primary family.}\label{tab:cost-learning-all-budgets}
\Description{A four-design by two-outcome grid comparing all frozen budget levels and the H, G, O and F arms.}
\end{figure*}
 The grid separates selected supervision from learned decisions. At 20\% allowance, history routing changes the selected positive-vote counts by $+9,-3,+11,+20$ in S1/S2/BASE/LARGER, while the label-visible optimum changes them by $+63,+63,+86,+260$. All four designs and arms use the same displayed allowance grid.
\begin{table*}[tp]
\centering
\normalsize
\caption{Twenty-percent contrasts under raw and original C-only rules, in flag pp. Calibrated intervals refit the score adjustment on each saved C draw. Calibrated centered tests are descriptive. The preceding primary table gives the original raw Holm family. Intervals quantify directional uncertainty, with error costs reported in the quality table.}\label{tab:cost-learning-calibration}
\begin{tabular}{lrrrr}
\toprule Design & Rule & Contrast & Flag pp [95\% interval] & Centered $p$\\\midrule
BASE & Raw & H--G & $+5.63\ [+2.66,+9.13]$ & 0.0020\\
BASE & Raw & O--F & $-3.23\ [-13.15,+4.90]$ & 0.5133\\
BASE & Common C & H--G & $+3.32\ [+0.00,+10.54]$ & 0.1534\\
BASE & Common C & O--F & $-1.57\ [-10.17,+5.06]$ & 0.6735\\
BASE & Per-model C & H--G & $+0.00\ [+0.00,+1.53]$ & 1.0000\\
BASE & Per-model C & O--F & $+0.00\ [+0.00,+0.20]$ & 1.0000\\
LARGER & Raw & H--G & $+6.65\ [-4.03,+20.30]$ & 0.3407\\
LARGER & Raw & O--F & $+17.54\ [+7.17,+28.55]$ & 0.0006\\
LARGER & Common C & H--G & $+3.69\ [+0.00,+10.12]$ & 0.1296\\
LARGER & Common C & O--F & $+41.74\ [+27.88,+50.81]$ & 0.0002\\
LARGER & Per-model C & H--G & $+1.94\ [-0.76,+4.57]$ & 0.1370\\
LARGER & Per-model C & O--F & $+0.55\ [+0.00,+6.57]$ & 0.6147\\
\bottomrule\end{tabular}
\end{table*}
 {\small
\begin{table*}[tp]
\centering
\small
\caption{Raw prediction quality for every design and condition. Each cell is FPR / FNR (percent) / soft BCE. Metrics are computed per model before averaging. Source-C contrasts are reported in the adjacent table.}\label{tab:cost-learning-quality-pivot}
\begin{tabular}{lrrrr}
\toprule Arm & S1 & S2 & BASE & LARGER\\\midrule
F & 96.23 / 1.19 / 0.809 & 92.24 / 0.00 / 0.796 & 62.83 / 4.88 / 0.733 & 45.37 / 14.63 / 0.658\\
G & 0.00 / 100.00 / 0.632 & 0.00 / 100.00 / 0.663 & 0.00 / 100.00 / 0.531 & 16.29 / 46.34 / 0.539\\
H & 2.03 / 95.24 / 0.611 & 0.00 / 100.00 / 0.633 & 3.19 / 77.24 / 0.529 & 22.26 / 37.40 / 0.547\\
O & 91.88 / 0.00 / 0.784 & 97.70 / 0.00 / 0.787 & 59.00 / 4.88 / 0.716 & 63.90 / 2.44 / 0.856\\
H-ref & 66.09 / 4.76 / 0.712 & 2.30 / 84.52 / 0.613 & 6.92 / 74.80 / 0.540 & 16.51 / 58.54 / 0.542\\
P1 & 0.00 / 100.00 / 0.621 & 0.00 / 100.00 / 0.628 & 0.00 / 100.00 / 0.529 & 14.59 / 47.97 / 0.534\\
\bottomrule\end{tabular}
\end{table*}
}
The adjacent C-only table covers every applicable design and rule. BASE's per-model-C arms are all-SAFE. LARGER O/F false-negative rates are 95.93/100\%. These error rates quantify the operating-point cost alongside any attenuation of directional contrasts.
 The same logistic calibration rules apply to every condition, using C alone: offsets bounded by [-8, 8], ridge $10^{-3}$, and 44 bisection steps. Common adjustments use original P1/P2 as reference models. Calibration is refitted in every saved C draw. S1/S2 have no C. The quality table reports BASE's all-SAFE predictions and LARGER O/F false-negative rates under calibration. Gross disagreement, newly acquired flags and lost flags show which items change, alongside the net difference in flag rates.

\subsection{Fixed editing placement controls}\label{app:placement-controls}
F-TIE A/B use two predetermined tie-breaking orders to choose among edits that maximize the positive training count under the same 20\% allowance. F-HIST A/B instead redistribute O20's final counts among items with the same panel size $k$ and starting positive count $c^0$. They implement those counts through 0-to-1 edits while keeping assignment P1 fixed. Let $c^{\rm final}$ be the final positive count and $c^O$ the auditor's count. These controls preserve the number of items undergoing each $(k,c^0,c^{\rm final})$ transition, the final $q$ histogram, total positives, actual edit cost and number of majority gains. They change which content receives each transition. O's counts satisfy $c^O\ge c^0$. Its actual costs and positive-count gains are both 63/63/86/260 in S1/S2/BASE/LARGER. All 16 maps and 48 training requests were fixed before model outputs were examined. Original O, P and $F_0$ models and comparison families are retained.
{\small
\begin{table*}[tp]
\centering
\small
\caption{New fixed-map primary family, raw UNSAFE decisions (pp): honest O20 minus the mean of two editing controls. Intervals reuse the original 5,000 group/stream weights. Maps are fixed. Holm adjustment covers this eight-test family. The earlier eight-test family keeps its separate adjustment. F-HIST matches actual edit cost and the complete $(k,c_0,c_{\rm final})$ histogram. F-TIE uses the same allowance.}
\label{tab:placement-primary}
\begin{tabular}{llrr}
\toprule Design & Editing family & O minus control [95\% interval] & Holm $p$\\\midrule

S1 & F-TIE & $-4.57\ [-11.20,+1.29]$ & 0.7798\\
S1 & F-HIST & $-5.14\ [-12.56,+1.86]$ & 0.7798\\
S2 & F-TIE & $-1.60\ [-4.74,+0.00]$ & 0.7798\\
S2 & F-HIST & $-1.83\ [-5.56,+0.00]$ & 0.7798\\
BASE & F-TIE & $-2.08\ [-27.01,+20.31]$ & 1.0000\\
BASE & F-HIST & $-18.19\ [-30.98,-1.14]$ & 0.0994\\
LARGER & F-TIE & $+1.25\ [-7.33,+10.96]$ & 1.0000\\
LARGER & F-HIST & $+12.79\ [+6.01,+20.75]$ & 0.0064\\
\bottomrule\end{tabular}
\end{table*}
}
 The newer eight-test family uses the original 5,000 group/stream draws and holds all maps fixed. LARGER's original O--$F_0$ contrast of +17.54 pp and new O--F-TIE contrast of +1.25 pp use different editing tie-breaks. O--F-HIST is +12.79 pp when both conditions have the same transition histogram and actual edit cost. BASE O--F-HIST is -18.19 pp, with Holm $p=0.0994$ in this family. These comparisons measure the effect of assigning label changes to different content under the specified rules. Their optimization objective is the selected training-label count, with neural outcomes measured after training. LARGER's O--F-HIST net change of 12.79 pp accompanies 19.07\% gross disagreement (15.93 pp acquisition, 3.14 pp loss). Its O--F-TIE net change of 1.25 pp accompanies 13.43\% gross disagreement.
\begin{table*}[tp]
\centering
\normalsize
\caption{New family contrasts under the unchanged C-only rules. Entries are O minus the fixed two-map mean, in flag pp. Intervals refit the score adjustment on each original saved source-C draw. Calibrated tests describe sensitivity to the scoring rule. The primary family remains the raw comparison. S1/S2 have no C. The centered/basic inverse interval is reported separately from the percentile interval.}\label{tab:placement-calibration}
\begin{tabular}{lllrr}
\toprule Design & Rule & Family & Point [percentile interval] & Basic inverse interval\\\midrule
BASE & Raw & TIE & $-2.08\ [-27.01,+20.31]$ & $[-24.47,+22.85]$\\
BASE & Raw & HIST & $-18.19\ [-30.98,-1.14]$ & $[-35.24,-5.40]$\\
BASE & Common C & TIE & $+0.78\ [-24.64,+24.81]$ & $[-23.24,+26.21]$\\
BASE & Common C & HIST & $-16.85\ [-29.58,+1.04]$ & $[-34.74,-4.12]$\\
BASE & Per-model C & TIE & $+0.00\ [+0.00,+0.36]$ & $[-0.36,+0.00]$\\
BASE & Per-model C & HIST & $+0.00\ [+0.00,+0.36]$ & $[-0.36,+0.00]$\\
LARGER & Raw & TIE & $+1.25\ [-7.33,+10.96]$ & $[-8.46,+9.83]$\\
LARGER & Raw & HIST & $+12.79\ [+6.01,+20.75]$ & $[+4.82,+19.56]$\\
LARGER & Common C & TIE & $+23.78\ [+8.18,+32.15]$ & $[+15.41,+39.38]$\\
LARGER & Common C & HIST & $+25.53\ [+15.71,+31.97]$ & $[+19.09,+35.35]$\\
LARGER & Per-model C & TIE & $+0.55\ [+0.00,+6.27]$ & $[-5.16,+1.11]$\\
LARGER & Per-model C & HIST & $+0.42\ [-0.37,+4.85]$ & $[-4.02,+1.20]$\\
\bottomrule\end{tabular}
\end{table*}
 {\small{\small
\begin{table*}[tp]
\centering
\small
\caption{Raw prediction quality for every design and condition. Each cell is FPR / FNR (percent) / soft BCE. Metrics are computed per model before averaging. Source-C contrasts are reported in the adjacent table.}\label{tab:placement-quality-pivot}
\begin{tabular}{lrrrr}
\toprule Arm & S1 & S2 & BASE & LARGER\\\midrule
Public & 10.87 / 70.24 / 0.629 & 0.00 / 100.00 / 0.627 & 0.00 / 100.00 / 0.529 & 19.86 / 43.50 / 0.550\\
O20 & 91.88 / 0.00 / 0.784 & 97.70 / 0.00 / 0.787 & 59.00 / 4.88 / 0.716 & 63.90 / 2.44 / 0.856\\
F0 & 96.23 / 1.19 / 0.809 & 92.24 / 0.00 / 0.796 & 62.83 / 4.88 / 0.733 & 45.37 / 14.63 / 0.658\\
TIE-A & 95.65 / 0.00 / 0.817 & 99.43 / 0.00 / 0.820 & 64.00 / 8.13 / 0.725 & 66.35 / 3.25 / 0.741\\
TIE-B & 99.71 / 0.00 / 0.824 & 100.00 / 0.00 / 0.836 & 60.06 / 11.38 / 0.704 & 59.11 / 5.69 / 0.730\\
HIST-A & 97.68 / 0.00 / 0.799 & 100.00 / 0.00 / 0.823 & 84.35 / 0.81 / 0.794 & 45.90 / 15.45 / 0.685\\
HIST-B & 99.13 / 0.00 / 0.809 & 100.00 / 0.00 / 0.838 & 74.65 / 1.63 / 0.754 & 55.48 / 11.38 / 0.725\\
\bottomrule\end{tabular}
\end{table*}
}
The adjacent C-only table covers every applicable design and rule. BASE's per-model-C arms are all-SAFE. LARGER O/F false-negative rates are 95.93/100\%. These error rates quantify the operating-point cost alongside any attenuation of directional contrasts.
 }
The training loss explains why assigning the same count values to different items can matter. Let $\theta$ be model parameters, $z_\theta(x_p)$ the unsafe logit, $w_p$ the fixed item weight, and $L(\theta,q)$ the weighted soft-label BCE with its common regularizer. For target vectors $q,q'$, $L(\theta,q)-L(\theta,q')=-\sum_p w_p(q_p-q'_p)z_\theta(x_p)$ for common contents, weights and regularization. With uniform item weights, equal histograms cancel any logit component shared by every item. Components that depend on content can still change the loss and its gradients. This is a standard loss identity, used here to interpret supervision placement. Quality and calibration are reported separately from raw placement effects.

\subsection{Honest native realizations and histogram controls}\label{app:honest-realizability}
This audit asks whether truthful assignments can reproduce each of the sixteen existing editing controls' exact per-item positive counts. A binary program minimizes newly selected person--item pairs while preserving eligibility, workloads, panel sizes, demographic composition and the target counts. Independent assignment witnesses certify all eight feasible optima. Every BASE/LARGER target is infeasible because at least one item lacks enough eligible target-supporting judgments under its quotas. Thus every case has a certificate. Optimal assignments certify the feasible costs, and item-level supply bounds certify the infeasible targets.
{\small
\begin{table*}[tp]
\centering
\small
\caption{Sixteen fixed complete editing endpoints: minimum honest replacements under all native quotas. Cells show direct flip cost / honest optimum. ``Infeasible'' denotes certified native infeasibility, independently supported by local stratum supply. Every target has a certified optimum or an infeasibility certificate. The audit uses current training judgments and reuses existing models when the complete learner inputs match.}
\label{tab:honest-realizability}
\begin{tabular}{lrrrr}
\toprule Design & F-TIE A & F-TIE B & F-HIST A & F-HIST B\\\midrule

S1 & $63/64$ & $63/64$ & $63/63$ & $63/66$\\
S2 & $63/66$ & $63/66$ & $63/65$ & $63/65$\\
BASE & $87/\text{Infeasible}$ & $87/\text{Infeasible}$ & $86/\text{Infeasible}$ & $86/\text{Infeasible}$\\
LARGER & $260/\text{Infeasible}$ & $260/\text{Infeasible}$ & $260/\text{Infeasible}$ & $260/\text{Infeasible}$\\
\bottomrule\end{tabular}
\end{table*}
}
 S1 F-HIST A has a truthful realization, or \emph{lift}, at the same cost of 63 new pairs. The other S1/S2 lifts cost 64--66, compared with flip cost 63. Matching content, supervision, weights, order, complete initial state and runtime identifies 24 truthful-realization conditions whose models are exact aliases of existing editing trajectories. These certificates show how truthful reassignment supplies identical learner inputs, at the separately reported costs.
The R-HIST controls instead search within truthful assignments that retain O20's actual cost and complete transition histogram. Two fixed item/final-count priorities select the controls. O20 itself shows that the constraint set is nonempty. All eight maps are fixed before outputs are examined. The 24 condition/stream roles require 21 new trajectories and three exact BASE aliases. Original O/P/H/G/F models are reused. The four O--mean R-HIST primary comparisons have their own Holm adjustment using the original 5,000 weights, separate from either earlier eight-test family.
{\small
\begin{table*}[tp]
\centering
\small
\caption{The new honest-control primary, O20 minus the mean of two fixed R-HIST maps. Raw UNSAFE-rate differences are in percentage points. Intervals reuse the saved group/whole-stream weights. The two maps are fixed. Centered two-sided $p$ values and Holm correction cover these four tests. Each earlier eight-test family retains its own adjustment.}
\label{tab:honest-control-primary}
\begin{tabular}{lrrr}
\toprule
Design & Raw pp [95\% interval] & Centered $p$ & Holm $p$ \\
\midrule

S1 & $-4.34\ [-10.88,+1.27]$ & 0.1348 & 0.5391 \\
S2 & $-0.46\ [-2.64,+1.11]$ & 0.5615 & 1.0000 \\
BASE & $+0.00\ [-2.91,+2.18]$ & 1.0000 & 1.0000 \\
LARGER & $+4.34\ [-2.61,+11.49]$ & 0.2494 & 0.7481 \\
\bottomrule\end{tabular}
\end{table*}
}

{\small
\begin{table*}[tp]
\centering
\small
\caption{Source-C sensitivity for the same fixed honest controls. Every saved C draw refits the original calibration rule. The mean-logit column is in logit units, not percentage points. These secondary contrasts do not define a new primary family. S1/S2 have no C set.}
\label{tab:honest-control-calibration}
\begin{tabular}{llrr}
\toprule
Design & Rule & Flag pp [95\% interval] & Mean logit [95\% interval] \\
\midrule

BASE & Common public offset & $+0.32\ [-2.53,+3.27]$ & $+0.01\ [-0.03,+0.05]$ \\
BASE & Per-model offset & $+0.00\ [+0.00,+0.05]$ & $+0.00\ [-0.00,+0.00]$ \\
LARGER & Common public offset & $+8.26\ [-0.30,+16.32]$ & $+0.17\ [-0.02,+0.36]$ \\
LARGER & Per-model offset & $-0.05\ [-0.91,+1.74]$ & $-0.04\ [-0.10,+0.02]$ \\
\bottomrule\end{tabular}
\end{table*}
}

{\small
\begin{table*}[tp]
\centering
\small
\caption{All three original streams, reported in order as Streams 1--3. E contains 146 items in each 350 split and 361 in each 990 design. Each flag count thresholds its own model at zero. The last column subtracts the two control rates within stream before averaging. All three conditions select the same total number of positive training judgments.}
\label{tab:honest-control-streams}
\begin{tabular}{lrrrrr}
\toprule
Design & T positives & O flags & A flags & B flags & O$-$mean pp \\
\midrule

S1 & 158 & 132/132/146 & 146/145/143 & 135/145/144 & -5.82/-8.90/+1.71 \\
S2 & 163 & 140/144/146 & 140/146/145 & 144/143/146 & -1.37/-0.34/+0.34 \\
BASE & 199 & 280/218/194 & 265/211/216 & 280/218/194 & +2.08/+0.97/-3.05 \\
LARGER & 597 & 254/221/266 & 210/212/242 & 211/251/262 & +12.05/-2.91/+3.88 \\
\bottomrule\end{tabular}
\end{table*}
}

{\small
\begin{table*}[tp]
\centering
\small
\caption{All honest-placement conditions: raw and calibrated soft BCE, with raw FPR/FNR (proportions). CO/MO are the original common/per-model source-C offsets. NA means that the design has no independent C set.}\label{tab:honest-control-quality}
\begin{tabular}{llllll}
\toprule
Model & Raw BCE & CO BCE & MO BCE & Raw FPR & Raw FNR\\\midrule

S1 O20 & 0.7844 & NA & NA & 0.9188 & 0.0000\\
S1 R-A & 0.7914 & NA & NA & 0.9884 & 0.0000\\
S1 R-B & 0.8230 & NA & NA & 0.9594 & 0.0000\\
S2 O20 & 0.7865 & NA & NA & 0.9770 & 0.0000\\
S2 R-A & 0.7992 & NA & NA & 0.9799 & 0.0000\\
S2 R-B & 0.7897 & NA & NA & 0.9856 & 0.0000\\
BASE O20 & 0.7157 & 0.6859 & 0.5306 & 0.5900 & 0.0488\\
BASE R-A & 0.7071 & 0.6773 & 0.5301 & 0.5889 & 0.0407\\
BASE R-B & 0.7157 & 0.6859 & 0.5306 & 0.5900 & 0.0488\\
LARGER O20 & 0.8563 & 0.6520 & 0.5308 & 0.6390 & 0.0244\\
LARGER R-A & 0.7593 & 0.5996 & 0.5293 & 0.5612 & 0.0569\\
LARGER R-B & 0.8208 & 0.6410 & 0.5273 & 0.6230 & 0.0407\\
\bottomrule\end{tabular}
\end{table*}
}

{\small
\begin{table*}[tp]
\centering
\small
\caption{Decision changes for honest controls (pp). Net changes and their CIs appear in the preceding primary and calibration tables. Acquisition counts newly flagged items, and loss counts items no longer flagged. Gross adds both, including changes that cancel in the net rate.}\label{tab:honest-control-paired}
\begin{tabular}{llll}
\toprule
Design/rule & Gross [95\%] & Acquisition [95\%] & Loss [95\%]\\\midrule

S1 Raw & $+5.48\ [+1.58,+10.99]$ & $+0.57\ [+0.00,+2.16]$ & $+4.91\ [+0.00,+10.92]$\\
S2 Raw & $+1.83\ [+0.21,+4.63]$ & $+0.68\ [+0.07,+2.02]$ & $+1.14\ [+0.00,+3.41]$\\
BASE Raw & $+2.03\ [+0.89,+3.36]$ & $+1.02\ [+0.00,+2.20]$ & $+1.02\ [+0.00,+3.00]$\\
BASE CO & $+2.54\ [+0.87,+4.02]$ & $+1.43\ [+0.00,+3.49]$ & $+1.11\ [+0.00,+2.72]$\\
BASE MO & $+0.00\ [+0.00,+0.07]$ & $+0.00\ [+0.00,+0.06]$ & $+0.00\ [+0.00,+0.00]$\\
LARGER Raw & $+7.94\ [+4.61,+12.07]$ & $+6.14\ [+1.93,+11.58]$ & $+1.80\ [+0.00,+4.73]$\\
LARGER CO & $+10.76\ [+6.34,+16.64]$ & $+9.51\ [+3.42,+16.43]$ & $+1.25\ [+0.00,+4.07]$\\
LARGER MO & $+0.60\ [+0.00,+3.97]$ & $+0.28\ [+0.00,+2.65]$ & $+0.32\ [+0.00,+1.77]$\\
\bottomrule\end{tabular}
\end{table*}
}

 All four truthful-control intervals span zero. In BASE, control B reuses O's model because their complete training inputs match. O minus control A changes the number of flagged items by 15, 7, -22 across the three streams. These changes cancel in the mean while retaining item-level decision differences. This family is reported separately from the earlier editing comparisons, with calibration, quality and gross/net decision changes.
 \clearpage
\section{Collection governance and its interaction with history}
\begingroup\let\GerrySavedSubsection\subsection\let\section\GerrySavedSubsection\let\subsection\subsubsection
\section{Bounded collection-policy diagnostics}\label{app:bounded-policy-audit}
These diagnostics ask how much the collector could change selected judgments if all current labels were known. They use released DICES training judgments under the original eligibility, workload, panel-size and demographic constraints. This is an auditor's feasibility calculation, separate from history-only assignment. The original models and hypotheses remain fixed.

\paragraph{Local supply and whole-vector failure.} For a target bit $t_p$, let $a_{pg}(t_p)$ count available eligible evaluators in stratum $g$ whose observed label equals $t_p$. The greatest target support permitted by that item's stratum quotas alone is $U_p(t_p)=\sum_g\min\{k_{pg},a_{pg}(t_p)\}$. For a three-person panel, $U_p(t_p)<2$ rules out the target majority on that item. Every target item needs enough eligible supporters. Meeting all targets together also requires an assignment that respects each person's workload:
\begin{table}[tbp]
\centering\small
\begin{tabular}{lrrrr}\toprule
Design & \shortstack{T\\items} & \shortstack{UNSAFE\\blocked} & \shortstack{SAFE\\blocked} & \shortstack{Both locally\\possible}\\\midrule
S1 &104&0&0&104\\
S2 &104&0&0&104\\
BASE &144&30&0&114\\
LARGER &432&87&0&345\\\bottomrule
\end{tabular}\end{table}
All original menu outcomes remain unchanged. This local bound limits the all-UNSAFE target to at most 114/144 BASE items. Attaining these item-level bounds simultaneously requires a jointly feasible assignment. An unattainable full target can therefore coexist with opportunities to change individual items.

\paragraph{Nested public reservations.} Let $m^P$ be the original P1 witness. Before optimization, fix two orderings of its person--item pairs using only identifiers and predetermined tie-breaking rules, without consulting judgments. At $\alpha\in\{0, 1/4, 1/2, 3/4, 1\}$ reserve the first $\lfloor\alpha B\rfloor$ edges, where $B=\sum_i r_i$. All original row, item and item-stratum quotas still apply to the completed assignment. Reserved counts may differ across people. Each final workload returns to its own original value. P1 supplies a feasible completion at every fraction.

Let $\ell_\alpha$ mark the reserved pairs, so $m\ge\ell_\alpha$ requires every reserved pair to remain selected. Write $\F_\alpha=\{m\in\F:m\ge\ell_\alpha\}$ for the remaining feasible assignments. Reserving more pairs in the same order only removes choices: $\F_\beta\subseteq\F_\alpha$ for $\beta\ge\alpha$. We measure the gap between the highest and lowest achievable percentages of positive selected judgments:
\[
W(\alpha)=\frac{100}{B}\left[\max_{m\in\F_\alpha}\sum_{ip}m_{ip}y_{ip}-\min_{m\in\F_\alpha}\sum_{ip}m_{ip}y_{ip}\right].
\]
The width is nonincreasing because each newly reserved pair restricts the feasible set. Integer min-cost flow uses evaluator supplies, item-stratum demands, unit edge capacities and costs $\pm y_{ip}$. All resulting assignments satisfy the original quotas and retain every reserved pair. Reusing zero-reservation optima and the full-reservation P1 witness gives 40 records from 56 flow solves.
\begin{center}\small
\begin{tabular}{lrrrrr}\toprule
&\multicolumn{5}{c}{Max--min selected-label width (support pp)}\\
Design &0\%&25\%&50\%&75\%&100\%\\\midrule
S1&99.68&74.52&49.84&23.88&0\\
S2&99.04&74.04&49.36&23.88&0\\
BASE&71.53&53.47&33.22&13.66&0\\
LARGER&71.30&53.70&35.49&16.74&0\\\bottomrule
\end{tabular}\end{center}
Entries average the two fixed orders. The audit verifies the integer solutions and their participation constraints. Reservation reduces the collector's choices while preserving the annotation budget. Width measures the possible change in collected judgments when the auditor knows all current judgments. Its endpoint is collected supervision, before model training. Full reservation fixes the training targets. Comparing subsequent models from a shared initial state isolates the assignment change, whereas varying initialization also measures training variability.
 \endgroup
\subsection{Learning with publicly fixed assignments}\label{app:reservation-material}\label{app:road-reservation21}
A public rule fixes 432 of the 864 Road assignments from P1, using an order chosen without judgments. The collector completes the remaining assignments under the original constraints. Each person still evaluates 18 items, and each item has three evaluators, even though the fixed pairs need not be evenly distributed across people. Four public completions and both directions of Hist and G are chosen before selected T judgments are retrieved. Original models with no fixed assignments are reused. ResNet's 24 new models use four epochs/144 updates. ConvNeXt's 24 use two epochs/72 updates. The primary compares upward-minus-downward mean model scores at 50\% versus 0\% reservation. Public-reference means cancel in this difference. The same source-group and stream resampling weights compare the two regimes. Fixing all pairs would reproduce P1 and its existing model.
{\small
\begin{table*}[tp]
\centering
\small
\caption{Primary reservation sensitivity: Hist up-minus-down raw score separation. The interval resamples E source groups and original streams using existing draws. Public means cancel from this contrast. Negative paired change means reduced directional separation.}\label{tab:road-reservation-separation}
\begin{tabular}{lrr}
\toprule Regime & Mean [95\% interval], points & Three matched streams\\\midrule

Original 0\% & $8.996\ [7.018, 11.041]$ & 7.922 / 9.808 / 9.259\\
New 50\% & $3.473\ [1.954, 4.987]$ & 4.247 / 2.150 / 4.021\\
50\% minus 0\% & $-5.524\ [-7.651,-3.611]$ & $-3.674/-7.658/-5.238$\\
\bottomrule\end{tabular}
\end{table*}
}

{\footnotesize
\begin{table*}[tp]
\centering
\footnotesize
\caption{All new reservation contrasts, actual signed method minus comparator. Public intervals additionally resample the four new completions. C/E-refit uses the same independent C rule for every model.}\label{tab:road-reservation-contrasts}
\begin{tabular}{lrr}
\toprule Contrast & Raw points [95\% interval] & Per-model-C points [95\% interval]\\\midrule

Hist up--public & $+2.356\ [0.634, 4.148]$ & $-0.048\ [-1.494, 1.481]$\\
Hist down--public & $-1.117\ [-2.425, 0.296]$ & $+0.313\ [-1.032, 1.649]$\\
Hist up--G up & $+2.294\ [1.175, 3.402]$ & $-0.155\ [-1.929, 1.592]$\\
Hist down--G down & $-1.440\ [-2.778,-0.209]$ & $+0.875\ [-0.682, 2.426]$\\
\bottomrule\end{tabular}
\end{table*}
}
 \begin{table*}[tp]
\centering\small
\caption{ConvNeXt collection restriction, using the eight existing masks and three original paired streams. Raw up-minus-down score separation (points). A negative paired change means reduced directional influence. Public means cancel from this registered primary.}
\label{tab:oct01-reservation}
\begin{tabular}{lr}
\toprule
Locked edges & Score points [95\% interval]\\\midrule
0\% (reused) & $+7.04\ [5.42, 8.89]$\\
50\% & $+2.76\ [1.73, 3.89]$\\
50\% minus 0\% & $-4.28\ [-5.12,-3.48]$\\
\bottomrule\end{tabular}
\par\smallskip\parbox{\linewidth}{\footnotesize The three paired changes are $-4.68,-3.78,-4.37$ points. Direction-specific and calibrated comparisons are reported in Appendix~\ref{app:reservation-material}.}
\end{table*}
 {\small
\begin{table*}[tp]
\centering
\small
\caption{All 50\%-reservation method and identity contrasts, using C3 for the calibrated rules. Each contrast identifies the intervention, comparator, output scale, and C-only scoring rule.}
\label{tab:oct01-all-reservation}
\begin{tabular}{llrr}
\toprule
Scoring & Contrast & Score points [95\%] & Net flag pp [95\%]\\\midrule

raw & $\mathrm{Hist}^{+}-P$ & $+1.65$ $[+0.81,+2.50]$ & $+6.68$ $[+4.19,+9.68]$\\
raw & $\mathrm{Hist}^{-}-P$ & $-1.10$ $[-1.88,-0.41]$ & $-4.83$ $[-8.20,-2.28]$\\
raw & $\mathrm{Hist}^{+}-G^{+}$ & $+2.03$ $[+1.38,+2.74]$ & $+7.88$ $[+5.14,+11.07]$\\
raw & $\mathrm{Hist}^{-}-G^{-}$ & $-1.54$ $[-2.10,-1.00]$ & $-6.86$ $[-10.09,-4.24]$\\
raw & $G^{+}-P$ & $-0.38$ $[-1.06,+0.22]$ & $-1.20$ $[-3.39,+0.50]$\\
raw & $G^{-}-P$ & $+0.44$ $[-0.20,+1.07]$ & $+2.03$ $[+0.26,+4.07]$\\
per model & $\mathrm{Hist}^{+}-P$ & $-0.30$ $[-1.52,+0.90]$ & $+3.16$ $[+1.14,+6.23]$\\
per model & $\mathrm{Hist}^{-}-P$ & $-0.17$ $[-1.04,+0.70]$ & $-2.96$ $[-4.99,-1.07]$\\
per model & $\mathrm{Hist}^{+}-G^{+}$ & $-0.52$ $[-1.66,+0.64]$ & $+3.23$ $[+1.32,+6.34]$\\
per model & $\mathrm{Hist}^{-}-G^{-}$ & $+0.15$ $[-1.07,+1.39]$ & $-3.74$ $[-6.45,-1.82]$\\
per model & $G^{+}-P$ & $+0.22$ $[-0.42,+0.85]$ & $-0.07$ $[-1.55,+1.58]$\\
per model & $G^{-}-P$ & $-0.32$ $[-1.12,+0.46]$ & $+0.78$ $[-0.57,+2.79]$\\
affine & $\mathrm{Hist}^{+}-P$ & $-0.38$ $[-1.29,+0.63]$ & $-0.92$ $[-3.83,+2.30]$\\
affine & $\mathrm{Hist}^{-}-P$ & $-0.15$ $[-1.11,+0.70]$ & $-0.35$ $[-3.33,+2.33]$\\
affine & $\mathrm{Hist}^{+}-G^{+}$ & $-0.62$ $[-1.51,+0.35]$ & $-0.79$ $[-3.50,+2.42]$\\
affine & $\mathrm{Hist}^{-}-G^{-}$ & $+0.24$ $[-0.99,+1.24]$ & $+0.91$ $[-2.90,+4.30]$\\
affine & $G^{+}-P$ & $+0.24$ $[-0.48,+0.90]$ & $-0.13$ $[-2.34,+1.90]$\\
affine & $G^{-}-P$ & $-0.38$ $[-1.16,+0.46]$ & $-1.26$ $[-3.44,+1.29]$\\
\bottomrule\end{tabular}
\end{table*}
}
 
\paragraph{How reservation changes the collector's choices.} Reserved pairs occupy workload and panel slots. They can fill an evaluator's workload or an item's panel, excluding otherwise favorable pairs from the completion. Even when both have room, selecting one pair changes which other pairs can satisfy the remaining quotas. The collector therefore chooses a jointly feasible completion. The shared scorer then learns from all reserved and collector-selected judgments. The unreserved pairs are the collector's remaining decision variables, subject to the quotas. Score separation measures the effect of the resulting assignments after training. The tables compare these learned effects with no reservation and with half the pairs reserved.

ConvNeXt's stream changes -4.68, -3.78, -4.37 give -4.28 [-5.12,-3.48] points. ResNet gives -5.52 [-7.65,-3.61]. Both 50\% reservation rules reduce raw separation while leaving a residual gap between the directional models. ResNet public MSE 0.05203 exceeds the historical-data constant predictor's 0.04965. Quality is compared with each regime's own public models. Assignment opportunity, selected training judgments and learned effects are measured separately.

\subsection{Content quotas with isomorphic sham controls}\label{app:conditional-road}\label{app:sign-audit}
Road's four content groups, also called buckets, are constructed from historical-item features alone. Let $b$ index these groups. The quota matrix $Q=(Q_{ib})$ gives the number of items person $i$ must evaluate within group $b$, as determined by original P1. A \emph{sham} control keeps these quotas and group sizes but changes which items belong to each group. Each fixed item permutation $\tau$ moves group memberships and public assignment masks, with $m'_{i,\tau(p)}=m_{ip}$. This maps feasible assignments one-to-one because Road has complete eligibility and constant panel size. For heterogeneous panels, the permutation must also preserve panel size, demographic composition and eligibility. The actual videos, judgments, forecasts and training order remain attached to their original item IDs. The feature rule, original sham and two additional shams use the original three complete ConvNeXt states and 72 updates. All directional and public assignments are fixed before selected T judgments are retrieved.
For each assignment rule, separation is the mean score under upward routing minus that under downward routing. Let $\rho$ index assignment rules, $+/-$ denote directions, and $\ell$ index training streams. The model's Road rating is $f_{\rho,\pm,\ell}(x_p)$, giving $S_\rho=100\,\operatorname{mean}_{\ell,p\in E}(f_{\rho,+,\ell}(x_p)-f_{\rho,-,\ell}(x_p))$. The original $D_0=S_{\rm feature}-S_{\rm sham0}$ and newer $D_{\rm new}=S_{\rm feature}-(S_{\rm shamA}+S_{\rm shamB})/2$ belong to separate comparison families. A negative value means less separation under the feature rule than under its control. The later interaction study uses all three shams together, without replacing either earlier comparison.
{\small
\begin{table*}[tp]
\centering
\small
\caption{Raw Road separation with three fixed sham maps. Scores use 100-point rating units. Flags use percentage points (pp) at the unchanged 0.5 threshold. Each interval reuses the original paired draws with the same 588 E items and three matched streams.}\label{tab:sign-audit-separation}
\begin{tabular}{lrr}
\toprule Regime & Raw score $S_\rho$ & Raw flag separation\\\midrule

Feature content & $4.99\,[3.56, 6.68]$ & $16.61\,[11.64, 22.05]$\\
Original sham 0 & $6.54\,[5.12, 8.06]$ & $20.98\,[16.29, 26.27]$\\
New sham A & $7.38\,[5.93, 8.89]$ & $22.96\,[17.97, 28.55]$\\
New sham B & $6.65\,[5.02, 8.49]$ & $23.13\,[18.23, 28.66]$\\\bottomrule\end{tabular}
\end{table*}
}

{\small
\begin{table*}[tp]
\centering
\small
\caption{Assignment-rule contrasts under the original C-only scoring rules. $D_{\rm new}$ compares feature separation with the mean of shams A and B. $D_0$ retains the original sham comparison. Negative values indicate reduced separation. Score and flag units differ. Intervals crossing zero describe uncertainty about direction.}\label{tab:sign-audit-rules}
\begin{tabular}{lrrr}
\toprule Rule & $D_0$ score & $D_{\rm new}$ score & $D_{\rm new}$ flag pp\\\midrule

Raw & $-1.55\,[-2.64,-0.57]$ & $-2.02\,[-2.89,-1.24]$ & $-6.43\,[-9.59,-3.39]$\\
Common offset & $-1.55\,[-2.64,-0.57]$ & $-2.02\,[-2.89,-1.24]$ & $-5.33\,[-8.81,-2.53]$\\
Per-model offset & $+0.86\,[-1.17, 2.88]$ & $+0.65\,[-1.10, 2.51]$ & $-2.44\,[-6.04, 0.80]$\\
Nonnegative affine & $+0.78\,[-1.30, 2.71]$ & $+0.62\,[-1.21, 2.26]$ & $-0.68\,[-5.75, 4.97]$\\\bottomrule\end{tabular}
\end{table*}
}

{\small
\begin{table*}[tp]
\centering
\small
\caption{Selected real T supervision and raw E quality across all four regimes. T ratings are on a 100-point scale. Public RMSE averages each of four masks and three matched streams, and each H column averages three streams. All E quality is against the fixed full-48 reference.}\label{tab:sign-audit-quality}
\begin{tabular}{lrrrrrr}
\toprule Regime & T public & T H-up & T H-down & E public RMSE & H-up RMSE & H-down RMSE\\\midrule

Feature content & 36.36 & 40.09 & 33.84 & 0.2045 & 0.2132 & 0.2055\\
Original sham 0 & 36.62 & 40.53 & 33.56 & 0.2065 & 0.2201 & 0.2053\\
New sham A & 36.01 & 40.69 & 32.75 & 0.2047 & 0.2213 & 0.2057\\
New sham B & 35.96 & 40.60 & 33.64 & 0.2072 & 0.2229 & 0.2053\\\bottomrule\end{tabular}
\end{table*}
}
 The original -1.55[-2.64,-0.57] and newer -2.02[-2.89,-1.24] contrasts show reduced raw separation, with a positive gap remaining under every rule. For quality, the upward history-guided model has higher RMSE than its own public models. Comparing public models across rules instead gives feature-minus-original-sham RMSE of -0.001973[-0.005348, 0.001029], with uncertain direction. The first comparison measures the prediction-quality cost of upward routing within an assignment rule. The second compares the public baselines under different rules. All shams and scoring rules appear above.

\subsection{Joint Road and text governance family}\label{app:final-iteration}
The Road interaction study asks whether the effect of content-group quotas depends on which content historical residuals concern. It tests each of four assignment rules under $\eta\in\{\mathrm{intact},\mathrm{HC}_A,\mathrm{HC}_B\}$: intact history and the two within-person historical-content controls. Its 48 new trajectories use three original complete ConvNeXt states, and 24 intact-history condition/stream roles reuse existing models. Let $S_\rho(\eta)$ be raw up--down separation under assignment rule $\rho$ and history condition $\eta$. We first compare feature and sham rules within each history condition: $D(\eta)=S_{\rm feature}(\eta)-\tfrac13\sum_{j=0,A,B}S_{{\rm sham}_j}(\eta)$. We then compare this rule difference under intact history with its mean under the content controls: $I_R=D({\rm intact})-[D({\rm HC_A})+D({\rm HC_B})]/2$.
The text governance study groups S1/S2 items by two-means clustering using H features alone. With the random seed fixed in advance, it selects among 10 restarts by within-cluster squared distance on H. Each person's group counts come from original P1. The two fixed sham permutations exchange only items with identical panel-size, demographic and eligibility requirements. This gives a one-to-one mapping of feasible assignments without moving content, judgments, forecasts or training order. Feature and each sham include Hist up/down and two public assignments under three complete training states: 72 trajectories. S1 uses the downward-routing reference experiment. For S2, Stream 1 uses the public-reference replay state, and Streams 2 and 3 use the supplementary reference experiment (Appendix~\ref{app:scope-ledger}). These recorded reference states define the comparisons. Earlier missing random states remain unavailable.
For text split $d\in\{\mathrm{S1},\mathrm{S2}\}$ and evaluation set $E_d$, let $z_{\rho,\pm,\ell}(x_p)$ be the unsafe logit, or log-odds, under assignment rule $\rho$, direction $+/-$ and stream $\ell$. Separation is $S_d^{\rm text}(\rho)=\operatorname{mean}_{\ell,p\in E_d}[z_{\rho,+,\ell}(x_p)-z_{\rho,-,\ell}(x_p)]$. The feature-minus-sham contrast is $D_d=S_d^{\rm text}({\rm feature})-[S_d^{\rm text}({\rm shamA})+S_d^{\rm text}({\rm shamB})]/2$. All 40 assignments pass current-label access and dummy-label checks before being fixed. Selected T judgments are retrieved afterwards. The joint primary family contains $I_R,D_{S1},D_{S2}$, with two-sided Holm(3), fixed maps and the original saved 2,000/5,000 weights. All three appear together in the main paper. The following tables give each training stream, history and assignment-rule combination, scoring rule and selected-supervision count.
The Road interaction compares rule differences across history conditions. Figure~\ref{fig:history-policy} shows the underlying separations. The interaction's interval is computed from its value in each saved paired draw, preserving dependence among the conditions.
\begin{table*}[tp]
\centering
\caption{Joint tests for the three prespecified governance comparisons. The original point estimates and pointwise 95\% percentile intervals are retained. Holm adjusts the three two-sided centered tests, not interval coverage.}
\label{tab:joint-inference}
\begin{tabular}{llrr}\toprule
Contrast & Unit & Estimate [95\% interval] & Holm $p$\\\midrule
Road history--assignment rule $I_R$ & Rating points & $-2.05\ [-3.36,-0.85]$ & $0.003$\\
Text feature--sham $D_{S1}$ & Unsafe logit & $+0.113\ [-0.010,0.273]$ & $0.219$\\
Text feature--sham $D_{S2}$ & Unsafe logit & $+0.175\ [-0.109,0.359]$ & $0.219$\\\bottomrule
\end{tabular}
\end{table*}
{\small
\begin{table*}[tp]
\centering
\small
\caption{Each training stream in the joint three-primary family. Stream numbers identify the shared training states within each experiment. The cohort and content-group maps remain fixed across streams. Units are listed for each contrast.}
\begin{tabular}{lrrrr}
\toprule Contrast & Unit & Stream 1 & Stream 2 & Stream 3\\\midrule

$I_R$ & points & -1.304 & -2.566 & -2.291\\
$D_{S1}$ & logit & +0.008 & +0.049 & +0.283\\
$D_{S2}$ & logit & +0.301 & +0.346 & -0.121\\
\bottomrule\end{tabular}
\end{table*}
}

{\small
\begin{table*}[tp]
\centering
\small
\caption{Road raw separation $S_\rho(\eta)$, rating points, for every assignment rule and history condition. $D(\eta)$ compares the feature rule with the three-sham mean. The interaction across history conditions is the prespecified primary.}
\begin{tabular}{lrrr}
\toprule Regime & Intact & HC A & HC B\\\midrule

feature & $+4.99\ [3.56, 6.68]$ & $-0.39\ [-0.96, 0.14]$ & $+0.80\ [-0.29, 1.84]$\\
sham 0 & $+6.54\ [5.12, 8.06]$ & $-0.55\ [-1.35, 0.17]$ & $+1.21\ [0.17, 2.29]$\\
sham A & $+7.38\ [5.93, 8.89]$ & $-0.80\ [-1.71, 0.07]$ & $+0.33\ [-0.65, 1.31]$\\
sham B & $+6.65\ [5.02, 8.49]$ & $-0.61\ [-1.38, 0.11]$ & $+0.51\ [-0.52, 1.50]$\\
\midrule $D(\eta)$ & $-1.87\ [-2.80,-1.03]$ & $+0.26\ [-0.53, 1.08]$ & $+0.11\ [-0.56, 0.81]$\\
\bottomrule\end{tabular}
\end{table*}
}

{\small
\begin{table*}[tp]
\centering
\small
\caption{Road interaction under every original scoring rule. All offsets/slopes use C alone and are refitted in each saved C draw. C48 is the separate high-information sensitivity, not pooled with C3. Flags use strict $f>0.5$ per model. Raw does not use a C reference and is printed once. The raw score interaction is the registered primary. The other intervals describe uncertainty for each specified scoring-rule comparison.}
\begin{tabular}{llrr}
\toprule C reference & Rule & Score interaction [95\%] & Flag interaction pp [95\%]\\\midrule

None & raw & $-2.05\ [-3.36,-0.85]$ & $-6.47\ [-10.36,-2.91]$\\
C3 & common offset & $-2.05\ [-3.36,-0.85]$ & $-5.79\ [-9.47,-2.02]$\\
C3 & per model offset & $+1.16\ [-1.04, 3.44]$ & $-2.10\ [-6.58, 2.10]$\\
C3 & nonnegative affine & $+1.14\ [-1.04, 3.14]$ & $+0.27\ [-5.20, 7.73]$\\
C48 & common offset & $-2.05\ [-3.36,-0.85]$ & $-5.69\ [-9.79,-2.10]$\\
C48 & per model offset & $+1.16\ [-1.04, 3.44]$ & $-2.59\ [-7.27, 2.15]$\\
C48 & nonnegative affine & $+1.15\ [-1.09, 3.22]$ & $+0.52\ [-5.68, 8.88]$\\
\bottomrule\end{tabular}
\end{table*}
}

{\small
\begin{table*}[tp]
\centering
\small
\caption{Text up--down separations and feature-minus-sham contrasts. Continuous separation is in unsafe logits, probability separation is on the 0--1 scale, and flag separation is in UNSAFE pp. C is NA for both splits. The prespecified primary uses logit separation. Probability and flag contrasts are secondary.}
\begin{tabular}{llrrrr}
\toprule Split & Quantity & Feature & Sham A & Sham B & Feature minus sham mean\\\midrule

S1 & logit & $+1.265\ [1.170, 1.369]$ & $+1.188\ [0.873, 1.678]$ & $+1.114\ [0.935, 1.370]$ & $+0.113\ [-0.010, 0.273]$\\
S1 & probability & $+0.276\ [0.256, 0.302]$ & $+0.249\ [0.189, 0.345]$ & $+0.237\ [0.203, 0.285]$ & $+0.032\ [0.006, 0.066]$\\
S1 & flag pp & $+62.79\ [36.86, 82.41]$ & $+48.63\ [29.87, 70.28]$ & $+49.09\ [38.15, 60.23]$ & $+13.93\ [-6.75, 29.86]$\\
S2 & logit & $+1.576\ [1.327, 1.814]$ & $+1.475\ [1.391, 1.577]$ & $+1.328\ [1.185, 1.451]$ & $+0.175\ [-0.109, 0.359]$\\
S2 & probability & $+0.314\ [0.266, 0.356]$ & $+0.282\ [0.267, 0.302]$ & $+0.250\ [0.232, 0.265]$ & $+0.048\ [-0.000, 0.080]$\\
S2 & flag pp & $+36.07\ [17.10, 54.66]$ & $+15.75\ [8.43, 24.58]$ & $+8.90\ [4.71, 14.12]$ & $+23.74\ [7.88, 39.43]$\\
\bottomrule\end{tabular}
\end{table*}
}

{\small
\begin{table*}[tp]
\centering
\small
\caption{Selected positive T judgments out of 312 for every text assignment. Public1 and Public2 are the two fixed reference assignments transported across the feature and sham constraints.}
\begin{tabular}{llrrrr}
\toprule Split & Regime & Up & Down & Public1 & Public2\\\midrule

S1 & feature & 132 & 96 & 114 & 110\\
S1 & shamA & 129 & 90 & 112 & 107\\
S1 & shamB & 129 & 96 & 113 & 115\\
S2 & feature & 129 & 77 & 96 & 95\\
S2 & shamA & 127 & 80 & 99 & 108\\
S2 & shamB & 119 & 80 & 93 & 100\\
\bottomrule\end{tabular}
\end{table*}
}
 The tables above report all three primaries across scoring rules and selected supervision. Text has no independent calibration set. Road uses the original C-only rules.

\clearpage
\balance
\subsection{What the quotas remove from a routing objective}\label{app:quota-algebra}
Exact counts prevent gains from asking a generally positive person more questions or adding evaluators to a generally positive item. Write the solver's coefficient as $u_{ip}=a_i+b_p+e_{ip}$, where $a_i$ depends only on the person, $b_p$ only on the item, and $e_{ip}$ on their combination. Quotas fix the additive total $\sum_i r_i a_i+\sum_p k_p b_p$ across assignments. Let $b(p)$ be item $p$'s bucket, $Q_{ib}$ its prescribed count for person $i$, and $c_{ib}$ a person--bucket coefficient. These quotas also fix the contribution $\sum_{ib}Q_{ib}c_{ib}$ from a component $ c_{i,b(p)}$.

These identities concern the solver input after clipping and quantization. Clipping can introduce interactions into initially additive coefficients. Let $B=\sum_i r_i$ be the selected edge count and $\|e\|_\infty=\max_{(i,p)\in\Omega}|e_{ip}|$. Two masks with $B$ selected edges have an objective difference of at most $2B\|e\|_\infty$. If each is restricted to $b\ge0$ new edges of a common baseline, the bound is $4b\|e\|_\infty$. Here $b$ is an edit allowance. Balanced sampling already fixes auxiliary totals\citep{devilleTille2004balanced}. Our experiments test the subsequent effects on selected judgments and learned predictions.

\subsection{From predicted judgments to collected supervision}\label{app:stage-diagnosis}
This diagnosis compares predicted and collected judgments after assignments are fixed, without changing selection or primary comparisons. Let $M^+,M^-$ be the upward and downward masks and $\delta m_{ip}=m^+_{ip}-m^-_{ip}$ their entrywise difference. Let $u_{ip}$ be the forecast after integer rounding and division by the original scale, returning it to judgment units. With recorded judgment $Y_{ip}=y_{ip}$, the true gap $A=\sum_{ip}Y_{ip}\delta m_{ip}$ and predicted gap $P=\sum_{ip}u_{ip}\delta m_{ip}$ differ by prediction error on selected pairs: $A-P=\sum_{ip}(Y_{ip}-u_{ip})\delta m_{ip}$. All 18 frozen pairs satisfy this identity and match saved per-item supervision differences, with maximum numerical residual $1.42\times10^{-14}$. Road measures sums of normalized ratings and text measures positive-judgment counts, before model training.
The text feature/shamA/shamB true gaps are 36/39/33 for S1 and 52/47/39 for S2. Predicted gaps are 10.24/12.25/12.24 and 9.15/10.77/10.55. Feature's true gap equals the sham mean in S1 and exceeds it in S2. These counts locate differences between assignment rules in the collected supervision, before training. The subsequent model-score differences appear in Table~\ref{tab:joint}. Road's intact history gives positive true gaps under all assignment rules. Content maps change prediction error and selected supervision. Original endpoints and uncertainty scopes remain unchanged.
For the learned models, let $I_E^{\rm raw},I_C^{\rm raw}$ denote the history--assignment-rule interaction on E/C raw scores, and $I_E^{\rm offset}$ its per-model-offset E value. The existing 2,000 Road resamples verify $I_E^{\rm offset}=I_E^{\rm raw}-I_C^{\rm raw}$ for C3/C48, with maximum error $3.20\times10^{-13}$ using the same resampling weights. Item and source identifiers have no overlap across assigned data roles. Original playback timing and the single assumed source alias remain as described in Appendix~\ref{sec:road-status}.
  \end{document}